%% file: weak_factor_arXiv.tex
\documentclass[12pt]{article}
\usepackage{amsmath}
\usepackage{graphicx}
\usepackage{enumerate}
\usepackage{xr-hyper}
\usepackage{natbib}
\usepackage{url} % not crucial - just used below for the URL 
\usepackage{subfig}
\usepackage{comment}

\newcommand{\blind}{1}

\makeatletter
\newcommand*{\addFileDependency}[1]{
  \typeout{(#1)}
  \@addtofilelist{#1}
  \IfFileExists{#1}{}{\typeout{No file #1.}}
}
\makeatother

\input{MyPack.tex}

\input{MyMath.tex}

\usepackage{subfig}

\theoremstyle{definition}

\newtheorem*{assumption*}{Assumption}

\theoremstyle{plain}
\newtheorem{theorem}{Theorem}[section]
\newtheorem{proposition}[theorem]{Proposition}
\newtheorem{lemma}[theorem]{Lemma}

\newtheorem{claim}{Claim}

\def \conP {\to_p}
\def \conD {\to_d}
\def \bbE {\mathbb{E}}	\def \calE {\mathcal{E}}

\allowdisplaybreaks

\usepackage{algorithm}
\usepackage{algpseudocode}



\newcommand{\bLambda}{\boldsymbol{\Lambda}}
\newcommand{\bSigma}{\boldsymbol{\Sigma}}
\newcommand{\bPhi}{\boldsymbol{\Phi}}
\newcommand{\bF}{\boldsymbol{F}}
\newcommand{\bX}{\boldsymbol{X}}
\newcommand{\bE}{\boldsymbol{E}}
\newcommand{\bM}{\boldsymbol{M}}
\newcommand{\bGamma}{\boldsymbol{\Gamma}}

\usepackage{minitoc}

\begin{document}

\doparttoc % Tell to minitoc to generate a toc for the parts
\faketableofcontents % Run a fake tableofcontents command for the partocs

\def\spacingset#1{\renewcommand{\baselinestretch}%
{#1}\small\normalsize} \spacingset{1}

%%%%%%%%%%%%%%%%%%%%%%%%%%%%%%%%%%%%%%%%%%%%%%%%%%%%%%%%%%%%%%%%%%%%%%%%%%%%%%

\if1\blind
{
	\title{High Dimensional Factor Analysis with Weak Factors\thanks{This research was supported in part by NSF Grants DMS-2015285 and DMS-2052955. Address for Correspondence: Department of Statistics, Columbia University, 1255 Amsterdam Avenue, New York, NY 10027. Email: \{jc5805, ming.yuan\}@columbia.edu.}
	}
	\author{Jungjun Choi and Ming Yuan\\
		Department of Statistics\\
		Columbia University}
	\maketitle
} \fi

\if0\blind
{
  \title{High Dimensional Factor Analysis with Weak Factors}
	\maketitle
} \fi

\bigskip

\begin{abstract}
%Approximate 
This paper studies the principal components (PC) estimator for high dimensional approximate factor models with weak factors in that the factor loading ($\bLambda^0$) scales sublinearly in the number $N$ of cross-section units, i.e., $\bLambda^{0\top}\bLambda^0 / N^\alpha$ is positive definite in the limit for some $\alpha \in (0,1)$. While the consistency and asymptotic normality of these estimates are by now well known when the factors are strong, i.e., $\alpha=1$, the statistical properties for weak factors remain less explored. Here, we show that the PC estimator maintains consistency and asymptotical normality for any $\alpha\in(0,1)$, provided suitable conditions regarding the dependence structure in the noise are met. This complements earlier result by \cite{onatski2012asymptotics} that the PC estimator is inconsistent when $\alpha=0$, and the more recent work by \cite{bai2023approximate} who established the asymptotic normality of the PC estimator when $\alpha \in (1/2,1)$. Our proof strategy integrates the traditional eigendecomposition-based approach for factor models with leave-one-out analysis similar in spirit to those used in matrix completion and other settings. This combination allows us to deal with factors weaker than the former and at the same time relax the incoherence and independence assumptions often associated with the later.
%We contribute to the literature on the  technique by allowing the dependence in the noise using the leave-neighbor-out estimator and not requiring the incoherence condition. As an extension, we show the asymptotic normality of the TW algorithm in \cite{bai2021matrix} under weak signals. Furthermore, we also study the more general case where $F^{0\top}F^0 / T^\theta$ has a positive definite limit with $\theta \in (0,1]$ where $F^0$ is the factors.
\end{abstract}

\vfill

\part{} % Start the document part

\newpage 

\spacingset{1.6}

\section{Introduction} \label{sec:intro}
Approximate factor models are widely used in diverse fields such as economics, finance, biology, and psychology, to name a few. In these models observations of $N$ cross-section units over $T$ time points are represented as the sum of two unobserved components, a common component driven by systematic factors and an idiosyncratic noise component:
\begin{equation}
\label{eq:model}
x_{it}=\lambda_i^{0\top} f_t^0 +\epsilon_{it},\qquad i=1,\ldots, N, \quad t=1,\ldots,T.
\end{equation}
For many modern applications, of particular interest is the high dimensional setting when both $N$ and $T$ are large. In response, many estimation methods and inferential tools for the latent factors $\bF^0:=(f_1^0,\ldots,f_T^0)^\top$, the loadings $\bLambda^0 =(\lambda_1^0,\ldots,\lambda_N^0)^\top$ and the common component $\bM^0 :=(\lambda_i^{0\top} f_t^0)_{1\le i\le N, 1\le t\le T}$ have been developed. See, e.g., \cite{stock1998diffusion,stock2002forecasting,forni2000generalized,bai2002determining,bai2003inferential,bai2008large,bai2019rank}.

Arguably, the most natural and popular techniques are based on the principal components (PC) and their use can be traced back at least to \cite{connor1986performance,connor1988risk}. Asymptotic properties of PC estimators for large dimensional factor model have also been well studied. See, e.g., \cite{bai2008large} for a recent survey. A common and crucial premise underlying this rich literature is that the factor structure is strong in the sense that both $\bF^{0\top}\bF^0 /T$ and $\bLambda^{0\top}\bLambda^0 /N$ are positive definite in the limit. Although this is a reasonable assumption for some applications, it could be problematic for many others. In the past several years, there has been growing interest in the case when the explanatory power of the factors is weak relative to idiosyncratic noise. See, e.g., \cite{onatski2012asymptotics,onatski2018asymptotics,giglio2021test,uematsu2022estimation,armstrong2022robust,anatolyev2022factor,bai2023approximate}.

To this end, consider a general weak factor structure where $\bF^{0\top}\bF^0 /T$ and $\bLambda^{0\top}\bLambda^0 / N^\alpha$ have positive definite limits for some $\alpha \in (0,1)$. The usual strong factor case corresponds to the choice of $\alpha=1$, under which both consistency and asymptotic normality of the PC estimator are now well known. See, e.g., \cite{bai2003inferential}. On the other hand, \cite{onatski2012asymptotics} showed that the PC estimator is inconsistent when $\alpha = 0$. More recently, \cite{bai2023approximate} established the asymptotic normality of the PC estimator when $\alpha \in (1/2,1)$. However, the inferential theory of the PC estimator when $\alpha \in (0,1/2]$ remains unknown. The main objective of this work is to fill this gap and investigate the consistency and asymptotic normality of PC estimators when $\alpha \in (0,1/2]$.

To fix ideas, let us focus the discussion here on the case when $N\asymp T$ although our main development is more general. Our results indicate that, in particular, if the idiosyncratic terms $\epsilon_{it}$s are cross-sectionally and temporally independent, then the PC estimators of both the factor $f_t^0$ and the common component $m_{it}^0$ are asymptotically normal whenever $\alpha>0$. On the other hand, the asymptotic normality of the estimator for the loading $\lambda_i^0$ may depend on its $\ell_2$ norm,  $\|\lambda_i^0\|$. Specifically, if $$\|\lambda_i^0\|\lesssim N^{(\alpha_i-1)/2}$$
for some $\alpha_i\le 1$, then its PC estimator is asymptotically normal if $\alpha > \alpha_i/2$. Note that if $\|\lambda_i^0\|$ is of the same order across all cross-section index $i$, in other words the loadings are incoherent, then $\alpha_i=\alpha$ and the asymptotic normality holds again whenever $\alpha>0$. Even if $\|\lambda_i^0\|$s are of different orders, as long as they are bounded, i.e., $\alpha_i=1$, we can still derive the inferential theory for all $i$ when $\alpha>1/2$. 

It is worth noting that in deriving the asymptotic normality of $m_{it}^0=\lambda_i^{0\top}f_t^0$, \cite{bai2023approximate} implicitly assume that $\|\lambda_i^0\|$ is bounded away from zero and infinity which amounts to setting $\alpha_i=1$. However, in light of the weak factor structure assumption, only a vanishing proportion of $\|\lambda_i^0\|$s, at most $N^\alpha$ out of $N$, can be bounded away from zero and hence their result can only be applied to small number of factor loadings if any. Our results, on the other hand, can be applied to more factor loadings.

We also investigate the impact of possible dependence among the noise $\epsilon_{it}$s. More specifically, if they are temporally independent but cross-sectionally dependent, we show that the PC estimator of $f_t^0$ is asymptotically normal for all $\alpha\in (0,1]$, whereas PC estimators of both $m_{it}^0$ and $\lambda_i^0$ are asymptotically normal if $\alpha>\max\{1/3,\alpha_i/2\}$. On the other hand, if $\epsilon_{it}$s are cross-sectionally independent but temporally dependent, then we show that PC estimators of both $f_t^0$ and $m_{it}^0$ are asymptotically normal when $\alpha>1/3$ whereas PC estimator of $\lambda_i^0$ is asymptotically normal when $\alpha>\alpha_i/2$. Moreover, if $\epsilon_{it}$s are cross-sectionally and temporally dependent, we show that PC estimators of both $m_{it}^0$ and $\lambda_i^0$ are asymptotically normal if $\alpha>\max\{1/3,\alpha_i/2\}$ while the PC estimator of $f_t^0$ is asymptotically normal when $\alpha > 1/3$.

These results offer an overall picture of the effect of the strength (or weakness) of the factor structure and the potential impact of the dependence structure of the noise terms. In general, to ensure the asymptotic normality of the PC estimates, weaker dependence among the noise is required for weaker factors. 

\begin{table}[htbp]
	\centering
	\begin{tabular}{c|ccc}
		\hline
		\hline
		\multirow{2}[4]{*}{Dependence in the noise} & \multicolumn{3}{c}{Target parameters} \\ [4pt]
		& Loadings ($\lambda^0_i$) & Factors ($f^0_t$) & Common ($m^0_{it}$) \\ [4pt]
		\hline
		Cr ind. \& Tm ind. & $\alpha > \alpha_i/2$  & $\alpha > 0$ & $\alpha > 0$ \\[4pt]
		Cr dp. \& Tm ind. & $\alpha > \max\{1/3,\alpha_i/2\}$ & $\alpha >  0 $ & $\alpha >  \max\{1/3,\alpha_i/2\}$\\[4pt]
  Cr ind. \& Tm dp. & $\alpha > \alpha_i/2$ & $\alpha >  1/3$ & $\alpha >  1/3$ \\[4pt]
		Cr dp. \& Tm dp. &   $\alpha > \max\{1/3,\alpha_i/2\}$  &   $\alpha > 1/3$  &   $\alpha >  \max\{1/3,\alpha_i/2\}$ \\[4pt]
		\hline
	\end{tabular}%
	\raggedright
	\\[2pt]
	\caption{Factor strength requirement to ensure asymptotic normality of PC estimates: Here, `Cr ind.' denotes the cross-sectional independence and `Tm dp.' denotes the temporal dependence. `Cr dp.' and `Tm ind.' are defined similarly.}
	\label{tab:tradeoff}%
\end{table}%

Moreover, a similar pattern can be founded in the condition for consistency. When $\epsilon_{it}$s are temporally independent, all PC estimators are consistent as long as $\alpha > 0 $. On the other hand, if $\epsilon_{it}$s are cross-sectionally independent but temporally dependent, then we show that the PC estimator of $f_t^0$ is consistent when $\alpha > 1/4$ and that of $m_{it}^0$ is consistent if $\alpha >  \max\{0,\alpha_i/4\}$, while that of $\lambda_i^0$ is consistent when $\alpha>0$. In addition, if $\epsilon_{it}$s are cross-sectionally and temporally dependent, we show that the PC estimator of $f_t^0$ is consistent when $\alpha > 1/4$ and that of $m_{it}^0$ is consistent when $\alpha >  \max\{1/7,\alpha_i/4\}$ whereas that of $\lambda_i^0$ is consistent if $\alpha>0$.

\begin{table}[htbp]
	\centering
	\begin{tabular}{c|ccc}
		\hline
		\hline
		\multirow{2}[4]{*}{Dependence in the noise} & \multicolumn{3}{c}{Target parameters} \\ [4pt]
		& \qquad Loadings ($\lambda^0_i$)  & \quad Factors ($f^0_t$) & \quad Common ($m^0_{it}$) \\ [4pt]
		\hline
		Cr ind. \& Tm ind. & $\alpha > 0$  & $\alpha > 0$ & $\alpha > 0$ \\[4pt]
		Cr dp. \& Tm ind. & $\alpha > 0$  & $\alpha >  0 $ & $\alpha >  0$\\[4pt]
  		Cr ind. \& Tm dp. & $\alpha > 0$ & $\alpha >  1/4$ & $\alpha >  \max\{0,\alpha_i/4\}$ \\[4pt]
		Cr dp. \& Tm dp. &   $\alpha >0 $  &   $\alpha > 1/4$  &   $\alpha >  \max\{1/7,\alpha_i/4\}$ \\[4pt]
		\hline
	\end{tabular}%
	\raggedright
	\\[2pt]
	\caption{Factor strength requirement to ensure consistency of PC estimates}
	\label{tab:tradeoff}%
\end{table}%

Our proof strategy combines the traditional approach based on the eigendecomposition of the covariance matrix \citep[see, e.g.,][]{bai2002determining,bai2003inferential,bai2023approximate} with the more recently developed leave-one-out analysis often used in the context of matrix completion \citep[see, e.g.,][]{abbe2020entrywise,ma2020implicit,chen2019inference,chen2020nonconvex,chen2020noisy}. The leave-one-out analysis allows us to derive higher order approximations to the estimation error than the traditional approach which can be used to handle weaker factors. On the other hand, the insights from the traditional approach enables us to do away with the incoherence conditions of the common component and independence assumption of the noise that are often associated with the leave-one-out type of analysis. The technical insights into the advantage of either method may be of independent interests and beneficial to other related problems.

The remainder of this paper is organized as follows. Section \ref{sec:idea} introduces our model and discusses important features of our proof technique in comparison with the traditional approach in \cite{bai2003inferential,bai2023approximate}. Section \ref{sec:asymptotic_indp} presents the asymptotic properties of the PC estimator for general weak factors when idiosyncratic noises are cross-sectionally and temporally independent. It shows the convergence rates of the estimator and the specific conditions for asymptotic normality. In addition, Section \ref{sec:dependent_noise} introduces the leave-neighbor-out technique which allows us to consider the case of dependent noises and studies the asymptotic properties of the PC estimator when the idiosyncratic noises are cross-sectionally or/and temporally dependent. Lastly, we conclude with a few remarks in Section \ref{sec:conclusion}. All proofs are relegated to the Appendix.

%In addition, as an extension of our inferential theory, Section \ref{sec:TWalg} establishes the asymptotic normality of the estimator of the TW algorithm in \cite{bai2021matrix} under the weaker loading assumption. Moreover, Section \ref{sec:general} studies the more general case where $F^{0\top}F^0/T^\theta$ has a positive definite limit with $\theta \in (0,1]$. Section \ref{sec:sim} examines the finite sample performance of our estimators using simulation studies.

In what follows, we use $\norm{\cdot}_{\rm F}$ and $\norm{\cdot}$ to denote the matrix Frobenius norm and spectral norm, respectively. For any vector $a$, $\norm{a}$ denotes its $\ell_2$ norm. $a \lesssim b $ and $b \gtrsim a$ mean $\abs{a}/\abs{b} \leq C$ for some constant $C > 0$. $a \asymp b$ means $a \lesssim b $ and $a \gtrsim b$. In addition, $[K] = \{1 , \dots, K \}$ and $\calO^{r \times r}$ is the set of $r\times r$ orthonormal matrices.

\section{Factor Model and Method of PC}\label{sec:idea}
Denote by $\bX=(x_{it})_{1\le i\le N, 1\le t\le T}$ and $\bE=(\epsilon_{it})_{1\le i\le N, 1\le t\le T}$. Then the approximate factor model \eqref{eq:model} can be expressed in matrix form as:
$$
\bX=\bLambda^0\bF^{0\top}+\bE.
$$
We shall assume that
\paragraph{Assumption A.} [Factors and Loadings]%\label{asp:moment_indp}
\begin{itemize}
\item[(i)] $T^{-1} \sum_{t=1}^T f_t^0 f_t^{0\top} \to_p \bSigma_{\bF}$ where $\bSigma_{\bF}$ is a $r \times r$ positive definite matrix;
\item[(ii)] $N^{-\alpha} \sum_{i=1}^N \lambda_i^0 \lambda_i^{0\top} \to \bSigma_{\bLambda}$ for some $\alpha \in (0,1]$ where $\bSigma_{\bLambda}$ is a $r \times r$ positive definite matrix;
\item[(iii)] The eigenvalues of $\bSigma_{\bLambda} \bSigma_{\bF}$ are distinct;
\item[(iv)] For all $t$, $\bbE \norm{f_t^0}^2  \leq C$ for some constant $C>0$. In addition, for each $i$, there is a parameter $\alpha_i\le 1$ such that for some constant $C'>0$,
$$
\norm{\lambda_i^0} \leq C' N^{(\alpha_i - 1)/2}.
$$
\end{itemize}

Here, $\alpha_i$ designates the order of $\lambda_i^0$. Under the setting $\sum_{i=1}^N \lambda_i^0\lambda_i^{0\top} \asymp N^\alpha$, some of the $\lambda_i^0$s, if not all, should decrease as $N$ increases when $\alpha < 1$. Because $\sum_{i=1}^N \norm{\lambda_i^0}^2 \asymp N^\alpha$ by Assumption A(ii), if the orders of $\lambda_i^0$s are the same across units, we have $\alpha_i = \alpha$ for all $i$. On the other hand, if the orders of $\lambda_i^0$s are heterogeneous, $\lambda_i^0$s would spread around the average order parameter `$\alpha$' due to Assumption A(ii).

Under Assumption A, the common component $\bM^0 = (m_{it}^0)_{1\le i\le N, 1\le t\le T}$ has reduced rank $r$ because the ranks of $\bLambda^0$ and $\bF^0$ are $r$. The method of PC proceeds to estimate $\bLambda^0$ and $\bF^0$ by minimizing the sum of the squared residuals:
\begin{equation}
\label{eq:PCest}
\min_{\bLambda\in \bbR^{N\times r}, \bF\in \bbR^{T\times r}} \left\{{1\over NT}\sum_{i=1}^N\sum_{t=1}^T(x_{it}-\lambda_i^\top f_t)^2\right\},
\end{equation}
subject to the normalization condition that $\bF^\top\bF/T=I_r$ and $\bLambda^\top\bLambda$ is diagonal where $\bF=(f_1,\ldots,f_T)^\top$ and $\bLambda=(\lambda_1,\ldots, \lambda_N)^\top$. The solution to \eqref{eq:PCest}, denoted by $(\widehat{\bF},\widehat{\bLambda})$, can also be expressed in terms of the singular values and vectors of $\bX$. More specifically, let
$$
\bX = \bfU\bfD\bfV^{\top} = \bfU_r \bfD_r \bfV_r^{\top} + \bfU_{N-r} \bfD_{N-r} \bfV_{N-r}^{\top}
$$
be its singular value decomposition where $\bfD_r$ is a diagonal matrix of the top-$r$ singular values, $\bfU_r$, $\bfV_r$ are the corresponding left and right singular values, respectively. Then $\widehat{\bLambda}=T^{-1/2}\bfU_r \bfD_r$ and $\widehat{\bF}=\sqrt{T} \bfV_r$.% This relationship is often exploited to study the asymptotic properties of PC estimates.

\paragraph{Eigenecomposition and Rotation.} %The traditional approach in \cite{bai2002determining,bai2003inferential,bai2023approximate} is based on the eigendecomposition.
Note that $\bX^\top \bX \widehat{\bF} = \widehat{\bF} \bfD_r^2$. We can derive from this identity that
$$
\widehat{\bF}- \bF^{0} \bfH_{\rm BN,0} = \bF^0 \bLambda^{0\top} \bE \widehat{\bF} \bfD_r^{-2} + \bE^{\top} \bLambda^0 \bF^{0\top}  \widehat{\bF}\bfD_r^{-2} + \bE^\top \bE \widehat{\bF}\bfD_r^{-2},
$$
where $\bfH_{\rm BN,0}=\bLambda^{0\top}\bLambda^0\bF^{0\top}\widehat{\bF}\bfD_r^{-2}$; and multiplying both sides of $\bX-\bLambda^0\bF^{0\top}=\bE$ with $\widehat{\bF}/T$ leads to
$$
\widehat{\bLambda} - \bLambda^0 \bfH^{-1}_{\rm BN,1} = \bE\bF^0 \bfH_{\rm BN,0} /T + \bE(\widehat{\bF} - \bF^0 \bfH_{\rm BN,0}) /T,
$$
where $\bfH_{\rm BN,1} = (\bF^{0\top} \widehat{\bF}/T)^{-1}$. See, e.g., \cite{bai2003inferential,bai2023approximate}. These decompositions are key to deriving the asymptotic properties of the PC estimates. For example, it follows immediately that, for any $1\le i\le N$,
$$
T^{1/2}(\widehat{\lambda}_i-\bfH^{-\top}_{\rm BN,1}\lambda_i^{0})=T^{-1/2}\bfH^{\top}_{\rm BN,0}\bF^{0\top}\bfe_i+T^{-1/2}(\widehat{\bF}-\bF^0\bfH_{\rm BN,0})^\top\bfe_i,
$$
where $\bfe_i=(\epsilon_{i1},\ldots,\epsilon_{iT})^\top$. The first term on the right hand side is asymptotically normal by central limit theorem and it therefore suffices to show that the second term is of order $o_p(1)$ to claim the asymptotic normality of PC estimate $\widehat{\lambda}_i$.

To this end, we note that
$$
\bfe_i^\top(\widehat{\bF}-\bF^0\bfH_{\rm BN,0})=\bfe_i^\top\bF^0 \bLambda^{0\top} \bE \widehat{\bF} \bfD_r^{-2} + \bfe_i^\top\bE^{\top} \bLambda^0 \bF^{0\top}  \widehat{\bF}\bfD_r^{-2} + \bfe_i^\top\bE^\top \bE \widehat{\bF}\bfD_r^{-2} .
$$
The conventional approach proceeds to bound each term on the right hand side. More specifically, it can be shown that (see, e.g., \cite{bai2023approximate})
$$
\norm{\bfe_i^\top \bE^{\top} \bLambda^0 \bF^{0\top}  \widehat{\bF}\bfD_r^{-2} } = O_p\left( \frac{T}{N^\alpha} + \frac{\sqrt{T}}{\sqrt{N^\alpha}} \right),
$$
and
$$
\norm{\bfe_i^\top \bE^\top \bE \widehat{\bF}\bfD_r^{-2}} = O_p\left( \frac{T}{N^\alpha} + \frac{N}{N^\alpha} \right).
$$
This, however, means that when $N\asymp T$, $\alpha > 1/2$ is needed to prove asymptotic normality of $\widehat{\lambda}_i$. Interestingly, this requirement is not inherent to the PC estimates themselves but rather due to the limitation of this particular proof technique. We now describe two main ideas that enable us to handle weaker factor structures.

\paragraph{Alternative Matching Matrix.} Our first observation is that the matching matrix $\bfH_{\rm BN}$ can be unduly affected by the noise $\bE$. To alleviate its impact, we shall seek an alternative matching matrix. To this end, consider a balanced version of singular vectors:
$$\bfY_r = \bfU_r \bfD_r^{1/2},\qquad {\rm and}\qquad \bfZ_r = \bfV_r \bfD_r^{1/2}.$$ 
It is clear that $ \widehat{\bLambda} = T^{-1/2}\bfY_r \bfD_r^{1/2}$ and $ \widehat{\bF} = T^{1/2}\bfZ_r \bfD_r^{-1/2}$.

Similarly, let $\bM^0=\bfU^0_r \bfD^0_r \bfV_r^{0\top}$ be its reduced singular value decomposition. Then $\bfY_r$ and $\bfZ_r$ can be viewed as estimates of $\bfY_r^0 = \bfU_r^0 (\bfD_r^{0})^{1/2}$ and $\bfZ_r^0 = \bfV_r^0 (\bfD_r^{0})^{1/2}$ respectively. More importantly, we can find a matrix $\bfH^0\in \bbR^{r\times r}$ that is independent of $\bE$ such that $\bF^0=T^{1/2}\bfZ_r^0\bfH^0$ and $\bLambda^0=T^{-1/2}\bfY_r^0(\bfH^{0})^{-\top}$. We then seek a refinement of $\bfH^0$ by rotating $(\bfY_r,\bfZ_r)$ to match $(\bfY_r^0, \bfZ_r^0)$:
$$
\bfO = \argmin_{\bfR \in \calO^{r\times r}} \left\|
\begin{bmatrix}
	\bfY_r \\
	\bfZ_r
\end{bmatrix}
\bfR - 
\begin{bmatrix}
	\bfY_r^0 \\
	\bfZ_r^0
\end{bmatrix}
\right\|_{\rm F}^2.
$$
Finally, we shall consider a matching matrix $\bfH=(\bfD_r^{1/2}\bfO\bfH^0)^{-1}$. This choice of matching matrix allows us to translate the estimation error of $\widehat{\bLambda}$ and $\widehat{\bF}$ into that of $\bfY_r$ and $\bfZ_r$:
$$
\widehat{\bLambda}-\bLambda^0\bfH^{-\top}=T^{-1/2}(\bfY_r\bfO-\bfY^0_r)\bfO^\top\bfD_r^{1/2},
$$
and
$$
\widehat{\bF}-\bF^0\bfH=T^{1/2}(\bfZ_r\bfO-\bfZ^0_r)\bfO^\top\bfD_r^{-1/2}.
$$

Note that
$$
\bfY_r=\bX\bfZ_r(\bfZ_r^\top \bfZ_r)^{-1}= \bE \bfZ_r(\bfZ_r^\top \bfZ_r)^{-1} + \bfY^0_r\bfZ^{0\top}_r\bfZ_r (\bfZ_r^\top \bfZ_r)^{-1}.
$$
We can write
\begin{gather}\label{eq:Ydecomposition}
\bfY_r \bfO - \bfY_r^0 = \bE \bfZ^0_r(\bfZ_r^{0\top} \bfZ_r^0)^{-1} + \underbrace{\bE (\widetilde{\bfZ}_r (\widetilde{\bfZ}_r^{\top} \widetilde{\bfZ}_r)^{-1} - \bfZ^0_r(\bfZ_r^{0\top} \bfZ_r^0)^{-1}) }_{\coloneqq R_{1}} + \underbrace{\bfY_r^0(\bfZ_r^{0\top}\widetilde{\bfZ}_r(\widetilde{\bfZ}_r^{\top} \widetilde{\bfZ}_r)^{-1} - I_r) }_{\coloneqq R_{2}},
\end{gather}
where $\widetilde{\bfZ}_r = \bfZ_r \bfO$. Similar to before, the first term on the right hand side is asymptotic normal and it suffices to show that the remaining two terms are of smaller order. The last term can be bounded by virtue of Davis-Kahan type of bounds for $\widetilde{\bfZ}_r-\bfZ_r^0$. Bounding the second term turns out to be the key when considering weaker factors ($\alpha\le 1/2$).

\paragraph{Leave-one-out Analysis.}
Recall that, with the new matching matrix, we have
$$
\sqrt{T}(\widehat{\lambda}_i - \bfH^{-1} \lambda_i^0 ) = \bfH^{-1} (\bF^{0\top} \bF^{0}/T)^{-1} \bF^{0\top} \bfe_i/\sqrt{T} + \bfD_r^{1/2} \bfO R_{1,i} +  D_r^{1/2}\bfO R_{2,i},
$$
where $R_{1,i}$ and $R_{1,i}$ are the transpose of $i$-th row of $R_1$ and $R_2$, respectively. Since $\norm{\bfD_r^{1/2}} = O_p(N^{\alpha/4} T^{1/4})$ and $\norm{\widetilde{\bfZ}_r - \bfZ_r^0} = O_p ( N^{-\alpha/4} T^{-1/4}\max\{\sqrt{N},\sqrt{T}\})$, a naive bound for the second term is:
$$
\norm{\bfD_r^{1/2} \bfO R_{1,i}} \leq \norm{\bfD_r^{1/2}}\norm{\bfe_i}\norm{((\widetilde{\bfZ}_r^{\top}\widetilde{\bfZ}_r)^{-1}\widetilde{\bfZ}^{\top}_r - (\bfZ_r^{0 \top} \bfZ_r^0)^{-1}\bfZ_r^{0 \top} )} = O_p\left( \frac{\max\{\sqrt{N},\sqrt{T}\}}{N^{\alpha/2}}\right).
$$
This bound however is not tight. 

Instead, we shall carry out a leave-one-out analysis to decouple the estimates and a particular noise term. Denote by $\bX^{(-i)}$ a $N\times T$ matrix whose $i$-th row is $(m^{0}_{it})_{1\leq t \leq T}$ and other rows are $(x_{jt})_{1\leq t \leq T}$ for all $j \neq i$. That is, $\bX^{(-i)}$ replaces the $i$-th row of $\bX$ with $(m^{0}_{it})_{1\leq t \leq T}$ to remove the noises of the unit $i$. We shall apply the aforementioned operations to $\bX^{(-i)}$ leading to corresponding balanced singular vectors $\bfY_r^{(-i)}$ and $\bfZ_r^{(-i)}$, rotation matrix $\bfO^{(-i)}$, matching matrix $\bfH^{(-i)}$ and etc..

We can then write
\begin{equation}
\label{eq:decompR1}
 R_{1,i}=\sum_{t=1}^T \epsilon_{it} \Delta_{1,t}^{(-i)} + \sum_{t=1}^T \epsilon_{it} \Delta_{2,t}^{(-i)}
\end{equation}
where 
$$\Delta_{1}^{(-i)} = \widetilde{\bfZ}^{(-i)}_r(\widetilde{\bfZ}^{(-i)\top}_r\widetilde{\bfZ}^{(-i)}_r)^{-1} - \bfZ_r^0(\bfZ_r^{0\top} \bfZ_r^0)^{-1}$$
and
$$\Delta_{2}^{(-i)} = \widetilde{\bfZ}_r(\widetilde{\bfZ}^\top_r\widetilde{\bfZ}_r)^{-1} - \widetilde{\bfZ}^{(-i)}_r(\widetilde{\bfZ}^{(-i)\top}_r\widetilde{\bfZ}^{(-i)}_r)^{-1},$$
$\Delta_{1,t}^{(-i)}$ and $\Delta_{2,t}^{(-i)}$ are the transpose of $t$-th row of $\Delta_{1}^{(-i)}$ and $\Delta_{2}^{(-i)}$, respectively. Note that $\Delta_{2,t}^{(-i)}$ is a higher order difference so that the second term on the right-hand side of \eqref{eq:decompR1} is typically negligible. The first term can now be bounded by exploiting the potential independence between $\bfe_i$ and $\Delta_{1,t}^{(-i)}$. For simplicity, consider the case when $\epsilon_{it}$s are cross-sectionally independent, then $\bfe_i$ is independent of $\Delta_1^{(-i)}$. This implies that
\begin{gather}\label{eq:usingindepedence}
 \bbE\left[\left. \left\|\sum_{t=1}^T \epsilon_{it} \Delta_{1,t}^{(-i)}\right\|^2 \right| \Delta_1^{(-i)} \right] \leq \|\Cov(\bfe_i)\| \|\Delta_1^{(-i)}\|_{\rm F}^2 = O_p\left( \frac{\max\{ N, T \}}{N^{3\alpha/2}T^{3/2}}\right).       
\end{gather}
Hence, the first term can be negligible even when $\alpha\le 1/2$.

%\end{assumption}
%
%\section{Asymptotic Properties}\label{sec:asymptotic}
%
%This section introduces asymptotic properties of the PC estimator. Here, we focus on the case where $F^{0\top}F^0/T$ are positive definite in the limit following \cite{onatski2012asymptotics} and \cite{bai2023approximate} to facilitate the comparison. In Section \ref{sec:general}, we consider the more general case where $F^{0\top}F^0/T^\theta$ has a positive definite limit with $\theta \in (0,1]$.

\section{Independent Noise}\label{sec:asymptotic_indp}
We now show how the ideas described in Section \ref{sec:idea} can be used to develop statistical properties for general weak factors. It is instructive to start with the case where the idiosyncratic noises $\epsilon_{it}$s are cross-sectionally and temporally independent.

\paragraph{Assumption B.} [Noise] %$\{\epsilon_{it}\}_{i \leq N, t\leq T}$ is independent across $i$ and $t$. Furthermore $\bbE[\epsilon_{it} | \bM^0] = 0$ and $\bbE[\epsilon_{it}^4] \leq C$ for some constant $C>0$.
\begin{itemize}
\item[(i)] $\bbE[\epsilon_{it} | 
\bM^0] = 0$ and $\bbE[\epsilon_{it}^2 | \bM^0] =  \bbE[\epsilon_{it}^2] \leq C$ for some constant $C>0$. In addition, $(\epsilon_{it})_{i \leq N, t\leq T}$ is independent across $i$ and $t$.
\item[(ii)] With probability converging to 1, $\norm{\bE} \lesssim \max\{\sqrt{N},\sqrt{T} \}$.
\end{itemize}

%Note that the finite fourth moment condition ensures that with probability converging to 1, $\norm{\bE} \lesssim \max\{\sqrt{N},\sqrt{T} \}$. 
We first consider the rate of convergences of the PC estimator defined in the previous section.

\begin{theorem}[Convergence rate of PC estimator]\label{thm:convergence_indp}
Suppose that Assumptions A and B are satisfied. If $\max\{N, T \}=o(N^\alpha T)$, then
\begin{eqnarray*}
\norm{ \widehat{f}_t - \bfH^{\top} f_t^0 } &=& O_p \left( \frac{1}{\sqrt{N^\alpha}} + \frac{\max\{N, T \}}{N^\alpha T} \right),\\
\norm{\widehat{\lambda}_i - \bfH^{-1} \lambda_i^0 } &=& O_p \left( \frac{1}{\sqrt{T}} + \sqrt{\frac{N^{\alpha_i}}{N}} \frac{\max\{N, T \}}{N^\alpha T} \right),\\
\norm{\widehat{m}_{it} - m^0_{it}} &=& O_p\left( \frac{1}{\sqrt{T}} + \sqrt{\frac{N^{\alpha_i}}{N}} \frac{1}{\sqrt{N^\alpha}}  + \sqrt{\frac{N^{\alpha_i}}{N}} \frac{\max\{N, T \}}{N^\alpha T} +  \frac{\max\{\sqrt{N}, \sqrt{T} \}}{\sqrt{N^\alpha} T} \right).
\end{eqnarray*}
\end{theorem}

The convergence rates of $\widehat{\lambda}_i$ and $\widehat{m}_{it}$ depend on the size of $\alpha_i$. As $\alpha_i$ decreases, the estimators converge to the corresponding parameters more quickly. In addition, since $N^{\alpha_i} \leq N$ for all $i$, the condition for the consistency of all three estimators for all $i$ and $t$ is $\alpha > 0$ and $\max\{N, T \}=o(N^\alpha T)$. Hence, if $N \asymp T$, the condition $\alpha > 0$ is sufficient for all estimators to be consistent. On the other hand, the traditional proof method requires $\alpha > 1/3$ for the consistency of $\widehat{f}_t$. Note that the condition $\alpha>0$ is also necessary in light of the results by \cite{onatski2012asymptotics}. 

Next, we present the asymptotic normality of the PC estimator. For this purpose, we need the following assumptions.

\paragraph{Assumption C.} [CLT for weak factors]
As $N,T \rightarrow \infty$,
$$ \frac{1}{\sqrt{N^\alpha}} \sum_{i=1}^N \lambda_i^0 \epsilon_{it} \conD \calN(0,\bPhi_{\bLambda,t}), \qquad{\rm and}\qquad  \frac{1}{\sqrt{T}} \sum_{t=1}^T f_t^0 \epsilon_{it} \conD \calN(0,\bPhi_{\bF,i}),$$
where $\bPhi_{\bLambda,t}$ and $\bPhi_{\bF,i}$ are $r \times r$ positive definite matrices.

For the first CLT, we use a normalization of $N^{\alpha/2}$ instead of $N^{1/2}$ to be consistent with Assumption A(ii). The next assumption presents specific conditions for the size of $\alpha$, $\alpha_i$, $N$ and $T$.

\paragraph{Assumption D.} [Parameter size]
\begin{itemize}
\item[(i)] For the inference of $\lambda_i^0$, we assume that
$$
\frac{\max\{N,T\}}{N^{\alpha}T} \rightarrow 0, \qquad {\rm and}\qquad \frac{\max\{N^2, T^2\}}{N^{(2\alpha - \alpha_i + 1)} T} \rightarrow 0.
$$
\item[(ii)] For the inference of $f_t^0$ and $m_{it}^0$, we assume that
$$
\frac{\max\{N^2, T^2\}}{N^\alpha T^2} \rightarrow 0.
$$
\end{itemize}

The following theorem provides the asymptotic normality of the PC estimator.

\begin{theorem}[CLT for PC estimator]\label{thm:clt_indp}
	Suppose that Assumptions A, B and C are satisfied.
	\begin{itemize}
		\item[(i)] If Assumption D(i) holds, then
		$$
		\sqrt{T} \left( \widehat{\lambda}_i - \bfH^{-1} \lambda_i^0 \right) \to_d \calN\left(0, \calQ^{-\top} \bPhi_{\bF,i} \calQ^{-1} \right),
		$$
		where $\calQ = \calD \calG^{\top} \bSigma_{\bLambda}^{-1/2}$, $\calD$ is the diagonal matrix with the square roots of the eigenvalues of $\bSigma_{\bLambda}^{1/2}\bSigma_{\bF}\bSigma_{\bLambda}^{1/2}$ and $\calG$ is an eigenvector of $\bSigma_{\bLambda}^{1/2}\bSigma_{\bF}\bSigma_{\bLambda}^{1/2}$.
		\item[(ii)] If Assumption D(ii) holds, then
		$$
		\sqrt{N^\alpha} \left( \widehat{f}_t - \bfH^{\top} f_t^0 \right)\to_d \calN\left(0, \calD^{-2} \calQ \bPhi_{\bLambda,t} \calQ^{\top} \calD^{-2} \right),
		$$
		\item[(iii)] If Assumption D(ii) holds and there are constants $c_1,c_2>0$ such that
		$$
		\norm{f^0_t} \geq c_1,\qquad {\rm and} \qquad \norm{\lambda^0_i} \geq c_2 N^{(\alpha_i - 1)/2},
		$$
		with probability tending to one, then
		$$
		\calV_{it}^{-1/2} \left( \widehat{m}_{it} - m^0_{it} \right) \to_d\calN\left(0,1 \right),
		$$
		where 
		$$\calV_{it} =
		\frac{1}{N^{\alpha}} \lambda_i^{0\top} \bSigma_{\bLambda}^{-1} \bPhi_{\bLambda,t} \bSigma_{\bLambda}^{-1} \lambda_i^0
		+ \frac{1}{T} f_t^{0 \top} \bSigma_{\bF}^{-1} \bPhi_{\bF,i} \bSigma_{\bF}^{-1} f_t^0.$$
	\end{itemize}
\end{theorem}

Note that to derive asymptotic normality of $m_{it}^0$, setting lower bounds for $\norm{\lambda_i^0}$ and $\norm{f_t^0}$ is necessary to avoid degenerate variances. In the weak factors setting, it is natural to allow the lower bound of $\norm{\lambda_i^0}$ to decrease as $N \rightarrow \infty$. This is to be contrast with \cite{bai2023approximate} who implicitly assume that $\norm{\lambda_i^0}$ is bounded away from zero while explicitly positing $\bLambda^{0\top} \bLambda^0 \asymp N^\alpha$. In other words, the asymptotic normality they established can only be applied to a vanishing proportion of $\lambda_i^0$s. Indeed, the normalizing constant ($\calV_{it}^{-1/2}$) derived under their assumptions, $O_p\left( \min\{ N^{\alpha/2} , T^{1/2} \}
\right)$, is too small in general. On the other hand, we show here that the correct normalizing constant should be of the order $O_p\left( \min\{ N^{(\alpha+1-\alpha_i)/2} , T^{1/2} \}\right)$. For instance, when $\alpha_i = \alpha$, the order of $\calV_{it}^{-1/2}$ becomes $O_p\left( \min\{ N^{1/2} , T^{1/2} \}\right)$.% So, under our assumption, showing that the residual terms are $o_p(\calV_{it}^{1/2})$ is more challenging.

When $N \asymp T$, for the asymptotic normality of the estimators for $f_t^0$ and $m_{it}^0$ to hold, the condition $\alpha > 0$ is sufficient. However, that of $\lambda_i^0$ requires an additional condition $2 \alpha > \alpha_i$. As noted above, if the orders of $\lambda_i^0$s are the same across units, it is satisfied since $\alpha_i = \alpha$ for all $i$. %Note that $\lambda_i^0$s can be different across units and only their orders are the same.

On the other hand, if the orders of $\lambda_i^0$s are heterogeneous, $\alpha_i$s would spread around `$\alpha$'. In this case, the inferential theory for $\lambda_{i}^0$ is still valid as long as $\alpha_{i}$ is not too much larger than `$\alpha$'. For example, when $\alpha = 1/3$, we can derive the inferential theory for $\lambda_{i}^0$ whose order is $\alpha_{i} < 2/3$. When $\alpha = 1/4$, we can derive the inferential theory for $\lambda_{i}^0$ whose order is $\alpha_{i} < 1/2$. Hence, for the typical $\lambda_{i}^0$ whose order $\alpha_i$ is not too different from `$\alpha$', we can still derive the inferential theory. In addition, we can get the inference of $\lambda_i^0$s whose $\alpha_i$ are the same as or smaller than `$\alpha$' as long as $\alpha > 0$. Since $\alpha_i$s are spread around `$\alpha$', we can expect that a large portion of $\alpha_i$s would be the same as or smaller than `$\alpha$'.

\section{Dependent Noise}\label{sec:dependent_noise}

Now we shall treat the case where the idiosyncratic noises are dependent temporally and/or cross-sectionally. Compared to the independent noise case, the main technical difficulty lies in the leave-one-out analysis: for example, the leave-one-out estimator which excludes noises of the time period $t$ in construction, is no longer independent of noises of the time period $t$ if the noises are temporally dependent. To this end, we shall consider a more general approach that leaves all neighbors of $t$ out.

\paragraph{Leave-neighbor-out Analysis.}
To address this issue, we consider the leave-neighbor-out estimator, which is constructed from hypothetical outcomes that exclude noises of the ‘neighbor’ of time period $t$ from true outcomes. Let $\calN_\delta(t) = (t-\delta, \dots, t, \dots, t + \delta )$ be the $\delta$-neighbor of the time period $t$. Denote by $\bX^{(-\calN_\delta(t))}$ a $N\times T$ matrix whose $s$-th columns with $s \in \calN_\delta(t)$ are $(m^{0}_{is})_{1\leq i \leq N}$ and other columns are $(x_{is'})_{1 \leq i \leq N}$ for all $s' \notin \calN_\delta(t)$. That is, $\bX^{(-\calN_\delta(t))}$ replaces the columns corresponding to $\calN_\delta(t)$ with $(m^{0}_{is})_{1\leq i \leq N}$ to remove the noises of the neighbor of time period $t$. Then, we apply the aforementioned operations to $\bX^{(-\calN_\delta(t))}$ leading to corresponding balanced singular vectors $\bfY_r^{(-\calN_\delta(t))}$, $\bfZ_r^{(-\calN_\delta(t))}$, rotation matrix $\bfO^{(-\calN_\delta(t))}$, and matching matrix $\bfH^{(-\calN_\delta(t))}$.

Similar to the above, the key task is to bound the following term:
\begin{align}\label{eq:usingneighbor}
	\norm{\sum_{i=1}^N \epsilon_{it} \Delta_{1,i}^{(-\calN_\delta(t))}}  &= \norm{\sum_{i=1}^N \left(\epsilon_{it} - \bbE \left[ \epsilon_{it} | (\bfe_{s})_{s \notin \calN_\delta(t)}, \bM^0 \right]\right) \Delta_{1,i}^{(-\calN_\delta(t))}}\\
	\nonumber &  + \norm{\sum_{i=1}^N \left(\bbE\left[\epsilon_{it} | (\bfe_{s})_{s \notin \calN_\delta(t)}, \bM^0 \right] - \bbE\left[\epsilon_{it}\right]\right) \Delta_{1,i}^{(-\calN_\delta(t))}} 
\end{align}
where $\bfe_{s} = (\epsilon_{1s},\dots,\epsilon_{Ns})^\top$ and
$$
\Delta_{1}^{(-\calN_\delta(t))} = \widetilde{\bfY}^{(-\calN_\delta(t))}_r(\widetilde{\bfY}^{(-\calN_\delta(t))\top}_r\widetilde{\bfY}^{(-\calN_\delta(t))}_r)^{-1} - \bfY_r^0(\bfY_r^{0\top} \bfY_r^0)^{-1}.
$$
The first term can be bounded as in \eqref{eq:usingindepedence} by conditioning on $\{(\bfe_{s})_{s \notin \calN_\delta(t)}, \bM^0\}$. On the other hand, the second term is an additional term in the leave-neighbor-out analysis. In the weak temporal dependence case where we can get a tight bound of $\sum_{i=1}^N \left(\bbE[\epsilon_{it} | (\bfe_{s})_{s \notin \calN_\delta(t)} ] - \bbE[\epsilon_{it}]\right)^2$ when $\delta$ grows slowly, the second term can also be bounded tightly. In this manner, we utilize the leave-neighbor-out estimator to allow for temporally dependent noises. Moreover, symmetrically, we can also allow for the cross-sectionally dependent noises.

\subsection{Temporal Dependence}\label{sec:asymptotic_dp}

We first consider the case when $\epsilon_{it}$s are temporally dependent but cross-sectionally independent. To this end, we shall assume the following.

\paragraph{Assumption B'.} [Noise]
\begin{itemize}
\item[(i)] $\bbE[\epsilon_{it}] = 0$, $\bbE[\epsilon_{it}^6] \leq C$ for a constant $C>0$. $\bE$ is independent of $\bM^0$ and $(\bfe_{i})_{i \in [N]}$ is independent across $i$;
\item[(ii)] With probability converging to 1, $\|\bE\| \lesssim \max\{\sqrt{N},\sqrt{T} \}$;
\item[(iii)] There is a constant $C>0$ such that for all $t \in [T]$, $\sum_{s=1}^T |\Cov(\epsilon_{it},\epsilon_{is})| \leq C$, and for all $i \in [N]$, $t,k \in [T]$, $\sum_{s=1}^T |\Cov(\epsilon_{it}\epsilon_{ik},\epsilon_{is}\epsilon_{ik})| \leq C$;
\item[(iv)] There is a sequence $\delta \to \infty$ such that $\delta \asymp (\log N)^\nu$ for some constant $\nu > 0$ and
	\begin{gather*}
		\sum_{j=1}^N \left( \bbE\left[\epsilon_{jt}\left| (\epsilon_{js})_{ s \in \calN_\delta(t)^c}\right. \right] - \bbE[\epsilon_{jt}] \right)^2= O_p(N^{1/3}),
	\end{gather*}
 where $\calN_\delta(t)^c = \{1,\dots, t-\delta-1\} \cup \{t+\delta +1 , \dots, T \}$.
\end{itemize}

For the convergence rate and inferential theory of $\widehat{\lambda}_i$, Assumptions B'(i) -- B'(iii) are sufficient. Assumptions B'(i) and B'(ii) are similar to the conditions for the noise in the independence case, and Assumption B'(iii) is a typical weak temporal dependence assumption. On the other hand, Assumption B'(iv) is an additional weak dependence condition for $\widehat{f}_t$ and $\widehat{m}_{it}$ to utilize the leave-neighbor-out analysis. It requires the dependence between $\epsilon_{it}$ and other $\epsilon_{is}$s on the outside of the neighbor of $t$ to be sufficiently weak. In particular, the following lemma shows that it holds for the usual autoregressive or moving average model.

\begin{lemma}[Examples of Assumption B'(iv)]\label{lem:example}
	\begin{itemize}
		\item[(i)] For each $i \in [N]$, let $\epsilon_{it}$ be a $MA(q)$ process with $q \lesssim (\log N)^\nu$ such that
		$$
		\epsilon_{it} = \sum_{k=0}^q {\phi}^{(i)}_{k} u_{i,t-k} ,
		$$
		where $(u_{is})_{s\in [T]}$ are serially independent white noises. Then, if $\delta = C \lceil (\ln N)^\nu \rceil$ for some large constant $C>0$, we have for all $i \in [N]$,
		$$
		\bbE\left[\epsilon_{it}\left| (\epsilon_{is})_{ s \in \calN_\delta(t)^c}\right. \right] - \bbE[\epsilon_{it}] = 0.
		$$
		\item[(ii)] For each $i$, $\epsilon_{it}$ is a stationary AR(p) process such that
		$$
		\epsilon_{it} = \phi^{(i)}_1 \epsilon_{i,t-1} + \cdots + \phi^{(i)}_p \epsilon_{i,t-p} + u_{it}, \ \ \text{ where } u_{it} \sim i.i.d. \ \  \calN (0,\sigma_{u,i}^2),
		$$
		and there is a constant $0< \vartheta <1$ such that $\max_{1\leq i \leq N,1\leq k\leq p}\abs{\psi^{(i)}_k}< \vartheta $, where $(\psi^{(i)}_1,\dots,\psi^{(i)}_p)$ are the roots of the characteristic polynomial
		$$
		\psi^{p} - \phi_1^{(i)} \psi^{p-1} - \cdots - \phi^{(i)}_{p-1} \psi -  \phi^{(i)}_{p} = 0.
		$$
		If $\delta = C \lceil \ln N \rceil$ for some constant $C >0$, we have
		$$
		\max_i \bbE\left[ \left( \bbE\left[\epsilon_{it}\left| (\epsilon_{is})_{ s \in \calN_\delta(t)^c}\right. \right] - \bbE[\epsilon_{it}]\right)^2  \right] \lesssim N^{-1}.
		$$
	\end{itemize}
\end{lemma}

We are now in position to state the statistical properties of the PC estimates. The following theorem provides the convergence rate of the PC estimator.

\begin{theorem}[Convergence rate of PC estimator]\label{thm:convergence_dp}
Suppose that $\max\{N, T \}(\log N)^{2\nu}=o(N^\alpha T)$. (i) If Assumptions A and B'(i) -- B'(iii) are satisfied, then
$$
		\norm{\widehat{\lambda}_i - \bfH^{-1} \lambda_i^0 } = O_p \left( \frac{1}{\sqrt{T}} + \sqrt{\frac{N^{\alpha_i}}{N}} \frac{\max\{N, T \}}{N^\alpha T} \right).
$$
	(ii) If Assumptions A and B' are satisfied, then
	\begin{eqnarray*}
		\norm{ \widehat{f}_t - \bfH^\top f_t^0 } &=& O_p \left( \frac{1}{\sqrt{N^\alpha}} + \frac{\max\{N, T \}N^{1/6}}{N^\alpha T} + \frac{\sqrt{N} \max\{N^{3/2}, T^{3/2} \}(\log N)^{\nu/2}}{N^{2\alpha} T^{3/2}} \right),\\
		\norm{\widehat{m}_{it} - m^0_{it}} &=& O_p\left( \frac{1}{\sqrt{T}} + \sqrt{\frac{N^{\alpha_i}}{N}} \frac{1}{\sqrt{N^\alpha}}  + \sqrt{\frac{N^{\alpha_i}}{N}} \frac{\max\{N, T \}N^{1/6}}{N^\alpha T} 
		\right.\\
		& &  \qquad   + \frac{\sqrt{N^{\alpha_i}} \max\{N^{3/2}, T^{3/2} \}(\log N)^{\nu/2}}{N^{2\alpha} T^{3/2}} +  \frac{\max\{N, T \}N^{1/6}}{N^\alpha T^{3/2}} \\
           & & \left. \qquad  + \frac{\sqrt{N} \max\{N^{3/2}, T^{3/2} \}(\log N)^{\nu/2}}{N^{2\alpha} T^{2}} \right) .
	\end{eqnarray*}
\end{theorem}

The convergence rate of $\widehat{\lambda}_i$ is the same as that in the independence case. On the other hand, the convergence rates of $\widehat{f}_t$ and $\widehat{m}_{it}$ are quite different from those in the independence case. To fix the idea, if we assume $N \asymp T$, the above results can be reduced to
\begin{eqnarray*}
	\norm{\widehat{\lambda}_i - \bfH^{-1} \lambda_i^0 } &=& O_p \left( \frac{1}{N^{1/2}} + \frac{1}{N^{(1-\alpha_i+ 2\alpha)/2 }} \right),\\
	\norm{ \widehat{f}_t - \bfH^\top f_t^0 } &=& O_p \left( \frac{1}{N^{\alpha/2} } + \frac{1}{N^{(\alpha - 1/6)} } + \frac{(\log N)^{\nu/2}}{N^{ (4\alpha - 1) / 2 } }
	\right),\\
	\norm{\widehat{m}_{it} - m^0_{it}} &=& O_p\left( \frac{1}{N^{1/2}} +  \frac{1}{N^{(1 + \alpha - \alpha_i)/2}}  + \frac{1}{N^{(\alpha - \alpha_i/2 + 1/3)}}
	+ \frac{(\log N)^{\nu/2}}{N^{( 4\alpha - \alpha_i )/2}} + \frac{(\log N)^{\nu/2}}{N^{2\alpha}} \right).
\end{eqnarray*}
Because $\alpha_i \leq 1$ for all $i$, the condition for the consistency of $\widehat{\lambda}_i$ is $\alpha > 0$ as in the independence case. On the other hand, that of $\widehat{f}_t $ becomes $\alpha > 1/4$ if we ignore the logarithmic terms. Moreover, that of $\widehat{m}_{it}$ becomes $\alpha > \max\{0,\alpha_i/4 \}$ in this case. Therefore, the conditions for the consistency of $\widehat{f}_t $ and $\widehat{m}_{it}$ become stronger compared to those in the independence case.

Next, we present the inferential theory for the PC estimator. For this purpose, we need the following assumption.

\paragraph{Assumption D'.} [Parameter size]
\begin{itemize}
\item[(i)] For the inference of $\lambda_i^0$, we assume that
$$
\frac{\max\{N,T\}}{N^{\alpha}T} \rightarrow 0, \qquad {\rm and}\qquad \frac{\max\{N^2, T^2\}}{N^{(2\alpha - \alpha_i + 1)} T} \rightarrow 0.
$$
If $N \asymp T$, it reduces to $\alpha > 0$ and $2\alpha - \alpha_i >0$.
\item[(ii)] For the inference of $f_t^0$, we assume that
\begin{align*}
		&\frac{\max\{N^3,T^3\}(\log N)^{\nu}}{N^{(3\alpha-1)}T^{3}} \rightarrow 0, \quad
		\frac{\max\{ N, T \} (\log N)^{\nu}}{N^{(2\alpha - 2)} T^{3}}\rightarrow 0, \quad
		\frac{\max\{N^2, T^2\} (\log N)^{\nu}}{N^\alpha T^{2}} \rightarrow 0 .
	\end{align*}
 If $N \asymp T$, it reduces to 
	$$
	\frac{(\log N)^{\nu}}{N^{(3\alpha - 1)}}  \rightarrow 0, \qquad {\rm and}\qquad \frac{(\log N)^{2\nu}}{ N^\alpha } \rightarrow 0.
	$$
\item[(iii)] For the inference of $m_{it}^0$, we assume that
 	\begin{align*}
		&\frac{ \max\{N^3,T^3 \} (\log N)^{\nu}}{N^{(3\alpha-1)} T^3  } \rightarrow 0, \qquad {\rm and}\qquad
		\frac{\max\{N^2 , T^2 \}(\log N)^{2\nu}}{ N^\alpha T^2 } \rightarrow 0.
	\end{align*}
	If $N \asymp T$, it reduces to
	\begin{align*}
		&\frac{  (\log N)^{\nu}}{N^{(3\alpha - 1)}  } \rightarrow 0, \qquad {\rm and}\qquad
		\frac{(\log N)^{2\nu}}{ N^\alpha } \rightarrow 0.
	\end{align*}
\end{itemize}

The condition for the inference of $\lambda_i^0$ is the same as that of the independence case. We can derive the inferential theory of $\lambda_i^0$ in the case of $\alpha  > 0$, as long as $\alpha_i$ is not too larger than `$\alpha$'. On the other hand, the condition for the inference of $f_t^0$ and $m_{it}^0$ is different from that of the independence case. If we consider the case of $N \asymp T$ and ignore logarithmic terms, we require $\alpha  > 1/3$ for the inference of $f_t^0$ and $m_{it}^0$.

Here, the requirement $\alpha  > 1/3$ shows that using the leave-neighbor-out analysis is not a free lunch. In deriving a tight bound of
$$
\Delta_{2}^{(-\calN_\delta(t))} = \widetilde{\bfY}_r(\widetilde{\bfY}^\top_r\widetilde{\bfY}_r)^{-1} - \widetilde{\bfY}^{(-\calN_\delta(t))}_r(\widetilde{\bfY}^{(-\calN_\delta(t))\top}_r\widetilde{\bfY}^{(-\calN_\delta(t))}_r)^{-1},
$$
one key challenge is to obtain a tight bound of the following term:
$$
\sum_{i=1}^N \epsilon_{is} Y^{(-\calN_\delta(t))}_{i} \quad \text{for all $s \in \calN_\delta(t) = (t - \delta, \dots, t , \dots, t + \delta )$},
$$
where $Y^{(-\calN_\delta(t))}_{i}$ is the transpose of $i$-th row of $\bfY^{(-\calN_\delta(t))}_r$. Note that if $\calN_\delta(t)= \{t\}$ and there is no temporal dependence in the noises, we can easily get a tight bound by the same token as in \eqref{eq:usingindepedence}. However, when noises are temporally dependent, this way does not work. Moreover, we cannot expliot the method in \eqref{eq:usingneighbor} because, e.g., when $s = t \pm \delta$, $s$ is too close to $\calN_\delta(t)^c$ to use the method in \eqref{eq:usingneighbor}. Specifically, we cannot get a tight bound of $\sum_{i=1}^N \left(\bbE[\epsilon_{is} | (\bfe_{s'} )_{s' \notin \calN_\delta(t)} ] - \bbE[\epsilon_{is}]\right)^2$ if $s = t \pm \delta$.

Instead, we resort to the expansion of $Y^{(-\calN(t))}_{i}$ similar to \eqref{eq:Ydecomposition} to obtain a tight bound of $\sum_{i=1}^N \epsilon_{is} Y^{(-\calN(t))}_{i}$. However, as using the expansion of $\widehat{\bF}$ to derive the asymptotic normality of $\widehat{\bLambda}$ requires stronger factors ($\alpha > 1/2$) in the conventional approach, using expansion of $Y^{(-\calN(t))}_{i}$ makes us to assume the stronger factors, $\alpha > 1/3$. Nonetheless, roughly speaking, we use the expansion in a more indirect way compared to the traditional method. This may be the reason why our condition, $\alpha > 1/3$, is still weaker than that of the conventional approach, $\alpha > 1/2$.

% In addition, our expansion of $Y^{(-\calN(t))}_{i}$ enjoys the small (higher-order) estimation error as long as noises are cross-sectionally independent because we can utilize the leave-one-out analysis.

\begin{theorem}[CLT for PC estimator]\label{thm:clt_dp}
Suppose that Assumptions A and C are satisfied.
\begin{itemize}
\item[(i)] If Assumptions B'(i) -- B'(iii), and D'(i) hold, then
$$
\sqrt{T} \left( \widehat{\lambda}_i - \bfH^{-1} \lambda_i^0 \right)\to_d \calN\left(0, \calQ^{-\top} \bPhi_{\bF,i} \calQ^{-1} \right).
$$
%where $\calQ = \calD \calG^{\top} \bSigma_{\bLambda}^{-1/2}$, $\calD$ is the diagonal matrix with the square roots of the eigenvalues of $\bSigma_{\bLambda}^{1/2}\bSigma_{\bF}\bSigma_{\bLambda}^{1/2}$ and $\calG$ is an eigenvector of $\bSigma_{\bLambda}^{1/2}\bSigma_{\bF}\bSigma_{\bLambda}^{1/2}$.
\item[(ii)] If Assumptions B' and D'(ii) hold, then
$$
\sqrt{N^\alpha} \left( \widehat{f}_t - \bfH^{\top} f_t^0 \right)\to_d \calN\left(0, \calD^{-2} \calQ \bPhi_{\bLambda,t} \calQ^{\top} \calD^{-2} \right).
$$
\item[(iii)] If Assumptions B' and  D'(iii) hold and there are constants $c_1,c_2>0$ such that
$$
\norm{f^0_t} \geq c_1,\qquad {\rm and} \qquad \norm{\lambda^0_i} \geq c_2 N^{(\alpha_i - 1)/2},
$$
with probability tending to one, then
$$
\calV_{it}^{-1/2} \left( \widehat{m}_{it} - m^0_{it} \right) \to_d\calN\left(0,1 \right).
$$
%where 
%$$\calV_{it} =
%\frac{1}{N^{\alpha}} \lambda_i^{0\top} \bSigma_{\bLambda}^{-1} \bPhi_{\bLambda,t} \bSigma_{\bLambda}^{-1} \lambda_i^0
%	+ \frac{1}{T} f_t^{0 \top} \bSigma_{\bF}^{-1} \bPhi_{\bF,i} \bSigma_{\bF}^{-1} f_t^0.$$
\end{itemize}
\end{theorem}

\subsection{Cross-Sectional Dependence Case}\label{sec:asymptotic_cross}

Next, we study the case where the idiosyncratic noises are cross-sectionally dependent. To this end, we first define the $\delta$-neighbor of the unit $i$. For each $i$, let $\calN_\delta(i)$ be a subset of $\{1, \dots, N \}$ that contains the $\delta$ number of units in order of correlation with the noise of $i$. In other words, $\calN_\delta(i)$ consists of $\delta$ units whose noise has a higher correlation with the noise of $i$. It would be a natural counterpart of $\calN_\delta(t)$ in the temporal dependence case.

\paragraph{Assumption A'.} [Factors and Loadings]
\begin{itemize}
\item[(i)] Assumptions A(i) -- A(iii) are satisfied;
\item[(ii)] For all $t \in [T]$, $\bbE\norm{f_t^0}^{2} \leq C$ for some constant $C>0$. Moreover, for each $i \in [N]$, there are parameters $\alpha_{1,i}, \alpha_{2,i} \leq 1$ so that for some constants $C_1,C_2> 0$,
	$$
	\norm{\lambda_i^0} \leq C_1 N^{(\alpha_{1,i} - 1)/2} ,\qquad {\rm and} \qquad \frac{1}{\abs{\calN_\delta(i)}} \sum_{j \in \calN_\delta (i)} \norm{\lambda_j^0}^2 \leq C_2 N^{(\alpha_{2,i} - 1)}.
	$$
\end{itemize}

Here, $\alpha_{2,i}$ designates the order of the average over the neighborhood of $\lambda_i^0$ while $\alpha_{1,i}$ designates the order of $\lambda_i^0$ itself. Note that, if the orders of $\lambda_i^0$s are the same across units, we have $\alpha_{1,i} = \alpha_{2,i} = \alpha$ for all $i$, because $\norm{\lambda_i^0}^2 \asymp N^{\alpha - 1}$ for all $i$ by Assumption A(ii). Otherwise, $\alpha_{1,i}$s and $\alpha_{2,i}$s would spread around the average order parameter `$\alpha$'.

\paragraph{Assumption B''.} [Noise]
\begin{itemize}
\item[(i)] $\bbE[\epsilon_{it}] = 0$, $\bbE[\epsilon_{it}^6] \leq C$ for a constant $C>0$. $\bE$ is independent of $\bM^0$ and $(\bfe_{t})_{t \in [T]}$ is independent across $t$;
\item[(ii)] With probability converging to 1, $\|\bE\| \lesssim \max\{\sqrt{N},\sqrt{T} \}$;
\item[(iii)] There is a constant $C>0$ such that for all $i \in [N]$, $\sum_{j=1}^N |\Cov(\epsilon_{it},\epsilon_{jt})| \leq C$, and for all $t \in [T]$, $i,l \in [N]$, $\sum_{j=1}^N |\Cov(\epsilon_{it}\epsilon_{lt},\epsilon_{jt}\epsilon_{lt})| \leq C$;
\item[(iv)] For each $i \in [N]$, there is a sequence $\delta \to \infty$ such that $\delta \asymp (\log N)^\omega $ for some constant $\omega > 0$ and
	\begin{gather*}
		\sum_{s=1}^T \left( \bbE\left[\epsilon_{is}\left| (\epsilon_{js})_{ j \in \calN_\delta(i)^c}\right. \right] - \bbE[\epsilon_{is}] \right)^2= O_p(T^{1/3}).
	\end{gather*}
\end{itemize}

Basically, this assumption is symmetric to Assumption B'. And similarly, for the convergence rate and inferential theory of $\widehat{f}_t$, Assumptions B''(i) -- B''(iii) are sufficient. On the other hand, Assumption B''(iv) is an additional condition for $\widehat{\lambda}_i$ and $\widehat{m}_{it}$. For example, if the noise of $i$ depends on the noises of at most $O((\log N)^{\omega^*})$ number of other units, it is satisfied with $\omega = \omega^*$ like the moving average process in the temporal dependence case.

The following theorem provides the convergence rate of the PC estimator. 

\begin{theorem}[Convergence rate of PC estimator]\label{thm:convergence_cross}
Suppose that $\max\{N, T \}(\log N)^{2\omega}=o(N^\alpha T)$. (i) If Assumptions A' and B''(i) -- B''(iii) are satisfied, then
	\begin{gather*}
		\norm{ \widehat{f}_t - \bfH^\top f_t^0 } = O_p \left( \frac{1}{\sqrt{N^\alpha}} + \frac{\max\{N, T \}}{N^\alpha T} \right).   
	\end{gather*}
	(ii) If Assumptions A' and B'' are satisfied, then
	\begin{eqnarray*}
		\norm{\widehat{\lambda}_i - \bfH^{-1} \lambda_i^0 } &=& O_p \left( \frac{1}{\sqrt{T}} + \left[ \sqrt{\frac{N^{\alpha_{1,i}}}{N}} + \sqrt{\frac{N^{\alpha_{2,i}}}{N}} \right] \frac{\max\{N, T \}(\log N)^{\omega}}{N^\alpha T } \right.\\
		& & \qquad \left.  + \frac{\max\{\sqrt{N}, \sqrt{T} \}}{N^{\alpha/2} T^{5/6} } + \frac{ \max\{N^{3/2}, T^{3/2} \}(\log N)^{\omega/2}}{N^{3\alpha/2} T^{3/2}}\right),\\
		\norm{\widehat{m}_{it} - m^0_{it}} &=& O_p\left(  \frac{1}{\sqrt{T}}  + \sqrt{\frac{N^{\alpha_{1,i}}}{N}} \frac{1}{\sqrt{N^\alpha}} +  \frac{ \max\{N^{3/2}, T^{3/2} \}(\log N)^{\omega/2}}{N^{3\alpha/2} T^{3/2}} + \frac{\max\{\sqrt{N}, \sqrt{T} \}}{N^{\alpha/2} T^{5/6} }
		\right.\\
		& & \qquad \left. +  \left[ \sqrt{\frac{N^{\alpha_{1,i}}}{N}} + \sqrt{\frac{N^{\alpha_{2,i}}}{N}} \right] \frac{\max\{N, T \}(\log N)^{\omega}}{N^\alpha T} \right) .
	\end{eqnarray*}
\end{theorem}

To fix the idea, if we assume that $N \asymp T$, the above results can be reduced to
\begin{eqnarray*}
	\norm{ \widehat{f}_t - \bfH^\top f_t^0 } &=& O_p \left( \frac{1}{N^{\alpha/2} } \right),\\
	\norm{\widehat{\lambda}_i - \bfH^{-1} \lambda_i^0 } &=& O_p \left( \frac{1}{N^{1/2}} + \frac{1}{N^{(\alpha/2 + 1/3)}} + \frac{(\log N)^{\omega}}{N^{(1-\max\{\alpha_{1,i},\alpha_{2,i}\}+ 2\alpha)/2 }} + \frac{(\log N)^{\omega/2}}{N^{3\alpha/2}} \right),\\
	\norm{\widehat{m}_{it} - m^0_{it}} &=& O_p\left( \frac{1}{N^{1/2}} + \frac{1}{N^{(\alpha/2 + 1/3)}} +  \frac{1}{N^{(1 + \alpha - \alpha_{1,i})/2}} + \frac{(\log N)^{\omega/2}}{N^{3\alpha/2}}  + \frac{(\log N)^{\omega}}{N^{(1 + 2\alpha - \max\{\alpha_{1,i},\alpha_{2,i}\})/2}}
	\right) .
\end{eqnarray*}
The condition for the consistency of $\widehat{f}_t$ is $\alpha > 0$ as in the independence case. In addition, since $\alpha_{1,i}, \alpha_{2,i} \leq 1$ for all $i$, the condition for the consistency of $\widehat{\lambda}_i$ is $\alpha > 0$ if we ignore logarithmic terms. Similarly, the condition for the consistency of $\widehat{m}_{it}$ is $\alpha > 0$. Therefore, if we ignore logarithmic terms, the conditions for all estimators are reduced to $\alpha > 0$ as in the independence case.

Next, we present the inferential theory for the PC estimator. For the asymptotic normality of the estimator, we require the following additional assumption.

\paragraph{Assumption D''.} [Parameter size]
\begin{itemize}
\item[(i)] For the inference of $\lambda_i^0$, we assume that
\begin{align*}
		\frac{\max\{N^3,T^3\}(\log N)^{\omega}}{N^{3\alpha}T^{2}} \rightarrow 0,\qquad
		\frac{\max\{N^2, T^2\} (\log N)^{2\omega}}{N^{(2\alpha - \max\{\alpha_{1,i},\alpha_{2,i}\} + 1)} T} \rightarrow 0,\qquad
		\frac{(\log N)^{2\omega}}{N^{\alpha}} \rightarrow 0.
	\end{align*}
	If $N \asymp T$, it reduces to 
	$$
	\frac{(\log N)^{\omega}}{N^{(3\alpha  - 1)}}  \rightarrow 0,\qquad
	\frac{(\log N)^{2\omega}}{N^{(2 \alpha - \max\{\alpha_{1,i},\alpha_{2,i}\})}} \rightarrow 0,\qquad
	\frac{(\log N)^{2\omega}}{N^{ \alpha}} \rightarrow 0.
	$$
\item[(ii)] For the inference of $f_t^0$, we assume that
$$
	\frac{\max\{N^2, T^2\}}{N^\alpha T^{2}} \rightarrow 0.
	$$
	If $N \asymp T$, it reduces to $\alpha  > 0$.
\item[(iii)] For the inference of $m_{it}^0$, we assume that
 	\begin{align*}
		 \frac{\max\{N^3,T^3\}(\log N)^{\omega}}{N^{3\alpha}T^{2}} \rightarrow 0,\qquad
		\frac{\max\{N^2 , T^2 \}}{N^\alpha T^2 } \rightarrow 0,\qquad
		\frac{\max\{N, T\} (\log N)^{2\omega}}{N^{(2\alpha - \alpha_{2,i} + 1)} } \rightarrow 0 .
	\end{align*}
	If $N \asymp T$, it reduces to
	\begin{align*}
		& \frac{(\log N)^{\omega}}{N^{(3\alpha  - 1)}}  \rightarrow 0, \qquad {\rm and} \qquad
		\frac{(\log N)^{2\omega}}{N^{(2\alpha - \alpha_{2,i} )} } \rightarrow 0 .
	\end{align*}
\end{itemize}

The condition for the inference of $f_t^0$ is the same as that of the independence case and $\alpha  > 0$ is enough. On the other hand, for the inference of $\lambda_i^0$, we require $\alpha  > \frac{1}{3}$ and $2 \alpha - \max\{\alpha_{1,i},\alpha_{2,i}\} > 0$ if we consider the case of $N \asymp T$ and ignore logarithmic terms. Moreover, for the inference of $m_{it}^0$, we need $\alpha  > \frac{1}{3}$ and $2 \alpha - \alpha_{2,i} > 0$. In the homogeneous order case where $\alpha_{1,i} = \alpha_{2,i} = \alpha$, the conditions for inference of $\lambda_i^0$ and $m_{it}^0$ can be reduced to $\alpha  > \frac{1}{3}$. Generally speaking, if $\alpha_{1,i}$ and $\alpha_{2,i}$ are not too far from $\alpha$, $\alpha  > \frac{1}{3}$ is enough for the inference of $\lambda_i^0$ and $m_{it}^0$.

\begin{theorem}[CLT for PC estimator]\label{thm:clt_cross}
Suppose that Assumptions A' and C are satisfied.
\begin{itemize}
\item[(i)] If Assumptions B'' and D''(i) hold, then
$$
\sqrt{T} \left( \widehat{\lambda}_i - \bfH^{-1} \lambda_i^0 \right)\to_d \calN\left(0, (\calQ^{\top})^{-1} \bPhi_{\bF,i} \calQ^{-1} \right).
$$
%where $\calQ = \calD \calG^{\top} \bSigma_{\bLambda}^{-1/2}$, $\calD$ is the diagonal matrix with the square roots of the eigenvalues of $\bSigma_{\bLambda}^{1/2}\bSigma_{\bF}\bSigma_{\bLambda}^{1/2}$ and $\calG$ is an eigenvector of $\bSigma_{\bLambda}^{1/2}\bSigma_{\bF}\bSigma_{\bLambda}^{1/2}$.
\item[(ii)] If Assumptions B''(i) -- B''(iii) and D''(ii) hold, then
$$
\sqrt{N^\alpha} \left( \widehat{f}_t - \bfH^{\top} f_t^0 \right)\to_d \calN\left(0, \calD^{-2} \calQ \bPhi_{\bLambda,t} \calQ^{\top} \calD^{-2} \right).
$$
\item[(iii)] If Assumptions B'' and  D''(iii) hold and there are constants $c_1,c_2>0$ such that
$$
\norm{f^0_t} \geq c_1,\qquad {\rm and} \qquad \norm{\lambda^0_i} \geq c_2 N^{(\alpha_{1,i} - 1)/2},
$$
with probability tending to one, then
$$
\calV_{it}^{-1/2} \left( \widehat{m}_{it} - m^0_{it} \right) \to_d\calN\left(0,1 \right).
$$
%where 
%$$\calV_{it} =
%\frac{1}{N^{\alpha}} \lambda_i^{0\top} \bSigma_{\bLambda}^{-1} \bPhi_{\bLambda,t} \bSigma_{\bLambda}^{-1} \lambda_i^0
%	+ \frac{1}{T} f_t^{0 \top} \bSigma_{\bF}^{-1} \bPhi_{\bF,i} \bSigma_{\bF}^{-1} f_t^0.$$
\end{itemize}
\end{theorem}

\subsection{General Dependence Case}\label{sec:asymptotic_general}

Lastly, we study the case where the idiosyncratic noises are cross-sectionally and temporally dependent. To this end, let $\calN_{\delta_1}(i)$ and $\calN_{\delta_2}(t)$ be the $\delta_1$-neighbor of the unit $i$ and $\delta_2$-neighbor of the time period $t$, respectively.

\paragraph{Assumption B'''.} [Noise]
\begin{itemize}
\item[(i)] $\bbE[\epsilon_{it}] = 0$, $\bbE[\epsilon_{it}^6] \leq C$ for a constant $C>0$. $\bE$ is independent of $\bM^0$;
\item[(ii)] With probability converging to 1, $\|\bE\| \lesssim \max\{\sqrt{N},\sqrt{T} \}$;
\item[(iii)] There is a constant $C>0$ such that for all $i,l \in [N]$ and $t,k \in [T]$, 
\begin{align*}
 &\sum_{j=1}^N \sum_{s=1}^T |\Cov(\epsilon_{it},\epsilon_{js})| \leq C ,\ \
 \sum_{j=1}^N \sum_{s=1}^T |\Cov(\epsilon_{it}\epsilon_{lt},\epsilon_{js}\epsilon_{ls})| \leq C,\ \
 \sum_{j=1}^N \sum_{s=1}^T |\Cov(\epsilon_{it}\epsilon_{ik},\epsilon_{js}\epsilon_{jk})| \leq C;
\end{align*}
\item[(iv)] For each $i \in [N]$ and $t \in [T]$, there are sequences $\delta_1,\delta_2 \to \infty$ such that $\delta_1 \asymp (\log N)^\omega $ for some constant $\omega > 0$ and $\delta_2 \asymp (\log N)^\nu $ for some constant $\nu > 0$,
	\begin{gather*}
		\sum_{s=1}^T \left( \bbE\left[\epsilon_{is}\left| (\bfe_{j})_{ j \in \calN_{\delta_1}(i)^c}\right. \right] - \bbE[\epsilon_{is}] \right)^2= O_p(T^{1/3}),\\
            \sum_{j=1}^N \left( \bbE\left[\epsilon_{jt}\left| (\bfe_{s})_{ s \in \calN_{\delta_2}(t)^c}\right. \right] - \bbE[\epsilon_{jt}] \right)^2= O_p(N^{1/3});
	\end{gather*}
\item[(v)] For each $i \in [N]$, we have
$$
  \max_{1\leq t \leq T}  \sum_{s=1}^T \abs{ \Cov\left(\epsilon_{it}, \epsilon_{is} \left| (\bfe_{j})_{ j \in \calN_{\delta_1}(j)^c} \right. \right)} =O_p \left( \max_{1\leq t \leq T} \Var \left(\epsilon_{it} \left| (\bfe_{j})_{ j \in \calN_{\delta_1}(t)^c} \right. \right) \right);
$$
In addition, for each $t \in [T]$, we have
$$
  \max_{1\leq i \leq N}  \sum_{j=1}^N \abs{ \Cov\left(\epsilon_{it}, \epsilon_{jt}\left| (\bfe_{s})_{ s \in \calN_{\delta_2}(t)^c} \right. \right)} =O_p \left( \max_{1\leq i \leq N}\Var \left(\epsilon_{it} \left| (\bfe_{s})_{ s \in \calN_{\delta_2}(t)^c} \right. \right) \right).
$$
\end{itemize}
Assumptions B'''(i) - (iv) are the generalizations of Assumptions B' and B''. If we assume cross-sectional or temporal independence, these assumptions reduce to Assumptions B' or B'', respectively. Here, Assumption B'''(iii) can be comparable to Assumption A3 in \cite{bai2023approximate}. To show Assumption A3 in \cite{bai2023approximate}, one may need Assumption B'''(iii). On the other hand, Assumption B'''(v) is a new condition. It requires the conditional weak dependence conditioning on noises of the outside of the neighbor. In the special case where noises are dependent on a finite number of other noises, it reduces to the unconditional weak dependence and is satisfied by Assumption B'''(iii).

Then, the following theorem provides the convergence rate of the PC estimator. 

\begin{theorem}[Convergence rate of PC estimator]\label{thm:convergence_general}
Suppose that $\max\{N, T \}(\log N)^{2\max\{\omega,\nu\}}=o(N^\alpha T)$. If Assumptions A' and B''' are satisfied, then
	\begin{align*}
		&\norm{\widehat{\lambda}_i - \bfH^{-1} \lambda_i^0 } = O_p \left( \frac{1}{\sqrt{T}} + \left[ \sqrt{\frac{N^{\alpha_{1,i}}}{N}} + \sqrt{\frac{N^{\alpha_{2,i}}}{N}} \right] \frac{\max\{N, T \}(\log N)^{\omega}}{N^\alpha T } \right.\\
		&  \qquad\qquad\qquad\qquad\qquad \left.  + \frac{\max\{\sqrt{N}, \sqrt{T} \}}{N^{\alpha/2} T^{5/6} } + \frac{ \max\{N^{3/2}, T^{3/2} \}(\log N)^{\omega/2}}{N^{3\alpha/2} T^{3/2}}\right),\\
  &\norm{ \widehat{f}_t - \bfH^\top f_t^0 } = O_p \left( \frac{1}{\sqrt{N^\alpha}} + \frac{\max\{N, T \}N^{1/6}}{N^\alpha T} + \frac{\sqrt{N} \max\{N^{3/2}, T^{3/2} \}(\log N)^{\nu/2}}{N^{2\alpha} T^{3/2}} \right),\\
  & \norm{\widehat{m}_{it} - m^0_{it}}\\
  & = O_p\left(  \frac{1}{\sqrt{T}}  + \sqrt{\frac{N^{\alpha_{1,i}}}{N}} \frac{1}{\sqrt{N^\alpha}} +  \frac{ \max\{N^{2}, T^{2} \}(\log N)^{(\omega+\nu)/2}}{N^{3\alpha/2} T^{2}}  +   \sqrt{\frac{N^{\alpha_{1,i}}}{N}}\frac{\max\{N, T \}N^{1/6}}{N^\alpha T}
		\right.\\
&  \qquad\quad \left. +   \sqrt{\frac{N^{\alpha_{2,i}}}{N}}  \frac{\max\{N, T \}(\log N)^{\omega}}{N^\alpha T}  + \frac{\sqrt{N^{\alpha_{1,i}}} \max\{N^{3/2}, T^{3/2} \}(\log N)^{\nu/2}}{N^{2\alpha} T^{3/2}} + \frac{\max\{N^{2/3}, T^{2/3} \}}{N^{\alpha/2} T } \right.\\
 &  \qquad\quad   +  \sqrt{\frac{N^{\alpha_{2,i}}}{N}}  \frac{\max\{N^2, T^2 \}N^{1/6}(\log N)^{\omega}}{N^{2\alpha} T^2 } +  \frac{\max\{N^{3/2}, T^{3/2} \}N^{1/6}}{N^{3\alpha/2} T^{11/6}} 
 \\
  &  \qquad\quad  +  \frac{\max\{N^{5/2}, T^{5/2} \}N^{1/6}(\log N)^{\omega/2}}{N^{5\alpha/2} T^{5/2}} +     \frac{\sqrt{N^{\alpha_{2,i}}}\max\{N^{5/2}, T^{5/2} \}(\log N)^{\nu/2+\omega}}{N^{3\alpha} T^{5/2}}  \\
   &  \left. \qquad\quad  
    +  \frac{\sqrt{N}\max\{N^{3}, T^{3} \}(\log N)^{(\omega+\nu)/2}}{N^{7\alpha/2} T^{3}}  +  \frac{\sqrt{N}\max\{N^{2}, T^{2} \}(\log N)^{\nu/2}}{N^{5\alpha/2} T^{7/3}}
   \right) .
  \end{align*}
\end{theorem}
To fix the idea, if we assume that $N \asymp T$, the above results can be reduced to
\begin{align*}
	&\norm{\widehat{\lambda}_i - \bfH^{-1} \lambda_i^0 } = O_p \left( \frac{1}{N^{1/2}} + \frac{1}{N^{(\alpha/2 + 1/3)}} + \frac{(\log N)^{\omega}}{N^{(1-\max\{\alpha_{1,i},\alpha_{2,i}\}+ 2\alpha)/2 }} + \frac{(\log N)^{\omega/2}}{N^{3\alpha/2}} \right),\\
 &\norm{ \widehat{f}_t - \bfH^\top f_t^0 } = O_p \left( \frac{1}{N^{\alpha/2} } + \frac{1}{N^{(\alpha - 1/6)} } + \frac{(\log N)^{\nu/2}}{N^{ (4\alpha - 1) / 2 } }
	\right),\\
	&\norm{\widehat{m}_{it} - m^0_{it}} \\
 &= O_p\left( \frac{1}{N^{1/2}} + \frac{1}{N^{(\alpha/2 + 1/3)}} +  \frac{1}{N^{(1 + \alpha - \alpha_{1,i})/2}} + \frac{(\log N)^{(\omega+\nu)/2}}{N^{3\alpha/2}} + \frac{(\log N)^{\omega}}{N^{(1 + 2\alpha -\alpha_{2,i})/2}}  + \frac{1}{N^{(\alpha - \alpha_{1,i}/2 + 1/3)}} \right.\\
& \left. \quad \qquad    + \frac{(\log N)^{\nu/2}}{N^{( 4\alpha - \alpha_{1,i} )/2}}  + \frac{(\log N)^{\omega}}{N^{(2\alpha -\alpha_{2,i}/2 + 1/3)}} + \frac{(\log N)^{(\omega+\nu)/2}}{N^{(5\alpha/2 - 1/6)}}  + \frac{(\log N)^{\omega+\nu/2}}{N^{(3\alpha - \alpha_{2,i}/2)}}  + \frac{(\log N)^{(\omega+\nu)/2}}{N^{(7\alpha/2 - 1/2)}} 
	\right) .
\end{align*}
Because $\alpha_{1,i}, \alpha_{2,i} \leq 1$ for all $i$, the condition for the consistency of $\widehat{\lambda}_i$ is $\alpha > 0$ if we ignore logarithmic terms. In addition, that of $\widehat{f}_t $ is $\alpha > 1/4$ when we ignore the logarithmic terms. On the other hand, the condition for the consistency of $\widehat{m}_{it}$ is somewhat complicated. For the consistency, we require $\alpha > \max\{1/7, \alpha_{1,i}/4 , \alpha_{2,i}/6 \}$ if we ignore the logarithmic terms.

\paragraph{Assumption D'''.} [Parameter size]
\begin{itemize}
\item[(i)] For the inference of $\lambda_i^0$, we assume that
\begin{align*}
		\frac{\max\{N^3,T^3\}(\log N)^{\omega}}{N^{3\alpha}T^{2}} \rightarrow 0,\qquad
		\frac{\max\{N^2, T^2\} (\log N)^{2\omega}}{N^{(2\alpha - \max\{\alpha_{1,i},\alpha_{2,i}\} + 1)} T} \rightarrow 0,\qquad
		\frac{(\log N)^{2\omega}}{N^{\alpha}} \rightarrow 0.
	\end{align*}
	If $N \asymp T$, it reduces to 
	$$
	\frac{(\log N)^{\omega}}{N^{(3\alpha  - 1)}}  \rightarrow 0,\qquad
	\frac{(\log N)^{2\omega}}{N^{(2 \alpha - \max\{\alpha_{1,i},\alpha_{2,i}\})}} \rightarrow 0,\qquad
	\frac{(\log N)^{2\omega}}{N^{ \alpha}} \rightarrow 0.
	$$
\item[(ii)] For the inference of $f_t^0$, we assume that
\begin{align*}
		&\frac{\max\{N^3,T^3\}(\log N)^{\nu}}{N^{(3\alpha-1)}T^{3}} \rightarrow 0, \quad
		\frac{\max\{ N, T \} (\log N)^{\nu}}{N^{(2\alpha - 2)} T^{3}}\rightarrow 0, \quad
		\frac{\max\{N^2, T^2\} (\log N)^{\nu}}{N^\alpha T^{2}} \rightarrow 0 .
	\end{align*}
 If $N \asymp T$, it reduces to 
	$$
	\frac{(\log N)^{\nu}}{N^{(3\alpha - 1)}}  \rightarrow 0, \qquad {\rm and}\qquad \frac{(\log N)^{2\nu}}{ N^\alpha } \rightarrow 0.
	$$
\item[(iii)] For the inference of $m_{it}^0$, we assume that
 	\begin{align*}
		&\frac{\max\{N^4,T^4\}(\log N)^{\omega + \nu}}{N^{3\alpha}T^{3}} \rightarrow 0,\qquad  \frac{\max\{N^{3/2},T^{3/2}\}}{N^{2\alpha}T}\rightarrow 0,\qquad
		 \frac{\max\{N, T\} (\log N)^{2\omega}}{N^{(2\alpha - \alpha_{2,i} + 1)} } \rightarrow 0.
	\end{align*}
	If $N \asymp T$, it reduces to
	\begin{align*}
		& \frac{(\log N)^{\omega+\nu}}{N^{(3\alpha - 1 )} } \rightarrow 0,\qquad 
		\frac{(\log N)^{2\omega}}{N^{(2\alpha - \alpha_{2,i} )} } \rightarrow 0 .
	\end{align*}
\end{itemize}

We then have the following asymptotic normality.

\begin{theorem}[CLT for PC estimator]\label{thm:clt_general}
	Suppose that Assumptions A', B''' and C are satisfied.
	\begin{itemize}
		\item[(i)] If Assumption D'''(i) hold, then
		$$
		\sqrt{T} \left( \widehat{\lambda}_i - \bfH^{-1} \lambda_i^0 \right)\to_d \calN\left(0, (\calQ^{\top})^{-1} \bPhi_{\bF,i} \calQ^{-1} \right).
		$$
%		where $\calQ = \calD \calG^{\top} \bSigma_{\bLambda}^{-1/2}$, $\calD$ is the diagonal matrix with the square roots of the eigenvalues of $\bSigma_{\bLambda}^{1/2}\bSigma_{\bF}\bSigma_{\bLambda}^{1/2}$ and $\calG$ is an eigenvector of $\bSigma_{\bLambda}^{1/2}\bSigma_{\bF}\bSigma_{\bLambda}^{1/2}$.
		\item[(ii)] If Assumption D'''(ii) hold, then
		$$
		\sqrt{N^\alpha} \left( \widehat{f}_t - \bfH^{\top} f_t^0 \right)\to_d \calN\left(0, \calD^{-2} \calQ \bPhi_{\bLambda,t} \calQ^{\top} \calD^{-2} \right).
		$$
		\item[(iii)] If Assumption D'''(iii) hold and there are constants $c_1,c_2>0$ such that
		$$
		\norm{f^0_t} \geq c_1,\qquad {\rm and} \qquad \norm{\lambda^0_i} \geq c_2 N^{(\alpha_{1,i} - 1)/2},
		$$
		with probability tending to one, then
		$$
		\calV_{it}^{-1/2} \left( \widehat{m}_{it} - m^0_{it} \right) \to_d\calN\left(0,1 \right).
		$$
%		where 
%		$$\calV_{it} =
%		\frac{1}{N^{\alpha}} \lambda_i^{0\top} \bSigma_{\bLambda}^{-1} \bPhi_{\bLambda,t} \bSigma_{\bLambda}^{-1} \lambda_i^0
%		+ \frac{1}{T} f_t^{0 \top} \bSigma_{\bF}^{-1} \bPhi_{\bF,i} \bSigma_{\bF}^{-1} f_t^0.$$
	\end{itemize}
\end{theorem}

The condition for the inference of $f_t^0$ is the same as that of the temporal dependence case and we require $\alpha  > 1/3$. Similarly, the condition for $\lambda_i^0$ is the same as that of the cross-sectional dependence case. For the inference of $\lambda_i^0$, we require $\alpha  > \frac{1}{3}$ and $2 \alpha - \max\{\alpha_{1,i},\alpha_{2,i}\} > 0$ if we consider the case of $N \asymp T$ and ignore logarithmic terms. Moreover, for the inference of $m_{it}^0$, we need $ \alpha  > \max\{1/3, \alpha_{2,i}/2 \}$. Hence, if $\alpha_{1,i}$ and $\alpha_{2,i}$ are not too far from $\alpha$, $\alpha  > \frac{1}{3}$ is enough for the inference of $\lambda_i^0$ and $m_{it}^0$.

Lastly, it is noteworthy that our requirement of $\alpha$ for inference ($\alpha>1/3$) is weaker than that in \cite{bai2023approximate}. However, this improvement is achieved at the cost of more restrictions in the dependence structure in the noise.

\section{Concluding Remarks}\label{sec:conclusion}

This paper investigates the asymptotic properties of the PC estimator for high dimensional approximate factor model with weak factors. Assuming that $\bLambda^{0\top}\bLambda^0 / N^\alpha$ has a positive definite limit, we establish the consistency and asymptotic normality of the PC estimator for $\alpha \in (0,1)$, under some conditions about the dependence structure in the noise. In particular, we show the asymptotic normality of the estimator when $\alpha \in (0,1/2]$, which has not yet been clarified in the literature. Our proof method combines the conventional approach based on the eigendecomposition of the covariance matrix with the more recently developed leave-one-out analysis. However, unlike the existing literature using the leave-one-out technique, we allow for the dependence in the noises by exploiting the leave-neighbor-out estimator and do not require the incoherence condition, which is a common assumption in the literature. The technical understanding of this generalization may have independent value and be useful for other related issues.

\bibliographystyle{apalike}
\bibliography{weakfactor}

%%%%%%%%%%%%%%%%%%%%%%%%%%%%%%%%%%%%%%%%%%%%%%%%%%%%

\newpage

\appendix

%{\LARGE 
%\begin{center}
%    APPENDIX
%\end{center}
%}

\addcontentsline{toc}{section}{Appendix} % Add the appendix text to the document TOC
\part{Appendix} % Start the appendix part
\parttoc % Insert the appendix TOC

\newpage 

\section{Expansions for the Balanced Singular Vectors}

We shall first present the expansions for the balanced singular vectors $\bfY_r$ and $\bfZ_r$ since these expansions serve as the main tool for deriving the asymptotic normality of the PC estimator.

\subsection{Independence Case}

First, we consider the case where the idiosyncratic noises are cross-sectionally and temporally independent. Denote the $i$-th row of $\bfY_r$, $\bfY_r^0$ and $\bfU_r^0$ by $Y_{i}^\top$, $Y_i^{0 \top}$ and $u^{0 \top}_i$, respectively. Denote the $t$-th row of $\bfZ_r$, $\bfZ_r^0$ and $\bfV_r^0$ by $Z_{t}^\top$, $Z_t^{0 \top}$ and $v^{0 \top}_t$, respectively. Furthermore, the largest and smallest nonzero singular values of $\bM^0$ are denoted by $\psi_{\max}$ and $\psi_{\min}$. The condition number is $\kappa = \psi_{\max}/\psi_{\min}$.

\paragraph{Assumption E.} [Singular vector]
\begin{itemize}
\item[(i)] For the expansion of $Y_{i}$, we assume there is $\rho_i > 0$ such that $\norm{u_i^0} = O_p (\rho_i)$.
\item[(ii)] For the expansion of $Z_{t}$, we assume there is $q_t > 0$ such that $\norm{v_t^0} = O_p (q_t)$.
\end{itemize}

\paragraph{Assumption F.} [Parameter size]
\begin{itemize}
\item[(i)] For the expansion of $Y_{i}$, we assume 
$$
\frac{ \kappa^{5/2} \max\{\sqrt{N},\sqrt{T}\}}{ \psi_{\min}} \conP 0, \qquad \frac{ \kappa^{6} \rho_i \max\{N,T\}}{ r^{1/2} \psi_{\min}} \conP 0, \qquad  \text{and} \qquad  r^{1/2} \kappa^2 \rho_i \conP 0 .
$$ 
\item[(ii)] For the expansion of $Z_{t}$, we assume 
$$
\frac{ \kappa^{5/2} \max\{\sqrt{N},\sqrt{T}\}}{ \psi_{\min}} \conP 0, \qquad \frac{\kappa^{6} q_t \max\{N,T\}}{ r^{1/2} \psi_{\min}} \conP 0, \qquad  \text{and} \qquad   r^{1/2} \kappa^2 q_t \conP 0 .
$$ 
\end{itemize}

\begin{proposition}\label{pro:pre_indp}
(i) Suppose that Assumptions B and E(i) are satisfied. In addition, assume that $\kappa \rho_i \conP 0$ and $\frac{ \max\{\sqrt{N},\sqrt{T}\}}{ \psi_{\min}} \conP 0$. Then, we have
$$
\bfO^{\top} Y_{i} - Y^{0}_i = (\bfZ_r^{0 \top} \bfZ_r^0)^{-1} \sum_{t=1}^T \epsilon_{it} Z_t^0 + \calR_{y,i},
$$
where 
\begin{align*}
\norm{\calR_{y,i}} = O_p \left( \frac{ \kappa^{11/2} \rho_i \max\{N,T\}}{\psi_{\min}^{3/2}}  + \frac{ \kappa^2 r^{1/2} \max\{\sqrt{N},\sqrt{T}\}}{\psi_{\min}^{3/2}} + \frac{ \kappa^{3/2} r  \rho_i }{\sqrt{\psi_{\min}}} 
 \right).   
\end{align*}
Additionally, if we assume Assumption F(i), we have $\norm{\calR_{y,i}} = o_p\left(\frac{ r^{1/2}}{\sqrt{\psi_{\max}}} \right)$.\\
(ii) Suppose that Assumptions B and E(ii) are satisfied. In addition, assume that $\kappa q_t \conP 0$ and $\frac{ \max\{\sqrt{N},\sqrt{T}\}}{ \psi_{\min}} \conP 0$. Then, we have
$$
\bfO^{\top} Z_{t} - Z^{0}_t = (\bfY_r^{0\top} \bfY_r^0)^{-1} \sum_{i=1}^N \epsilon_{it} Y_i^0 + \calR_{z,t},
$$
where 
\begin{align*}
\norm{\calR_{z,t}} = O_p \left( \frac{ \kappa^{11/2} q_t \max\{N,T\}}{\psi_{\min}^{3/2}}  + \frac{ \kappa^2 r^{1/2} \max\{\sqrt{N},\sqrt{T}\}}{\psi_{\min}^{3/2}} + \frac{ \kappa^{3/2} r  q_t }{\sqrt{\psi_{\min}}} 
 \right).   
\end{align*}
Additionally, if we assume Assumption F(ii), we have $\norm{\calR_{z,t}} = o_p\left(\frac{ r^{1/2}}{\sqrt{\psi_{\max}}} \right)$.
\end{proposition}

\subsection{Temporal Dependence Case}

Next, we consider the case where the idiosyncratic noises are temporally dependent.

\paragraph{Assumption E'.} [Singular vector]
\begin{itemize}
\item[(i)] For the expansion of $Y_{i}$, we assume there is $\rho_i > 0$ such that $\norm{u_i^0} = O_p (\rho_i)$.
\item[(ii)] For the expansion of $Z_{t}$, we assume there are $q_{1,t},q_{2,t},q_{3,t} > 0$ such that
\begin{gather*}
\norm{v_t^0} = O_p (q_{1,t}) , \qquad 
\frac{1}{|\calN_\delta(t)|} \sum_{s \in \calN_\delta(t)} \norm{v_s^0}^2 = O_p (q^2_{2,t}),\\
\frac{1}{|\calN_\delta(t)| N} \sum_{s \in \calN_\delta(t)} \sum_{i=1}^N \sum_{k \in \calN_\delta (t)^c}|\Cov(\epsilon_{is},\epsilon_{ik})| \norm{v_k^0} = O_p\left( q_{3,t} \right),
\end{gather*}
where $\delta \asymp (\log N)^{\nu}$ for some constant $\nu > 0$.
\end{itemize}

\paragraph{Assumption F'.} [Parameter size]
\begin{itemize}
\item[(i)] For the expansion of $Y_{i}$, we assume 
$$
\frac{ \kappa^{5/2} \max\{\sqrt{N},\sqrt{T}\}}{ \psi_{\min}} \conP 0, \qquad \frac{ \kappa^{6} \rho_i \max\{N,T\}}{ r^{1/2} \psi_{\min}} \conP 0, \qquad \text{and} \qquad r^{1/2} \kappa^2 \rho_i \conP 0 .
$$ 
\item[(ii)] For the expansion of $Z_{t}$, we assume 
\begin{align*}
& \frac{\kappa^{5/2} \sqrt{N} \max\{N^{3/2},T^{3/2}\}(\log N)^{\nu/2}}{\psi_{\min}^3} \conP 0,  \qquad
\frac{\kappa^{3} \max\{\sqrt{N},\sqrt{T}\}N^{1/6}}{ \psi_{\min}} \conP 0, \\
& r^{1/2} \kappa^2 q_{1,t} \conP 0 , \qquad
\frac{\kappa^{6} \max\{N,T\}(\log N)^{\nu}\max\{q_{1,t},q_{2,t}\}}{ r^{1/2} \psi_{\min}} \conP 0 , \\
&\frac{\kappa^{5/2} N \max\{\sqrt{N},\sqrt{T}\}(\log N)^{\nu}q_{3,t}}{ r^{1/2} \psi_{\min}^2} \conP 0.
\end{align*}
\end{itemize}

\begin{proposition}\label{pro:pre_dp}
(i) Suppose that Assumptions B'(i) -- (iii) and E'(i) are satisfied. In addition, assume that $\kappa \rho_i \conP 0$ and $\frac{ \max\{\sqrt{N},\sqrt{T}\}}{ \psi_{\min}} \conP 0$. Then, we have
$$
\bfO^{\top} Y_{i} - Y^{0}_i = (\bfZ_r^{0 \top} \bfZ_r^0)^{-1} \sum_{t=1}^T \epsilon_{it} Z_t^0 + \calR_{y,i},
$$
where 
\begin{align*}
\norm{\calR_{y,i}} = O_p \left( \frac{ \kappa^{11/2} \rho_i \max\{N,T\}}{\psi_{\min}^{3/2}}  + \frac{ \kappa^2 r^{1/2} \max\{\sqrt{N},\sqrt{T}\}}{\psi_{\min}^{3/2}} + \frac{ \kappa^{3/2} r  \rho_i }{\sqrt{\psi_{\min}}} 
 \right).   
\end{align*}
Additionally, if we assume Assumption F'(i), we have $\norm{\calR_{y,i}} = o_p\left(\frac{ r^{1/2}}{\sqrt{\psi_{\max}}} \right)$.\\
(ii) Suppose that Assumptions A, B' and E'(ii) are satisfied. In addition, assume that $(\log N)^{\nu/2}\kappa  q_{2,t} \conP 0$ and $\frac{ \max\{\sqrt{N},\sqrt{T}\}}{ \psi_{\min}} \conP 0$. Then, we have
$$
\bfO^{\top} Z_{t} - Z^{0}_t = (\bfY_r^{0\top} \bfY_r^0)^{-1} \sum_{i=1}^N \epsilon_{it} Y_i^0 + \calR_{z,t},
$$
where 
\begin{align*}
||\calR_{z,t}|| & = O_p\left( 
 \frac{\kappa^{5/2} \sqrt{N} \max\{\sqrt{N},\sqrt{T}\} (\log N)^{\nu}q_{2,t}}{\psi_{\min}^{3/2}} + \frac{ r^{1/2}\kappa^{5/2} \max\{\sqrt{N},\sqrt{T}\} N^{1/6}}{\psi_{\min}^{3/2}} \right. \\
& \qquad \quad    + \frac{ \kappa^{2} N \max\{\sqrt{N},\sqrt{T} \} (\log N)^{\nu} q_{3,t} }{\psi_{\min}^{5/2}} + \frac{  \kappa^{2} \sqrt{N} \max\{N^{3/2},T^{3/2} \} (\log N)^{\nu/2}}{\psi_{\min}^{7/2}}\\
 &\left. \qquad \quad +\frac{ \kappa^{11/2} q_{1,t} \max\{N,T\}}{\psi_{\min}^{3/2}} + \frac{ r \kappa^{3/2} q_{1,t} }{\sqrt{\psi_{\min}}} \right).
 \end{align*}
Additionally, if we assume Assumption F'(ii), we have $\norm{\calR_{z,t}} = o_p\left(\frac{ r^{1/2}}{\sqrt{\psi_{\max}}} \right)$.
\end{proposition}

\subsection{Cross-Sectional Dependence Case}

Next, we consider the case where the idiosyncratic noises are cross-sectionally dependent. Basically, it is symmetric to the temporal dependence case.

\paragraph{Assumption E''.} [Singular vector]
\begin{itemize}
\item[(i)] For the expansion of $Y_{i}$, we assume there are $\rho_{1,i},\rho_{2,i},\rho_{3,i} > 0$ such that
\begin{gather*}
\norm{u_i^0} = O_p (\rho_{1,i}) , \qquad 
\frac{1}{|\calN_\delta(i)|} \sum_{j \in \calN_\delta(i)} \norm{u_j^0}^2 = O_p (\rho^2_{2,i}),\\
\frac{1}{|\calN_\delta(i)| T} \sum_{j \in \calN_\delta(i)} \sum_{t=1}^T \sum_{k \in \calN_\delta (i)^c}|\Cov(\epsilon_{jt},\epsilon_{kt})| \norm{u_k^0} = O_p\left( \rho_{3,i} \right),
\end{gather*}
where $\delta \asymp (\log N)^{\omega}$ for some $\omega > 0$.\
\item[(ii)] For the expansion of $Z_{t}$, we assume there is $q_t > 0$ such that $\norm{v_t^0} = O_p (q_t)$.
\end{itemize}

\paragraph{Assumption F''.} [Parameter size]
\begin{itemize}
\item[(i)] For the expansion of $Y_{i}$, we assume 
\begin{align*}
& \frac{\kappa^{5/2} \sqrt{T} \max\{N^{3/2},T^{3/2}\}(\log N)^{\omega/2}}{\psi_{\min}^3} \conP 0,  \qquad
\frac{\kappa^{3} \max\{\sqrt{N},\sqrt{T}\}T^{1/6}}{ \psi_{\min}} \conP 0, \\
& r^{1/2} \kappa^2 \rho_{1,i} \conP 0 , \qquad
\frac{\kappa^{6} \max\{N,T\}(\log N)^{\omega}\max\{\rho_{1,i},\rho_{2,i}\}}{ r^{1/2} \psi_{\min}} \conP 0 , \\
&\frac{\kappa^{5/2} T \max\{\sqrt{N},\sqrt{T}\}(\log N)^{\omega} \rho_{3,i}}{ r^{1/2} \psi_{\min}^2} \conP 0.
\end{align*}
\item[(ii)] For the expansion of $Z_{t}$, we assume 
$$
\frac{ \kappa^{5/2} \max\{\sqrt{N},\sqrt{T}\}}{ \psi_{\min}} \conP 0, \qquad \frac{ \kappa^{6} q_t \max\{N,T\}}{ r^{1/2} \psi_{\min}} \conP 0, \qquad \text{and}  \qquad r^{1/2} \kappa^2   q_t \conP 0 .
$$ 
\end{itemize}

\begin{proposition}\label{pro:pre_cross}
(i) Suppose that Assumptions A', B'' and E''(i) are satisfied. In addition, assume that $(\log N)^{\omega/2} \kappa  \rho_{2,i} \conP 0$ and $\frac{ \max\{\sqrt{N},\sqrt{T}\}}{ \psi_{\min}} \conP 0$. Then, we have
$$
\bfO^{\top} Y_{i} - Y^{0}_i = (\bfZ_r^{0 \top} \bfZ_r^0)^{-1} \sum_{t=1}^T \epsilon_{it} Z_t^0 + \calR_{y,i},
$$
where 
\begin{align*}
\norm{\calR_{y,i}} & = O_p\left( 
 \frac{\kappa^{5/2} \sqrt{T} \max\{\sqrt{N},\sqrt{T}\} (\log N)^{\omega} \rho_{2,i}}{\psi_{\min}^{3/2}} + \frac{ r^{1/2}\kappa^{5/2} \max\{\sqrt{N},\sqrt{T} \} T^{1/6} }{\psi_{\min}^{3/2}} \right. \\
& \qquad \quad    + \frac{ \kappa^{2} T \max\{\sqrt{N},\sqrt{T} \} (\log N)^{\omega} \rho_{3,i} }{\psi_{\min}^{5/2}} + \frac{\kappa^{2} \sqrt{T} \max\{N^{3/2},T^{3/2} \} (\log N)^{\omega/2}}{\psi_{\min}^{7/2}}\\
 &\left. \qquad \quad +\frac{ \kappa^{11/2} \rho_{1,i} \max\{N,T\}}{\psi_{\min}^{3/2}} + \frac{ r \kappa^{3/2} \rho_{1,i} }{\sqrt{\psi_{\min}}} \right).
 \end{align*}
Additionally, if we assume Assumption F''(i), we have $\norm{\calR_{y,i}} = o_p\left(\frac{ r^{1/2}}{\sqrt{\psi_{\max}}} \right)$.\\
(ii) Suppose that Assumptions B''(i) - (iii), E''(ii) are satisfied. In addition, assume that $\kappa q_t \conP 0$ and $\frac{ \max\{\sqrt{N},\sqrt{T}\}}{ \psi_{\min}} \conP 0$. Then, we have
$$
\bfO^{\top} Z_{t} - Z^{0}_t = (\bfY_r^{0\top} \bfY_r^0)^{-1} \sum_{i=1}^N \epsilon_{it} Y_i^0 + \calR_{z,t},
$$
where 
\begin{align*}
\norm{\calR_{z,t}} = O_p \left( \frac{ \kappa^{11/2} q_t \max\{N,T\}}{\psi_{\min}^{3/2}}  + \frac{ \kappa^2 r^{1/2} \max\{\sqrt{N},\sqrt{T}\}}{\psi_{\min}^{3/2}} + \frac{ \kappa^{3/2} r  q_t }{\sqrt{\psi_{\min}}} 
 \right).   
\end{align*}
Additionally, if we assume Assumption F''(ii), we have $\norm{\calR_{z,t}} = o_p\left(\frac{ r^{1/2}}{\sqrt{\psi_{\max}}} \right)$.
\end{proposition}

\subsection{General Dependence Case}

Lastly, we consider the case where the idiosyncratic noises are cross-sectionally and temporally dependent.

\paragraph{Assumption E'''.} [Singular vector]
\begin{itemize}
\item[(i)] For the expansion of $Y_{i}$, we assume there are $\rho_{1,i},\rho_{2,i},\rho_{3,i} > 0$ such that
\begin{gather*}
\norm{u_i^0} = O_p (\rho_{1,i}) , \qquad 
\frac{1}{|\calN_{\delta_1}(i)|} \sum_{j \in \calN_{\delta_1}(i)} \norm{u_j^0}^2 = O_p (\rho^2_{2,i}),\\
\frac{1}{|\calN_{\delta_1}(i)| T} \sum_{j \in \calN_{\delta_1}(i)} \sum_{t=1}^T \sum_{k \in \calN_{\delta_1} (i)^c}|\Cov(\epsilon_{jt},\epsilon_{kt})| \norm{u_k^0} = O_p\left( \rho_{3,i} \right),
\end{gather*}
where $\delta_1 \asymp (\log N)^{\omega}$ for some $\omega > 0$.\
\item[(ii)] For the expansion of $Z_{t}$, we assume there are $q_{1,t},q_{2,t},q_{3,t} > 0$ such that
\begin{gather*}
\norm{v_t^0} = O_p (q_{1,t}) , \qquad 
\frac{1}{|\calN_{\delta_2}(t)|} \sum_{s \in \calN_{\delta_2}(t)} \norm{v_s^0}^2 = O_p (q^2_{2,t}),\\
\frac{1}{|\calN_{\delta_2}(t)| N} \sum_{s \in \calN_{\delta_2}(t)} \sum_{i=1}^N \sum_{k \in \calN_{\delta_2}(t)^c}|\Cov(\epsilon_{is},\epsilon_{ik})| \norm{v_k^0} = O_p\left( q_{3,t} \right),
\end{gather*}
where $\delta_2 \asymp (\log N)^{\nu}$ for some constant $\nu > 0$.
\end{itemize}

\paragraph{Assumption F'''.} [Parameter size]
\begin{itemize}
\item[(i)] For the expansion of $Y_{i}$, we assume 
\begin{align*}
& \frac{\kappa^{5/2} \sqrt{T} \max\{N^{3/2},T^{3/2}\}(\log N)^{\omega/2}}{\psi_{\min}^3} \conP 0,  \qquad
\frac{\kappa^{3} \max\{\sqrt{N},\sqrt{T}\}T^{1/6}}{ \psi_{\min}} \conP 0, \\
& r^{1/2} \kappa^2 \rho_{1,i} \conP 0 , \qquad
\frac{\kappa^{6} \max\{N,T\}(\log N)^{\omega}\max\{\rho_{1,i},\rho_{2,i}\}}{ r^{1/2} \psi_{\min}} \conP 0 , \\
&\frac{\kappa^{5/2} T \max\{\sqrt{N},\sqrt{T}\}(\log N)^{\omega} \rho_{3,i}}{ r^{1/2} \psi_{\min}^2} \conP 0.
\end{align*}
\item[(ii)] For the expansion of $Z_{t}$, we assume 
\begin{align*}
& \frac{\kappa^{5/2} \sqrt{N} \max\{N^{3/2},T^{3/2}\}(\log N)^{\nu/2}}{\psi_{\min}^3} \conP 0,  \qquad
\frac{\kappa^{3} \max\{\sqrt{N},\sqrt{T}\}N^{1/6}}{ \psi_{\min}} \conP 0, \\
& r^{1/2} \kappa^2 q_{1,t} \conP 0 , \qquad
\frac{\kappa^{6} \max\{N,T\}(\log N)^{\nu}\max\{q_{1,t},q_{2,t}\}}{ r^{1/2} \psi_{\min}} \conP 0 , \\
&\frac{\kappa^{5/2} N \max\{\sqrt{N},\sqrt{T}\}(\log N)^{\nu}q_{3,t}}{ r^{1/2} \psi_{\min}^2} \conP 0.
\end{align*}
\end{itemize}

\begin{proposition}\label{pro:pre_general}
(i) Suppose that Assumptions A', B''' and E'''(i) are satisfied. In addition, assume that $(\log N)^{\omega/2} \kappa  \rho_{2,i} \conP 0$ and $\frac{ \max\{\sqrt{N},\sqrt{T}\}}{ \psi_{\min}} \conP 0$. Then, we have
$$
\bfO^{\top} Y_{i} - Y^{0}_i = (\bfZ_r^{0 \top} \bfZ_r^0)^{-1} \sum_{t=1}^T \epsilon_{it} Z_t^0 + \calR_{y,i},
$$
where 
\begin{align*}
\norm{\calR_{y,i}} & = O_p\left( 
 \frac{\kappa^{5/2} \sqrt{T} \max\{\sqrt{N},\sqrt{T}\} (\log N)^{\omega} \rho_{2,i}}{\psi_{\min}^{3/2}} + \frac{ r^{1/2}\kappa^{5/2} \max\{\sqrt{N},\sqrt{T} \} T^{1/6} }{\psi_{\min}^{3/2}} \right. \\
& \qquad \quad    + \frac{ \kappa^{2} T \max\{\sqrt{N},\sqrt{T} \} (\log N)^{\omega} \rho_{3,i} }{\psi_{\min}^{5/2}} + \frac{\kappa^{2} \sqrt{T} \max\{N^{3/2},T^{3/2} \} (\log N)^{\omega/2}}{\psi_{\min}^{7/2}}\\
 &\left. \qquad \quad +\frac{ \kappa^{11/2} \rho_{1,i} \max\{N,T\}}{\psi_{\min}^{3/2}} + \frac{ r \kappa^{3/2} \rho_{1,i} }{\sqrt{\psi_{\min}}} \right).
 \end{align*}
Additionally, if we assume Assumption F'''(i), we have $\norm{\calR_{y,i}} = o_p\left(\frac{ r^{1/2}}{\sqrt{\psi_{\max}}} \right)$.\\
(ii) Suppose that Assumptions A', B''' and E'''(ii) are satisfied. In addition, assume that $(\log N)^{\nu/2}\kappa  q_{2,t} \conP 0$ and $\frac{ \max\{\sqrt{N},\sqrt{T}\}}{ \psi_{\min}} \conP 0$. Then, we have
$$
\bfO^{\top} Z_{t} - Z^{0}_t = (\bfY_r^{0\top} \bfY_r^0)^{-1} \sum_{i=1}^N \epsilon_{it} Y_i^0 + \calR_{z,t},
$$
where 
\begin{align*}
||\calR_{z,t}|| & = O_p\left( 
 \frac{\kappa^{5/2} \sqrt{N} \max\{\sqrt{N},\sqrt{T}\} (\log N)^{\nu}q_{2,t}}{\psi_{\min}^{3/2}} + \frac{ r^{1/2}\kappa^{5/2} \max\{\sqrt{N},\sqrt{T}\} N^{1/6}}{\psi_{\min}^{3/2}} \right. \\
& \qquad \quad    + \frac{ \kappa^{2} N \max\{\sqrt{N},\sqrt{T} \} (\log N)^{\nu} q_{3,t} }{\psi_{\min}^{5/2}} + \frac{  \kappa^{2} \sqrt{N} \max\{N^{3/2},T^{3/2} \} (\log N)^{\nu/2}}{\psi_{\min}^{7/2}}\\
 &\left. \qquad \quad +\frac{ \kappa^{11/2} q_{1,t} \max\{N,T\}}{\psi_{\min}^{3/2}} + \frac{ r \kappa^{3/2} q_{1,t} }{\sqrt{\psi_{\min}}} \right).
 \end{align*}
Additionally, if we assume Assumption F'''(ii), we have $\norm{\calR_{z,t}} = o_p\left(\frac{ r^{1/2}}{\sqrt{\psi_{\max}}} \right)$.
\end{proposition}

\section{Proof of Propositions}

\begin{comment}
From now on, for simplicity, we prove all results by conditioning on the event A where
\begin{gather}\label{eq:event}
A = \left\{ \norm{\bE} \lesssim \max\{\sqrt{N} , \sqrt{T} \} \ll \psi_{\min} \right\}.    
\end{gather}
Note that, for any sequence of random vectors $\{ \omega_{NT}\}$, $P(\{ \omega_{NT} \leq \tau \} \cap A^c) \leq P(A^c) = o(1)$ and
\begin{align*}
P( \omega_{NT} \leq \tau ) &= P(\{ \omega_{NT} \leq \tau \} \cap A) + P(\{ \omega_{NT} \leq \tau \} \cap A^c) = P(\{ \omega_{NT} \leq \tau \} \cap A) + o(1) \\
& =  P(\{ \omega_{NT} \leq \tau \} | A) P(A) + o(1) = P(\{ \omega_{NT} \leq \tau \} | A) + o(1)
\end{align*}
since $P(A) \rightarrow 1$. Hence, it is enough to show the results (convergence in distribution, bounds in $O_p(\cdot)$, $o_p(\cdot)$) by conditioning on the event A.
\end{comment}

\subsection{Proof of Proposition \ref{pro:pre_indp}}

Here, we only prove Proposition \ref{pro:pre_indp} (i) since we can prove  Proposition \ref{pro:pre_indp} (ii) symmetrically. Proposition \ref{pro:pre_indp} (i) comes from the following two lemmas where $R_{1,i}$ and $R_{2,i}$ are defined in \eqref{eq:Ydecomposition}.

\begin{lemma}\label{lem:bound_r1}
Under the assumption for Proposition \ref{pro:pre_indp} (i) or Proposition \ref{pro:pre_dp} (i), 
$$
\norm{R_{1,i}} = O_p \left( \frac{ \kappa^3 \rho_i \sqrt{T} \max\{\sqrt{N},\sqrt{T}\}}{\psi_{\min}^{3/2}}  + \frac{\kappa^2 r^{1/2} \max\{\sqrt{N},\sqrt{T}\}}{\psi_{\min}^{3/2}} \right).
$$

\end{lemma}

\begin{lemma}\label{lem:bound_r2}
Under the assumption for Proposition \ref{pro:pre_indp} (i) or Proposition \ref{pro:pre_dp} (i), 
$$
\norm{R_{2,i}} =O_p\left( \frac{ \kappa^{11/2} \rho_i \max\{N,T\}}{\psi_{\min}^{3/2}} + \frac{ r \kappa^{3/2} \rho_i }{\sqrt{\psi_{\min}}} \right).
$$
\end{lemma}
\bigskip

\noindent \textbf{Proof of Lemma \ref{lem:bound_r1}.} We start from the following decomposition:
$$
\norm{R_{1,i}} \leq \norm{ \bfe_i^\top \left(\widetilde{\bfZ}^{(-i)}_r(\widetilde{\bfZ}^{(-i)\top}_r\widetilde{\bfZ}^{(-i)}_r)^{-1} - \bfZ_r^0(\bfZ_r^{0\top} \bfZ_r^0)^{-1}\right)} + \norm{\bfe_i^\top \left(\widetilde{\bfZ}_r(\widetilde{\bfZ}_r^\top \widetilde{\bfZ}_r)^{-1} - \widetilde{\bfZ}^{(-i)}_r(\widetilde{\bfZ}^{(-i)\top}_r\widetilde{\bfZ}^{(-i)}_r)^{-1}  \right)}
$$
where $\widetilde{\bfZ}^{(-i)}_r = \bfZ^{(-i)}_r \bfO^{(-i)}$. We bound the first term. Define $\Delta_1^{(-i)} = \widetilde{\bfZ}^{(-i)}_r(\widetilde{\bfZ}^{(-i)\top}_r\widetilde{\bfZ}^{(-i)}_r)^{-1} - \bfZ_r^0(\bfZ_r^{0\top} \bfZ_r^0)^{-1}$ and the $t$-th row of it as $\Delta_{1,t}^{(-i)
\top}$. Then, because $(\epsilon_{it})_{t \in [T]}$ are independent of $\Delta_1^{(-i)}$ by construction, we have by Assumption B (or B') that
\begin{align}\label{eq:leaveoneoutbound}
\nonumber \bbE\left[\left. \norm{\bfe_i^\top \Delta_1^{(-i)}}^2 \right|\Delta_1^{(-i)} ,\bM^0 \right] &= \bbE\left[\left. \left\|\sum_{t=1}^T \epsilon_{it} \Delta_{1,t}^{(-i)}\right\|^2 \right|\Delta_1^{(-i)},\bM^0 \right]\\ 
\nonumber &= \sum_{l=1}^r |\Delta^{(-i)\top}_{1,l}\Cov(\bfe_i) \Delta^{(-i)}_{1,l} | \leq \sum_{l=1}^r \norm{\Delta^{(-i)}_{1,l}}^2 \norm{\Cov(\bfe_i)}\\
& \leq \norm{\Cov(\bfe_i)} \norm{\Delta_1^{(-i)}}_F^2 \lesssim \norm{\Delta_1^{(-i)}}_F^2
\end{align}
where $ \bfe_i = [\epsilon_{i1}, \dots, \epsilon_{iT}]^\top$. By the perturbation theory for pseudo-inverses (e.g., Lemma 12 of \cite{chen2019inference}) with Lemma \ref{lem:maintechlem} (ii), we have
\begin{align*}
\norm{\Delta_1^{(-i)}} &\lesssim \max\left\{\norm{\widetilde{\bfZ}^{(-i)}_r(\widetilde{\bfZ}^{(-i)\top}_r\widetilde{\bfZ}^{(-i)}_r)^{-1}}^2, \norm{ \bfZ_r^0(\bfZ_r^{0\top} \bfZ_r^0)^{-1}}^2 \right\} \norm{\widetilde{\bfZ}^{(-i)}_r - \bfZ_r^0} \\
& = O_p \left( \frac{ \kappa^2 \max\{\sqrt{N},\sqrt{T}\} }{\psi_{\min}\sqrt{\psi_{\min}}} \right) 
\end{align*}
since $\max\{||\widetilde{\bfZ}^{(-i)}_r(\widetilde{\bfZ}^{(-i)\top}_r\widetilde{\bfZ}^{(-i)}_r)^{-1}||^2, || \bfZ_r^0(\bfZ_r^{0\top} \bfZ_r^0)^{-1}||^2 \} = O_p \left( \frac{1}{\psi_{\min}} \right)$. Hence, we have
\begin{gather*}
\bbE\left[\left. \norm{\bfe_i^\top \Delta_1^{(-i)}}^2 \right| \Delta_1^{(-i)},\bM^0\right] = O_p \left( \frac{ r \kappa^4 \max\{N,T\} }{\psi_{\min}^3} \right),
\end{gather*}
and by Lemma 6.1 of \cite{chernozhukov2018double}, we have 
\begin{gather*}
\norm{\bfe_i^\top  \left(\widetilde{\bfZ}^{(-i)}_r(\widetilde{\bfZ}^{(-i)\top}_r\widetilde{\bfZ}^{(-i)}_r)^{-1} - \bfZ_r^0(\bfZ_r^{0\top} \bfZ_r^0)^{-1}\right)} = O_p\left(\frac{ r^{1/2} \kappa^2 \max\{\sqrt{N},\sqrt{T}\} }{\psi_{\min}\sqrt{\psi_{\min}}}\right).    
\end{gather*}

Next, we bound the second term. Define $\Delta_2^{(-i)} = \widetilde{\bfZ}_r(\widetilde{\bfZ}^\top_r\widetilde{\bfZ}_r)^{-1} - \widetilde{\bfZ}^{(-i)}_r(\widetilde{\bfZ}^{(-i)\top}_r\widetilde{\bfZ}^{(-i)}_r)^{-1}$. By the the perturbation bound for pseudo-inverses with Lemma \ref{lem:maintechlem} (iv), we have 
\begin{align*}
\norm{\bfe_i^\top\Delta_2^{(-i)}}
&\leq \norm{\bE}\norm{\Delta_2^{(-i)}} = O_p \left( \frac{\kappa^3 \rho_i \sqrt{T} \max\{\sqrt{N},\sqrt{T}\}}{\psi_{\min}^{3/2}}  + \frac{\kappa^3 r^{1/2} \max\{\sqrt{N},\sqrt{T}\}}{\psi_{\min}^{3/2}} \right).
\end{align*}
\bigskip

\noindent \textbf{Proof of Lemma \ref{lem:bound_r2}.} First, note that
\begin{align*}
\norm{R_{2,i}} &= \norm{Y_i^{0\top} \left( \bfZ_r^{0\top} \widetilde{\bfZ}_r(\widetilde{\bfZ}_r^\top\widetilde{\bfZ}_r)^{-1} -  \widetilde{\bfZ}_r^{\top} \widetilde{\bfZ}_r(\widetilde{\bfZ}_r^\top\widetilde{\bfZ}_r)^{-1} \right)} \\
&\leq \norm{Y_i^{0}} \norm{ (\bfZ_r^0 - \widetilde{\bfZ}_r)^\top \widetilde{\bfZ}_r }  \norm{(\widetilde{\bfZ}_r^\top \widetilde{\bfZ}_r)^{-1} } \\
&= O_p\left( \frac{\kappa^{1/2} \rho_i }{\sqrt{\psi_{\min}}}\right) \norm{ (\bfZ_r^0 - \widetilde{\bfZ}_r)^\top \widetilde{\bfZ}_r }
\end{align*}
since $||(\widetilde{\bfZ}_r^\top \widetilde{\bfZ}_r)^{-1}|| = O_p \left( \frac{1}{\psi_{\min}} \right)$ and $||u_i^0|| = O_p\left( \rho_i \right)$. Let $(\Delta_y,\Delta_z) = (\widetilde{\bfY}_r - \bfY_r^0,\widetilde{\bfZ}_r - \bfZ_r^0)$. Note that $(\bfZ_r^0 - \widetilde{\bfZ}_r)^\top \widetilde{\bfZ}_r = \Delta_z^\top  \bfZ_r^0 + \Delta_z^\top  \Delta_z$. Using the decomposition of $\Delta_z$ and Claim \ref{clm:symmetric}, we get
\begin{align*}
 \Delta_z^\top \bfZ_r^0 &= \left[ \bfZ_r^0(\bfY_r^{0\top}\widetilde{\bfY}_r (\widetilde{\bfY}_r^\top \widetilde{\bfY}_r)^{-1} - I_r) + \bE^\top\widetilde{\bfY}_r (\widetilde{\bfY}_r^\top \widetilde{\bfY}_r)^{-1} \right]^\top \bfZ_r^0\\
 & = -(\widetilde{\bfY}_r^\top \widetilde{\bfY}_r)^{-1} \bfY_r^{0\top} \Delta_y \bfD_r^0  \underbrace{-(\widetilde{\bfY}_r^\top \widetilde{\bfY}_r)^{-1} \Delta_y^\top  \Delta_y \bfD_r^0 + (\widetilde{\bfY}_r^\top \widetilde{\bfY}_r)^{-1} \widetilde{\bfY}_r^\top \bE \bfZ_r^0}_{\coloneqq P} \\
&=  -(\widetilde{\bfY}_r^\top \widetilde{\bfY}_r)^{-1} \left(\Delta_z^\top \bfZ_r^0 - \frac{1}{2}(\Delta_y^\top \Delta_y - \Delta_z^\top \Delta_z) \right) \bfD_r^0 + P,
\end{align*}
and
 $(\widetilde{\bfY}_r^\top \widetilde{\bfY}_r)\Delta_z^\top \bfZ_r^0 + \Delta_z^\top \bfZ_r^0\bfD_r^0 = (\widetilde{\bfY}_r^\top \widetilde{\bfY}_r)P + \frac{1}{2}(\Delta_y^\top \Delta_y - \Delta_z^\top \Delta_z)  \bfD_r^0 $. 
 
 \begin{claim}\label{clm:symmetric}
$\Delta_z^\top \bfZ_r^0 - \bfY_r^{0\top}\Delta_y = \frac{1}{2}(\Delta_y^\top \Delta_y - \Delta_z^\top \Delta_z).$
 \end{claim}
 
Then, applying the Sylvester equation (e.g., Lemma 15 of \cite{chen2019inference}), we have with probability converging to $1$ that
\begin{align*}
\norm{\Delta_z^\top \bfZ_r^0} &\lesssim \frac{1}{\psi_{\min}} \norm{(\widetilde{\bfY}_r^\top \widetilde{\bfY}_r)P + \frac{1}{2}(\Delta_y^\top \Delta_y - \Delta_z^\top \Delta_z)  \bfD_r^0}\\
& \leq \frac{1}{\psi_{\min}} \norm{ \widetilde{\bfY}_r^\top \bE \bfZ_r^0 - \Delta_y^\top  \Delta_y \bfD_r^0 + \frac{1}{2}(\Delta_y^\top \Delta_y - \Delta_z^\top \Delta_z)  \bfD_r^0}\\
&\leq \frac{1}{\psi_{\min}} \norm{ \widetilde{\bfY}_r^\top \bE \bfZ_r^0 - \frac{1}{2}(\Delta_z^\top \Delta_z + \Delta_y^\top \Delta_y)  \bfD_r^0},
\end{align*}
and
\begin{align*}
 \norm{(\bfZ_r^0 - \widetilde{\bfZ}_r)^\top\widetilde{\bfZ}_r} \leq \norm{\Delta_z^\top \bfZ_r^0} + \norm{\Delta_z^\top \Delta_z}
 \leq \frac{1}{\psi_{\min}} \norm{ \widetilde{\bfY}_r^\top \bE \bfZ_r^0} + \kappa \left(\norm{\Delta_z^\top \Delta_z} + \norm{\Delta_y^\top \Delta_y}\right).
 \end{align*}
First, from Lemma \ref{lem:maintechlem} (ii), we know $\kappa(||\Delta_z^\top \Delta_z|| + ||\Delta_y^\top \Delta_y||) = O_p\left(\frac{ \kappa^5 \max\{N,T\} }{\psi_{\min}}\right)$. Then, it is enough to bound $|| \widetilde{\bfY}_r^\top \bE \bfZ_r^0||$. Note that $|| \widetilde{\bfY}_r^\top \bE \bfZ_r^0|| \leq ||\bfY_r^{0\top} \bE \bfZ_r^0|| + || \Delta_y^\top \bE \bfZ_r^0||$. In addition, we have
\begin{align*}
\bbE\left[\left. \norm{\bfY_r^{0\top} \bE \bfZ_r^0}_F^2 \right| \bM^0 \right] 
&= \sum_{k=1}^r \sum_{l=1}^r \bbE\left[\left.\left(\sum_{i=1}^N \sum_{t=1}^T \epsilon_{it} Y_{i,k}^0Z_{t,l}^0\right)^2\right|\bM^0\right]\\
& = \left(\sum_{i=1}^N\sum_{k=1}^r \left(Y_{i,k}^{0}\right)^2 \right) \sum_{l=1}^r \left|\sum_{t=1}^T \sum_{s=1}^T \bbE\left[\left. \epsilon_{it}\epsilon_{is} \right| \bM^0 \right] Z_{t,l}^0  Z_{s,l}^0 \right| \\
& = \norm{\bfY_r^0}_F^2 O_p \left(  r \kappa \psi_{\min} \right)\\
& =  O_p \left( r^2 \kappa^2 \psi_{\min}^2 \right)
\end{align*}
since $||\bfY_r^0||_F^2 \leq r \kappa \psi_{\min}$ and for all $l \in [r]$,
\begin{align*}
\left|\sum_{t=1}^T \sum_{s=1}^T \bbE \left[ \left. \epsilon_{it}\epsilon_{is} \right| \bM^0 \right] Z_{t,l}^0  Z_{s,l}^0 \right| 
\leq \left|  \bfZ_{r,l}^{0\top} \Cov(\bfe_i|\bM^0)  \bfZ_{r,l}^{0}  \right| 
\leq \norm{\bfZ_{r,l}^{0}}^2 \norm{ \Cov(\bfe_i)} = O_p( \kappa \psi_{\min}).
\end{align*}
So, we have $||\bfY_r^{0\top} \bE \bfZ_r^0|| = O_p\left( r \kappa \psi_{\min} \right)$. In addition, 
$$
\norm{\Delta_y^\top \bE \bfY_r^0} \leq \norm{\Delta_y} \norm{\bE} \norm{\bfZ_r^0} =  O_p\left(  \kappa^{5/2} \max\{N,T \} 
 \right).
$$
Hence, $|| \widetilde{\bfY}_r^\top \bE \bfZ_r^0|| = O_p\left(  r \kappa \psi_{\min} +  \kappa^{5/2} \max\{N,T \}  \right)$, $||(\bfZ_r^0 - \widetilde{\bfZ}_r)^\top\widetilde{\bfZ}_r|| = O_p\left( r \kappa  + \frac{ \kappa^{5} \max\{N,T \} }{\psi_{\min}}  \right)$, and 
$$
\norm{R_{2,i}} = O_p\left( \frac{ \kappa^{11/2} \rho_i \max\{N,T\}}{\psi_{\min}^{3/2}} + \frac{ r \kappa^{3/2} \rho_i }{\sqrt{\psi_{\min}}} \right). 
$$
\bigskip

\noindent\textbf{Proof of Claim \ref{clm:symmetric}.}
Since $\bfY_r^{0\top}\bfY_r^0 = \bfZ_r^{0\top}\bfZ_r^0$, $\bfY_r^\top \bfY_r = \bfZ_r^\top \bfZ_r$, and $\widetilde{\bfY}_r^\top \widetilde{\bfY}_r = \widetilde{\bfZ}_r^\top \widetilde{\bfZ}_r$, we have
\begin{align}\label{eq:symmetric}
\nonumber\Delta_z^\top \bfZ_r^0 - \bfY_r^{0\top}\Delta_y 
&= (\widetilde{\bfZ}_r -\bfZ_r^0)^\top \bfZ_r^0 - \bfY_r^{0\top}(\widetilde{\bfY}_r -\bfY_r^0) \\
\nonumber&= \widetilde{\bfZ}_r^\top \bfZ_r^0 - \bfY_r^{0\top} \widetilde{\bfY}_r \\
\nonumber&  = \widetilde{\bfZ}_r^\top (\bfZ_r^0 - \widetilde{\bfZ}_r) + \widetilde{\bfZ}_r^\top \widetilde{\bfZ}_r - \widetilde{\bfY}_r^\top \widetilde{\bfY}_r -  (\bfY_r^0 - \widetilde{\bfY}_r)' \widetilde{\bfY}_r \\
&= \Delta_y^\top  \bfY_r^0 + \Delta_y^\top \Delta_y - \bfZ_r^{0\top} \Delta_z - \Delta_z^\top \Delta_z.
\end{align}
In addition, by Lemma 35 of \cite{ma2020implicit}, we have $\widetilde{\bfY}_r^\top \bfY_r^0 + \widetilde{\bfZ}_r^\top \bfZ_r^0 \succeq 0$ and it implies that $\Delta_y^\top \bfY_r^0 + \Delta_z^\top \bfZ_r^0$ is a symmetric matrix. Then, since $ \Delta_z^\top \bfZ_r^0 - \bfY_r^{0\top}\Delta_y=  \bfZ_r^{0\top}\Delta_z - \Delta_y^\top \bfY_r^0$, we have $\Delta_z^\top \bfZ_r^0 - \bfY_r^{0\top}\Delta_y = \frac{1}{2}(\Delta_y^\top \Delta_y - \Delta_z^\top \Delta_z)$ from \eqref{eq:symmetric}. $\square$
\bigskip

\paragraph{Technical Lemmas.}

First, we introduce several notations. Let $\boldsymbol{\calF}_r^{(-i)} = \frac{1}{\sqrt{2}} \begin{bmatrix}
\bfY^{(-i)}_r \\
\bfZ^{(-i)}_r
\end{bmatrix}$, 
$\boldsymbol{\calF}_r^{0} = \frac{1}{\sqrt{2}} \begin{bmatrix}
\bfY_r^0 \\
\bfZ_r^0
\end{bmatrix}$,
$\bfW_r^{(-i)} = \frac{1}{\sqrt{2}} \begin{bmatrix}
\bfU^{(-i)}_r \\
\bfV^{(-i)}_r
\end{bmatrix}$, and
$\bfW_r^0 = \frac{1}{\sqrt{2}} \begin{bmatrix}
\bfU_r^0 \\
\bfV_r^0
\end{bmatrix}$. We define $\bar{\bX} = \begin{bmatrix}
0 & \bX \\
\bX^\top & 0
\end{bmatrix}$, $\bar{\bX}^{(-i)} = \begin{bmatrix}
0 & \bX^{(-i)} \\
\bX^{(-i)\top} & 0
\end{bmatrix}$, and $\bar{\bM}^0= \begin{bmatrix}
0 & \bM^0 \\
\bM^{0\top} & 0
\end{bmatrix}$. In addition, the rotation matrices related to the leave-one-out estimator are defined as
$\bfO^{(-i)} = \argmin_{\bfR \in \calO^{r\times r}} \left\|
\begin{bmatrix}
\bfY^{(-i)}_r \\
\bfZ^{(-i)}_r
\end{bmatrix}
\bfR - 
\begin{bmatrix}
\bfY_r^0 \\
\bfZ_r^0
\end{bmatrix}
\right\|_F$, $\bfQ^{(-i)} = \argmin_{\bfR \in \calO^{r\times r}} \left\|
\begin{bmatrix}
\bfU^{(-i)}_r \\
\bfV^{(-i)}_r
\end{bmatrix}
\bfR - 
\begin{bmatrix}
\bfU_r^0 \\
\bfV_r^0
\end{bmatrix}
\right\|_F$,
$\bfB^{(-i)} = \argmin_{\bfR \in \calO^{r\times r}} \left\|
\begin{bmatrix}
\bfU^{(-i)}_r \\
\bfV^{(-i)}_r
\end{bmatrix}
\bfR - 
\begin{bmatrix}
\bfU_r \\
\bfV_r
\end{bmatrix}
\right\|_F$,
$\bfR^{(-i)} = \argmin_{\bfR \in \calO^{r\times r}} \left\|
\begin{bmatrix}
\bfY_r \\
\bfZ_r
\end{bmatrix}
\bfO - 
\begin{bmatrix}
\bfY^{(-i)}_r \\
\bfZ^{(-i)}_r
\end{bmatrix}
\bfR
\right\|_F$.
\bigskip

\begin{lemma}\label{lem:maintechlem}
We have\\
(i) $ ||\bar{\bX}^{(-i)} - \bar{\bM}^0||, ||\bar{\bX} - \bar{\bM}^0||  = O_p\left( \max\{\sqrt{N},\sqrt{T}\} \right) = o_p\left( \psi_{\min} \right)$;\\
(ii) $||\boldsymbol{\calF}_r^{(-i)} \bfO^{(-i)} - \boldsymbol{\calF}_r^0||,||\boldsymbol{\calF}_r \bfO - \boldsymbol{\calF}_r^0|| = O_p \left( \frac{ \kappa^2 \max\{\sqrt{N},\sqrt{T}\} }{\sqrt{\psi_{\min}}} \right) $;\\
(iii) $
||\boldsymbol{\calF}_r \bfO - \boldsymbol{\calF}_r^{(-i)}\bfR^{(-i)}||_F = O_p\left(  \frac{ \kappa^2 \sqrt{T} \rho_i}{\sqrt{\psi_{\min}}} + \frac{ \kappa \sqrt{r}}{\sqrt{\psi_{\min}}}  \right)$;\\
(iv) $
||\boldsymbol{\calF}_r\bfO - \boldsymbol{\calF}_r^{(-i)}\bfO^{(-i)}||_F = O_p\left(  \frac{ \kappa^3 \sqrt{T} \rho_i}{\sqrt{\psi_{\min}}} + \frac{ \kappa^2 \sqrt{r}}{\sqrt{\psi_{\min}}}  \right)$.
\end{lemma}

\noindent\textbf{Proof of Lemma \ref{lem:maintechlem}.} (i) Because $\bX^{(-i)} - \bM^0 = \bE^{(-i)}$ where $\bE^{(-i)}$ replaces the $i$-th row of $\bE$ with zeros, we can easily know that $||\bar{\bX}^{(-i)} - \bar{\bM}^0|| =  ||\bX^{(-i)} - \bM^0|| \leq ||\bE|| = O_p \left( \max\{\sqrt{N},\sqrt{T}\} \right)$. Similarly, we have $||\bar{\bX} - \bar{\bM}^0|| = O_p \left( \max\{\sqrt{N},\sqrt{T}\} \right)$.\\
(ii) Consider the following decomposition:
\begin{align*}
&\norm{\boldsymbol{\calF}_r^{(-i)} \bfO^{(-i)} - \boldsymbol{\calF}_r^0} \\
&= \left\|\bfW_r^{(-i)} (\bfD_r^{(-i)})^{1/2} ( \bfO^{(-i)} - \bfQ^{(-i)})
+ \bfW_r^{(-i)}(  (\bfD_r^{(-i)})^{1/2}  \bfQ^{(-i)} -   \bfQ^{(-i)} (\bfD_r^{0})^{1/2})\right. \\
& \qquad \left. +  (\bfW_r^{(-i)} \bfQ^{(-i)} - \bfW_r^0)(\bfD_r^{0})^{1/2}  \right\| \\
&\leq \norm{\bfD_r^{(-i)1/2}} \norm{\bfO^{(-i)} - \bfQ^{(-i)}} + \norm{\bfD_r^{(-i)1/2}  \bfQ^{(-i)} -   \bfQ^{(-i)} (\bfD_r^{0})^{1/2}}
+ \norm{(\bfD_r^{0})^{1/2}}\norm{\bfW_r^{(-i)} \bfQ^{(-i)} - \bfW_r^0}.
\end{align*}
Note that $\bfW_r^{(-i)}\bfD_r^{(-i)} \bfW_r^{(-i)\top}$ is the top-$r$ eigenvalue decomposition of $\bar{\bX}^{(-i)}$. So, by Weyl's inequality with the result (i), we have with probability converging to 1,
$$
\frac{3}{4}\psi_r(\bM^0) \leq \psi_r(\bfD_r^{(-i)}) \leq \psi_1(\bfD_r^{(-i)}) \leq 2 \psi_1(\bM^0).
$$
Then, by Lemmas B.2, B.3, and B.4 of \cite{chen2020nonconvex},
\begin{align*}
 &\norm{\bfO^{(-i)} - \bfQ^{(-i)}} = O_p\left( \frac{\kappa^{3/2}}{\psi_{\min}} \right) \norm{\bar{\bX}^{(-i)} - \bar{\bM}^0} = O_p \left( \frac{\kappa^{3/2} \max\{\sqrt{N},\sqrt{T}\} }{\psi_{\min}} \right),\\  
 &\norm{(\bfD_r^{(-i)})^{1/2}  \bfQ^{(-i)} -   \bfQ^{(-i)} (\bfD_r^{0})^{1/2}} = O_p \left( \frac{\kappa}{\sqrt{\psi_{\min}}} \right) \norm{\bar{\bX}^{(-i)} - \bar{\bM}^0 } = O_p \left(\frac{\kappa \max\{\sqrt{N},\sqrt{T}\} }{\sqrt{\psi_{\min}}}\right),\\
 &\norm{\bfW_r^{(-i)} \bfQ^{(-i)} - \bfW_r^0} = O_p \left( \frac{1}{\psi_{\min}} \right)  \norm{\bar{\bX}^{(-i)} - \bar{\bM}^0} = O_p \left( \frac{ \max\{\sqrt{N},\sqrt{T}\} }{\psi_{\min}} \right).
\end{align*}
Hence, we can get $||\boldsymbol{\calF}_r^{(-i)} \bfO^{(-i)} - \boldsymbol{\calF}_r^0|| = O_p \left( \frac{ \kappa^2 \max\{\sqrt{N},\sqrt{T}\} }{\sqrt{\psi_{\min}}} \right)$. By the same token, we can get the result for $||\boldsymbol{\calF}_r \bfO - \boldsymbol{\calF}_r^0||$.\\
(iii) By definition, we have
$
||
\boldsymbol{\calF}_r
\bfO - 
\boldsymbol{\calF}_r^{(-i)}
\bfR^{(-i)}
||_F
\leq 
||
\boldsymbol{\calF}_r^{(-i)}\bfB^{(-i)}
-\boldsymbol{\calF}_r
||_F
$. Note that
\begin{align*}
\left\|\boldsymbol{\calF}_r^{(-i)}\bfB^{(-i)}-\boldsymbol{\calF}_r\right\|_F 
&\leq \norm{\bfW_r^{(-i)}((\bfD_r^{(-i)})^{1/2}\bfB^{(-i)} - \bfB^{(-i)}\bfD_r^{1/2})}_F + 
 \norm{(\bfW_r^{(-i)}\bfB^{(-i)} -\bfW_r)\bfD_r^{1/2}}_F\\
&\leq \norm{(\bfD_r^{(-i)})^{1/2} \bfB^{(-i)} - \bfB^{(-i)}\bfD_r^{1/2}}_F + \norm{\bfW_r^{(-i)}\bfB^{(-i)} - \bfW_r}_F \norm{\bfD_r^{1/2}}.
\end{align*}
By Lemma B.3 of \cite{chen2020nonconvex}, we have
$$
\norm{\bfD_r^{(-i)1/2}  \bfB^{(-i)} -   \bfB^{(-i)} \bfD_r^{1/2}} = O_p \left( \frac{\kappa}{\sqrt{\psi_{\min}}} \right) \norm{(\bar{\bX}^{(-i)} - \bar{\bX})\bfW_r^{(-i)}}_F .
$$ In addition, by Davis-Kahan theorem, we have 
$$
\norm{\bfW_r^{(-i)}\bfB^{(-i)} - \bfW_r}_F = O_p \left( \frac{\kappa}{\psi_{\min}} \right) \norm{(\bar{\bX}^{(-i)} - \bar{\bX})\bfW_r^{(-i)} }_F .
$$ 
So, we have
$$
\left\|\boldsymbol{\calF}_r^{(-i)}\bfB^{(-i)}-\boldsymbol{\calF}_r\right\|_F = O_p \left(  \frac{\kappa}{\sqrt{\psi_{\min}}} \right) \norm{(\bar{\bX}^{(-i)} - \bar{\bX})\bfW_r^{(-i)}}_F .
$$
Note that
\begin{align*}
&(\bar{\bX}^{(-i)} - \bar{\bX})\bfW_r^{(-i)} = 
\begin{bmatrix}
0 & \bE^{(i)} \\
\bE^{(i)\top} & 0
\end{bmatrix}
\frac{1}{\sqrt{2}} \begin{bmatrix}
\bfU^{(-i)}_r \\
\bfV^{(-i)}_r
\end{bmatrix}, \ \ \text{  where  } \bE^{(i)} = \begin{bmatrix}
0 & \cdots & 0 \\
\vdots & \vdots & \vdots \\
\epsilon_{i1} & \cdots & \epsilon_{iT}\\
\vdots & \vdots & \vdots \\
0 & \cdots & 0
\end{bmatrix},\\
& \left[(\bar{\bX}^{(-i)} - \bar{\bX})\bfW_r^{(-i)} \right]^\top = 
\frac{1}{\sqrt{2}} \begin{bmatrix}
0 & \cdots & 0 & \sum_{t=1}^T \epsilon_{it} v_{t}^{(-i)} & 0 & \cdots & 0 & \epsilon_{i1}  u_{i}^{(-i)} & \cdots & \epsilon_{iT}  u_{i}^{(-i)}
\end{bmatrix}
\end{align*}
where $u_{i}^{(-i)\top}$ and $v_{t}^{(-i)\top}$ are the $i$-th row of $\bfU_r^{(-i)}$ and the $t$-th row of $\bfV_r^{(-i)}$, respectively. Hence, we have
$$
\norm{(\bar{\bX}^{(-i)} - \bar{\bX})\bfW_r^{(-i)}}_F \lesssim \left\|\sum_{t=1}^T \epsilon_{it} v_{t}^{(-i)}\right\| + \left\|[\epsilon_{i1}  u_{i}^{(-i)} , \cdots  , \epsilon_{iT}  u_{i}^{(-i)}]\right\|_F .
$$ 
By using the same assertion in \eqref{eq:leaveoneoutbound}, the first term is bounded like 
$$
\bbE\left[\left.\left\|\sum_{t=1}^T \epsilon_{it} v_{t}^{(-i)}\right\|^2 \right|\bfV_{r}^{(-i)},\bM^0\right]\lesssim ||\bfV_{r}^{(-i)}||_F^2 = r. 
$$
In addition, by Lemma \ref{lem:loovector}, the second term is bounded like
$$
\bbE\left[\left. \norm{[\epsilon_{i1}  u_{i}^{(-i)},  \cdots , \epsilon_{iT}  u_{i}^{(-i)}]}_F^2 \right| \bfU_{r}^{(-i)},\bM^0 \right] = \sum_{t=1}^T \bbE\left[\epsilon_{it}^2\right] \norm{u_{i}^{(-i)}}^2 \lesssim T \norm{u_{i}^{(-i)}}^2 = O_p \left( \kappa^2 T \norm{u_{i}^{0}}^2 \right) .
$$
Then, we have $||(\bar{\bX}^{(-i)} - \bar{\bX})\bfW_r^{(-i)}||_F = O_p\left(\kappa \sqrt{T} \rho_i +  \sqrt{r} \right)$ and 
$$
\norm{\boldsymbol{\calF}_r\bfO - \boldsymbol{\calF}_r^{(-i)}\bfR^{(-i)}}_F
\leq \norm{\boldsymbol{\calF}_r^{(-i)}\bfB^{(-i)}-\boldsymbol{\calF}_r}_F = O_p\left(  \frac{\kappa^2 \sqrt{T} \rho_i}{\sqrt{\psi_{\min}}} + \frac{ \kappa \sqrt{r}}{\sqrt{\psi_{\min}}}  \right).
$$
(iv) By the result in (ii), we have with probability converging to 1,
$$
\norm{\boldsymbol{\calF}_r \bfO - \boldsymbol{\calF}_r^0}\norm{ \boldsymbol{\calF}_r^0} \lesssim
 r \kappa^{5/2} \max \{ \sqrt{N},\sqrt{T} \}  \ll \psi_{\min},
 $$
 and
 \begin{align*}
\norm{\boldsymbol{\calF}_r\bfO - \boldsymbol{\calF}_r^{(-i)}\bfR^{(-i)}} \norm{\boldsymbol{\calF}_r^0} &\leq \norm{\boldsymbol{\calF}_r\bfO - \boldsymbol{\calF}_r^{(-i)}\bfO^{(-i)}}\norm{ \boldsymbol{\calF}_r^0 }\\
& \leq  \left( \norm{\boldsymbol{\calF}_r\bfO - \boldsymbol{\calF}_r^0} + \norm{\boldsymbol{\calF}_r^{(-i)}\bfO^{(-i)} - \boldsymbol{\calF}_r^0}\right)   \norm{ \boldsymbol{\calF}_r^0 }\\
&\lesssim  r \kappa^{5/2} \max \{ \sqrt{N},\sqrt{T} \} \\
&\ll \psi_{\min}.
\end{align*}
Hence, by applying Lemma 22 of \cite{chen2020noisy}, we have with probability converging to 1,
$$
\norm{\boldsymbol{\calF}_r\bfO - \boldsymbol{\calF}_r^{(-i)}\bfO^{(-i)}}_F \leq 5 \kappa \norm{\boldsymbol{\calF}_r\bfO - \boldsymbol{\calF}_r^{(-i)}\bfR^{(-i)}}_F. \ \ \square
$$

\begin{lemma}\label{lem:loovector}
When $\norm{u_i^0} \ll 1/\kappa$ and $\norm{\bE} \ll \psi_{\min}$, we have $||u_{i}^{(-i)}|| \lesssim \kappa ||u_i^0||$.
\end{lemma}

\noindent\textbf{Proof of Lemma \ref{lem:loovector}.} Denote the matrix derived by zeroing out the $i$-th row and column of $\bar{\bX}^{(-i)}$ and the corresponding leading $r$ eigenvectors by $\bar{\bX}^{(-i),zero} = \begin{bmatrix}
0 & \bX^{(-i),zero} \\
\bX^{(-i),zero\top} & 0
\end{bmatrix}$ and $\bfW_r^{(-i),zero} = \frac{1}{\sqrt{2}} \begin{bmatrix}
\bfU^{(-i),zero}_r \\
\bfV^{(-i),zero}_r
\end{bmatrix}$, respectively. Here, $\bfU^{(-i),zero}_r \bfD^{(-i),zero}_r \bfV^{(-i),zero\top}_r$ is the top-r singular value decomposition of $\bX^{(-i),zero}$
where $\bX^{(-i),zero}$ is the matrix derived by zeroing out the $i$-th row of $\bX^{(-i)}$. Note that 
$\bfW_r^{(-i),zero}\bfD_r^{(-i),zero}\bfW_r^{(-i),zero\top}$ is the top-$r$ eigenvalue decomposition of $\bar{\bX}^{(-i),zero}$.

First, we confirm that $||u^{(-i),zero}_{i}|| = 0$. Note that entries on $i$-th row of $\bar{\bX}^{(-i),zero}$ is zeros. So, if there is an eigenvector whose entries on $i$-th row is not zero, the corresponding eigenvalue must be zero. However, the top-$r$ eigenvalues of $\bar{\bX}^{(-i),zero}$ are bigger than $\frac{1}{2}\psi_{\min}$ by Weyl's theorem because 
$$
 \norm{\bar{\bX}^{(-i),zero} - \bar{\bM}^0} = \norm{\bX^{(-i),zero} - \bM^0} \leq \norm{\bM^0_{i,\cdot}} + \norm{\bE^{(-i)}} < \frac{1}{4} \psi_{\min}.
$$
It follows from $\norm{\bM^0_{i,\cdot}} = \left(\sum_{t=1}^T (m_{it}^{0})^2\right)^{1/2} \leq \kappa \psi_{\min} \norm{u_i^0} < \frac{1}{8} \psi_{\min}$ by the assumption $\norm{u_i^0} \ll 1/\kappa$, and $\norm{\bE^{(-i)}} \leq  \norm{\bE} < \frac{1}{8} \psi_{\min} $. Then, since $\bfW^{(-i),zero}_{r}$ is the collection of top-$r$ eigenvectors, we have $W^{(-i),zero}_{i} = \frac{1}{\sqrt{2}}u^{(-i),zero}_{i} = 0$.

Next, we bound $|| \bfW_r^{(-i)} sgn(\bfW_r^{(-i)\top} \bfW_r^{(-i),zero}) - \bfW_r^{(-i),zero}||_F$. By Davis-Kahan theorem, we have
$$
\norm{ \bfW_r^{(-i)} sgn(\bfW_r^{(-i)\top} \bfW_r^{(-i),zero}) - \bfW_r^{(-i),zero}}_F
\leq \frac{2\sqrt{2}}{\psi_{\min}} \norm{(\bar{\bX}^{(-i),zero} - \bar{\bX}^{(-i)}) \bfW_r^{(-i),zero}}_F.
$$
For $l \neq i$, we have
$$
(\bar{\bX}^{(-i),zero} - \bar{\bX}^{(-i)})_{l,\cdot} \bfW_r^{(-i),zero} = (\bar{\bX}^{(-i),zero} - \bar{\bX}^{(-i)})_{l,i} W^{(-i),zero}_{i} = 0
$$
because $W^{(-i),zero}_{i} = 0$. So, we have
\begin{align*}
\norm{(\bar{\bX}^{(-i),zero} - \bar{\bX}^{(-i)})_{i,\cdot} \bfW_r^{(-i),zero}}_2 = \norm{ \bar{\bM}^{0}_{i,\cdot} \bfW_r^{(-i),zero}}_2 \leq \norm{ \bar{\bM}^{0}_{i,\cdot}}_2 = \left(\sum_{t=1}^T (m_{it}^{0})^2\right)^{1/2} \leq \kappa \psi_{\min} \norm{u_i^0}
\end{align*}
and $|| \bfW_r^{(-i)} sgn(\bfW_r^{(-i)\top} \bfW_r^{(-i),zero}) - \bfW_r^{(-i),zero}||_F \leq 2\sqrt{2} \kappa ||u_i^0||$. Therefore,
\begin{align*}
\norm{u_{i}^{(-i)}} &= \sqrt{2} \norm{e_i^\top \bfW_r^{(-i)}} \\
&= \sqrt{2} \norm{e_i^\top \bfW_r^{(-i)}sgn(\bfW_r^{(-i)\top} \bfW_r^{(-i),zero})}\\
& \leq \sqrt{2} \norm{e_i^\top \bfW_r^{(-i),zero}} + \sqrt{2} \norm{e_i^\top (\bfW_r^{(-i)} sgn(\bfW_r^{(-i)\top} \bfW_r^{(-i),zero}) - \bfW_r^{(-i),zero})}\\
& \leq  \sqrt{2} \norm{ \bfW_r^{(-i)} sgn(\bfW_r^{(-i)\top} \bfW_r^{(-i),zero}) - \bfW_r^{(-i),zero}}_F\\
&\lesssim \kappa \norm{u_i^0},
\end{align*}
where $e_i$ is the $i$-th column of the $(N + T) \times (N + T)$ identity matrix $I_{(N+T)}$. $\square$

\subsection{Proof of Proposition \ref{pro:pre_dp}}

\paragraph{Proof of Proposition \ref{pro:pre_dp} (i).}

Note that Lemmas \ref{lem:bound_r1} and \ref{lem:bound_r2} allow the weak dependence of error terms across time (Assumptions B'(i) -- (iii)). Hence, Proposition \ref{pro:pre_dp} (i) simply follows from Lemmas \ref{lem:bound_r1} and \ref{lem:bound_r2}.

\paragraph{Proof of Proposition \ref{pro:pre_dp} (ii).}

By the same token as \eqref{eq:Ydecomposition}, we have
\begin{align*}
\bfO^\top Z_{t} - Z^0_{t} = (\bfY_r^{0\top}\bfY_r^0)^{-1}\bfY_r^{0\top} \bfe_t + \underbrace{\left((\widetilde{\bfY}_r^\top\widetilde{\bfY}_r)^{-1}\widetilde{\bfY}_r - (\bfY_r^{0\top} \bfY_r^0)^{-1} \bfY_r^0 \right)\bfe_t}_{\coloneqq R_{1,t}} + \underbrace{\left( (\widetilde{\bfY}_r^\top\widetilde{\bfY}_r)^{-1}\widetilde{\bfY}^{\top}_r  \bfY^{0}_r - I_r \right)Z_t^0 }_{\coloneqq R_{2,t}},
\end{align*}
where $\bfe_t = [\epsilon_{1t}, \dots, \epsilon_{Nt}]$.

\begin{lemma}\label{lem:bound_r1_dp}
Under the assumption for Proposition \ref{pro:pre_dp} (ii),
\begin{align*}
\norm{R_{1,t}} & = O_p\left( 
 \frac{ \kappa^{5/2} \sqrt{N} \max\{\sqrt{N},\sqrt{T}\} (\log N)^{\nu}q_{2,t}}{\psi_{\min}^{3/2}} + \frac{r^{1/2}\kappa^{5/2} \max\{\sqrt{N},\sqrt{T}\} N^{1/6}}{\psi_{\min}^{3/2}} \right. \\
& \qquad \qquad  \left.  + \frac{ \kappa^{2} N \max\{\sqrt{N},\sqrt{T} \} (\log N)^{\nu} q_{3,t} }{\psi_{\min}^{5/2}} + \frac{ \kappa^{2} \sqrt{N} \max\{N^{3/2},T^{3/2} \} (\log N)^{\nu/2}}{\psi_{\min}^{7/2}}
 \right).
\end{align*}

\end{lemma}

\begin{lemma}\label{lem:bound_r2_dp}
Under the assumption for Proposition \ref{pro:pre_dp} (ii),
\begin{align*}
&\norm{R_{2,t}} =O_p\left( \frac{ \kappa^{11/2} q_{1,t} \max\{N,T\}}{\psi_{\min}^{3/2}} + \frac{ r \kappa^{3/2} q_{1,t} }{\sqrt{\psi_{\min}}} \right).
\end{align*}
\end{lemma}

\noindent \textbf{Proof of Lemma \ref{lem:bound_r1_dp}.} First, we define the leave-neighbor-out estimator. Denote the neighbor of $t$ by $\calN_\delta(t) = \{t-\delta,\cdots,t,\cdots,t+\delta\}$. Here, we set $\delta = C \lceil(\log N)^{\nu}\rceil$ for some large constant $C \geq 0$. Then, we denote by $\bX^{(-\calN_\delta(t))}$ the matrix that replaces the columns in $\calN_\delta(t)$ of $\bX$ with $(m^0_{is})_{i \in [N], s \in \calN_\delta(t)}$ to remove the noises $(\epsilon_{is})_{i \in [N],s \in \calN_\delta(t)}$. The corresponding top-r singular value decomposition is denoted by $\bfU_r^{(-\calN_\delta(t))}\bfD_r^{(-\calN_\delta(t))}\bfV_r^{(-\calN_\delta(t))\top}$ and $(\bfY_r^{(-\calN_\delta(t))}, \bfZ_r^{(-\calN_\delta(t))}) =  (\bfU_r^{(-\calN_\delta(t))} (\bfD_r^{(-\calN_\delta(t))})^{1/2},\bfV_r^{(-\calN_\delta(t))} (\bfD_r^{(-\calN_\delta(t))})^{1/2})$. In addition, the corresponding rotation matrix is $\bfO^{(-\calN_\delta(t))} = \argmin_{\bfR \in \calO^{r\times r}} \left\|
\begin{bmatrix}
\bfY^{(-\calN_\delta(t))}_r \\
\bfZ^{(-\calN_\delta(t))}_r
\end{bmatrix}
\bfR - 
\begin{bmatrix}
\bfY_r^0 \\
\bfZ_r^0
\end{bmatrix}
\right\|_F$. Then, consider the following decomposition:
\begin{align*}
\norm{\calR_{1,t}} &\leq  ||\underbrace{\bfe_t^\top \left(\widetilde{\bfY}^{(-\calN_\delta(t))}_r(\widetilde{\bfY}^{(-\calN_\delta(t))\top}_r\widetilde{\bfY}^{(-\calN_\delta(t))}_r)^{-1} - \bfY_r^0(\bfY_r^{0\top} \bfY_r^0)^{-1}\right)}_{\coloneqq a_1}||\\
&\ \ + ||\underbrace{\bfe_t^\top \left(\widetilde{\bfY}_r(\widetilde{\bfY}_r^\top\widetilde{\bfY}_r)^{-1} - \widetilde{\bfY}^{(-\calN_\delta(t))}_r(\widetilde{\bfY}^{(-\calN_\delta(t))\top}_r\widetilde{\bfY}^{(-\calN_\delta(t))}_r)^{-1}  \right)}_{\coloneqq a_2}|| 
\end{align*}
where $\widetilde{\bfY}^{(-\calN_\delta(t))}_r = \bfY^{(-\calN_\delta(t))}_r \bfO^{(-\calN_\delta(t))}$. We bound $||a_1||$. Define 
$$
\Delta_1^{(-\calN_\delta(t))} = \widetilde{\bfY}^{(-\calN_\delta(t))}_r(\widetilde{\bfY}^{(-\calN_\delta(t))\top}_r\widetilde{\bfY}^{(-\calN_\delta(t))}_r)^{-1} - \bfY_r^0(\bfY_r^{0\top} \bfY_r^0)^{-1}
$$
and the $i$-th row of it as $\Delta_{1,i}^{(-\calN_\delta(t))\top}$. Then, the first term can be represented as
\begin{align*}
\sum_{i=1}^N \epsilon_{it} \Delta_{1,i}^{(-\calN_\delta(t))}  &= \sum_{i=1}^N \left(\epsilon_{it} - \bbE[\epsilon_{it} | (\epsilon_{js})_{j \in [N],s \in (\calN_\delta(t))^c} ]\right) \Delta_{1,i}^{(-\calN_\delta(t))}\\
&\ \ + \sum_{i=1}^N \left(\bbE[\epsilon_{it} | (\epsilon_{js})_{j \in [N],s \in (\calN_\delta(t))^c} ] - \bbE[\epsilon_{it}]\right) \Delta_{1,i}^{(-\calN_\delta(t))}.
\end{align*}
Then, because conditioning on $\{ (\epsilon_{js})_{j \in [N], s \in (\calN_\delta(t))^c}, \bM^0 \}$, we can treat $\Delta_{1,i}^{(-\calN_\delta(t))}$ as a constant and $\bbE[\epsilon_{it} | (\epsilon_{js})_{j \in [N], s \in (\calN_\delta(t))^c},\bM^0 ] = \bbE[\epsilon_{it} | (\epsilon_{js})_{j \in [N],s \in (\calN_\delta(t))^c} ]$, we have that
\begin{align*}
& \bbE\left[\left. \left\|\sum_{i=1}^N \left(\epsilon_{it} - \bbE[\epsilon_{it} |  (\epsilon_{js})_{j \in [N], s \in (\calN_\delta(t))^c} ]\right) \Delta_{1,i}^{(-\calN_\delta(t))}\right\|^2 \right| (\epsilon_{jt})_{j \in [N], s \in (\calN_\delta(t))^c}, \bM^0 \right]\\ 
&= \sum_{l=1}^r |\Delta^{(-\calN_\delta(t))\top}_{1,l}\Cov(\bfe_t|  (\epsilon_{js})_{j \in [N], s \in (\calN_\delta(t))^c}) \Delta^{(-\calN_\delta(t))}_{1,l} | \\
&\leq \sum_{l=1}^r \norm{\Delta^{(-\calN_\delta(t))}_{1,l}}^2 \norm{\Cov(\bfe_t |  (\epsilon_{js})_{j \in [N], s \in (\calN_\delta(t))^c})} \\
 &= \norm{\Cov(\bfe_t |  (\epsilon_{js})_{j \in [N], s \in (\calN_\delta(t))^c})} \norm{\Delta_1^{(-\calN_\delta(t))}}_F^2 \\
 &= O_p\left( N^{1/3} \right) \norm{\Delta_1^{(-\calN_\delta(t))}}_F^2
\end{align*}
by Lemma \ref{lem:conditionalvariance}. In addition, by the perturbation theory for pseudo-inverses with Lemma \ref{lem:maintechlem_ext} (ii), we have
\begin{align*}
\norm{\Delta_1^{(-\calN_\delta(t))}} &\lesssim \max\left\{\norm{\widetilde{\bfY}^{(-\calN_\delta(t))}_r(\widetilde{\bfY}^{(-\calN_\delta(t))\top}_r\widetilde{\bfY}^{(-\calN_\delta(t))}_r)^{-1}}^2, \norm{ \bfY_r^0(\bfY_r^{0\top} \bfY_r^0)^{-1} }^2 \right\} \norm{\widetilde{\bfY}^{(-\calN_\delta(t))}_r - \bfY_r^0} \\
&  = O_p\left(\frac{ \kappa^2 \max\{\sqrt{N},\sqrt{T}\} }{\psi_{\min}\sqrt{\psi_{\min}}}\right).
\end{align*}
Hence, we have 
$$
 \left\|\sum_{i=1}^N \left(\epsilon_{it} - \bbE[\epsilon_{it} |  (\epsilon_{js})_{j \in [N], s \in (\calN_\delta(t))^c} ]\right) \Delta_{1,i}^{(-\calN_\delta(t))}\right\| 
 = O_p\left(\frac{ r^{1/2} \kappa^2 \max\{\sqrt{N},\sqrt{T}\} N^{1/6}}{\psi_{\min}\sqrt{\psi_{\min}}}\right).
$$
In addition, we have 
\begin{align*}
\left\| \sum_{i=1}^N \left(\bbE[\epsilon_{it} | (\epsilon_{js})_{j \in [N], s \in (\calN_\delta(t))^c} ] - \bbE[\epsilon_{it}]\right) \Delta_{1,i}^{(-\calN_\delta(t))} \right\|^2 &= \left\| \sum_{i=1}^N \left(\bbE[\epsilon_{it} | (\epsilon_{is})_{s \in (\calN_\delta(t))^c} ] - \bbE[\epsilon_{it}]\right) \Delta_{1,i}^{(-\calN_\delta(t))} \right\|^2 \\
&\leq \left( \sum_{i=1}^N \left(\bbE[\epsilon_{it} | (\epsilon_{is})_{s \in (\calN_\delta(t))^c} ] - \bbE[\epsilon_{it}] \right)^2 \right) \norm{\Delta_1^{(-\calN_\delta(t))}}_F^2.
\end{align*}
By Assumption B'(iv), we have $\sum_{i=1}^N \left(\bbE[\epsilon_{it} | (\epsilon_{is})_{s \in (\calN_\delta(t))^c} ] - \bbE[\epsilon_{it}] \right)^2 = O_p(N^{1/3})$. Moreover, by Lemma \ref{lem:maintechlem_ext} (ii), we have
$||\Delta_1^{(-\calN_\delta(t))}||_F^2 = O_p \left( \frac{ r \kappa^4 \max\{N,T\} }{\psi_{\min}^3} \right)$.
Hence, we have
\begin{gather}\label{eq:differencefromcross}
  \left\| \sum_{i=1}^N \left(\bbE[\epsilon_{it} | (\epsilon_{js})_{j \in [N],s \in (\calN_\delta(t))^c} ] - \bbE[\epsilon_{it}]\right) \Delta_{1,i}^{(-\calN_\delta(t))} \right\| 
 = O_p\left(\frac{ r^{1/2} \kappa^2 \max\{\sqrt{N},\sqrt{T}\} N^{1/6}}{\psi_{\min}\sqrt{\psi_{\min}}}\right).   
\end{gather}
Next, we bound $||a_2||$. Define 
$$
\Delta_2^{(-\calN_\delta(t))} = \widetilde{\bfY}_r(\widetilde{\bfY}^\top_r\widetilde{\bfY}_r)^{-1} - \widetilde{\bfY}^{(-\calN_\delta(t))}_r(\widetilde{\bfY}^{(-\calN_\delta(t))\top}_r\widetilde{\bfY}^{(-\calN_\delta(t))}_r)^{-1}.
$$ 
By the the perturbation bound for pseudo-inverses with Lemma \ref{lem:maintechlem_ext} (iv), we have 
\begin{align*}
 \norm{a_2}  & = \norm{\bfe_t^\top \Delta_2^{(-\calN_\delta(t))}} \\
&\leq \norm{\bE}\norm{\Delta_2^{(-\calN_\delta(t))}}\\
&= O_p \left( \frac{ \max\{\sqrt{N} , \sqrt{T} \}}{\psi_{\min}} \right)\norm{\widetilde{\bfY}_r - \widetilde{\bfY}^{(-\calN_\delta(t))}_r}  \\
& = O_p\left( 
 \frac{ \kappa^3 \sqrt{N} \max\{\sqrt{N},\sqrt{T}\} (\log N)^{\nu}q_{2,t}}{\psi_{\min}^{3/2}} + \frac{ r^{1/2}\kappa^{5/2} \max\{\sqrt{N},\sqrt{T}\} (\log N)^{\nu/2}}{\psi_{\min}^{3/2}} \right. \\
& \qquad \quad  \left.  + \frac{  \kappa^{2} N \max\{\sqrt{N},\sqrt{T} \} (\log N)^{\nu} q_{3,t} }{\psi_{\min}^{5/2}} + \frac{ \kappa^{2} \sqrt{N} \max\{N^{3/2},T^{3/2} \} (\log N)^{\nu/2}}{\psi_{\min}^{7/2}}
 \right). \ \ \square
\end{align*}

\noindent \textbf{Proof of Lemma \ref{lem:bound_r2_dp}.}
This proof is same as the proof of Lemma \ref{lem:bound_r2}. Hence, we omit it. $\square$

\paragraph{Technical Lemmas}

We introduce several notations. Let $\boldsymbol{\calF}_r^{(-\calN_\delta(t))} = \frac{1}{\sqrt{2}} \begin{bmatrix}
\bfY^{(-\calN_\delta(t))}_r \\
\bfZ^{(-\calN_\delta(t)))}_r
\end{bmatrix}$, 
$\boldsymbol{\calF}_r^{0} = \frac{1}{\sqrt{2}} \begin{bmatrix}
\bfY_r^0 \\
\bfZ_r^0
\end{bmatrix}$,
$\bfW_r^{(-\calN_\delta(t))} = \frac{1}{\sqrt{2}} \begin{bmatrix}
\bfU^{(-\calN_\delta(t))}_r \\
\bfV^{(-\calN_\delta(t))}_r
\end{bmatrix}$, and
$\bfW_r^0 = \frac{1}{\sqrt{2}} \begin{bmatrix}
\bfU_r^0 \\
\bfV_r^0
\end{bmatrix}$. We define $\bar{\bX} = \begin{bmatrix}
0 & \bX \\
\bX^\top & 0
\end{bmatrix}$, $\bar{\bX}^{(-\calN_\delta(t))} = \begin{bmatrix}
0 & \bX^{(-\calN_\delta(t))} \\
\bX^{(-\calN_\delta(t))\top} & 0
\end{bmatrix}$, and $\bar{\bM}^0= \begin{bmatrix}
0 & \bM^0 \\
\bM^{0\top} & 0
\end{bmatrix}$. In addition, the rotation matrices related to the leave-neighbor-out estimator are defined as
\begin{align*}
&\bfO^{(-\calN_\delta(t))} = \argmin_{\bfR \in \calO^{r\times r}} \left\|
\begin{bmatrix}
\bfY^{(-\calN_\delta(t))}_r \\
\bfZ^{(-\calN_\delta(t))}_r
\end{bmatrix}
\bfR - 
\begin{bmatrix}
\bfY_r^0 \\
\bfZ_r^0
\end{bmatrix}
\right\|_F, \ \ 
\bfQ^{(-\calN_\delta(t))} = \argmin_{\bfR \in \calO^{r\times r}} \left\|
\begin{bmatrix}
\bfU^{(-\calN_\delta(t))}_r \\
\bfV^{(-\calN_\delta(t))}_r
\end{bmatrix}
\bfR - 
\begin{bmatrix}
\bfU_r^0 \\
\bfV_r^0
\end{bmatrix}
\right\|_F ,\\  
&\bfB^{(-\calN_\delta(t))} = \argmin_{\bfR \in \calO^{r\times r}} \left\|
\begin{bmatrix}
\bfU^{(-\calN_\delta(t))}_r \\
\bfV^{(-\calN_\delta(t))}_r
\end{bmatrix}
\bfR - 
\begin{bmatrix}
U_r \\
V_r
\end{bmatrix}
\right\|_F, \ \
\bfR^{(-\calN_\delta(t))} = \argmin_{\bfR \in \calO^{r\times r}} \left\|
\begin{bmatrix}
\bfY_r \\
\bfZ_r
\end{bmatrix}
\bfO - 
\begin{bmatrix}
\bfY^{(-\calN_\delta(t))}_r \\
\bfZ^{(-\calN_\delta(t))}_r
\end{bmatrix}
\bfR
\right\|_F.
\end{align*}
\smallskip

\begin{lemma}\label{lem:maintechlem_ext}
We have\\
(i) $||\bar{\bX}^{(-\calN_\delta(t))} - \bar{\bM}^0||, ||\bar{\bX} - \bar{\bM}^0|| = O_p \left(  \max\{\sqrt{N},\sqrt{T}\} \right) = o_p \left( \psi_{\min} \right)$;\\
(ii) $||\boldsymbol{\calF}_r^{(-\calN_\delta(t))} \bfO^{(-\calN_\delta(t))} - \boldsymbol{\calF}_r^0||,||\boldsymbol{\calF}_r \bfO - \boldsymbol{\calF}_r^0|| = O_p \left( \frac{ \kappa^2 \max\{\sqrt{N},\sqrt{T}\} }{\sqrt{\psi_{\min}}} \right) $;
\begin{align*}
(iii) \ \ ||
\boldsymbol{\calF}_r
\bfO - 
\boldsymbol{\calF}_r^{(-\calN_\delta(t))}
\bfR^{(-\calN_\delta(t))}
||_F 
&= O_p\left( \frac{ \kappa^2 \sqrt{N} (\log N)^{\nu} q_{2,t} }{\psi^{1/2}_{\min}}
+ \frac{ r^{1/2} \kappa^{3/2} (\log N)^{\nu/2}}{\psi^{1/2}_{\min}} \right. \\
&\qquad \quad  \left. + \frac{ \kappa N q_{3,t} (\log N)^{\nu}}{\psi^{3/2}_{\min}}
+ \frac{ \kappa \sqrt{N} \max\{N,T \} (\log N)^{\nu/2}}{\psi_{\min}^{5/2}}
 \right);
\end{align*}
\begin{align*}
(iv) \ \ ||
\boldsymbol{\calF}_r
\bfO - 
\boldsymbol{\calF}_r^{(-\calN_\delta(t))}
\bfO^{(-\calN_\delta(t))}
||_F 
&= O_p\left( \frac{ \kappa^3 \sqrt{N} (\log N)^{\nu} q_{2,t} }{\psi^{1/2}_{\min}}
+ \frac{ r^{1/2} \kappa^{5/2} (\log N)^{\nu/2}}{\psi^{1/2}_{\min}} \right. \\
&\qquad \quad  \left. + \frac{ \kappa^2 N q_{3,t} (\log N)^{\nu}}{\psi^{3/2}_{\min}}
+ \frac{ \kappa^{2} \sqrt{N} \max\{N,T \} (\log N)^{\nu/2}}{\psi_{\min}^{5/2}}
 \right).
\end{align*}
\end{lemma}

\noindent\textbf{Proof of Lemma \ref{lem:maintechlem_ext}.}
(i) Because $\bX^{(-\calN_\delta(t))} - \bM^0 = \bf\bE^{(-\calN_\delta(t))}$ where $\bf\bE^{(-\calN_\delta(t))}$ replaces the columns in $\calN_\delta(t)$ of $\bE$ with zeros, we can easily know that $||\bar{\bX}^{(-\calN_\delta(t))} - \bar{\bM}^0|| = ||\bX^{(-\calN_\delta(t))} - \bM^0|| \leq ||\bE|| = O_p \left( \max\{\sqrt{N},\sqrt{T}\} \right)$. Similarly, we have $||\bar{\bX} - \bar{\bM}^0|| = O_p \left( \max\{\sqrt{N},\sqrt{T}\} \right)$.\\
(ii) This proof is the same as that of Lemma \ref{lem:maintechlem} (ii). Hence, we omit it.\\
(iii) By definition, $
||
\boldsymbol{\calF}_r
\bfO - 
\boldsymbol{\calF}_r^{(-\calN_\delta(t))}
\bfR^{(-\calN_\delta(t))}
||_F
\leq 
||
\boldsymbol{\calF}_r^{(-\calN_\delta(t))}\bfB^{(-\calN_\delta(t))}
-\boldsymbol{\calF}_r
||_F
$. Note that
\begin{align*}
&\left\|\boldsymbol{\calF}_r^{(-\calN_\delta(t))}\bfB^{(-\calN_\delta(t))}-\boldsymbol{\calF}_r\right\|_F \\
&\leq \norm{\bfW_r^{(-\calN_\delta(t))}((\bfD_r^{(-\calN_\delta(t))})^{1/2}\bfB^{(-\calN_\delta(t))} - \bfB^{(-\calN_\delta(t))}\bfD_r^{1/2})}_F 
  + \norm{(\bfW_r^{(-\calN_\delta(t))}\bfB^{(-\calN_\delta(t))} - \bfW_r)\bfD_r^{1/2}}_F\\
&\leq \norm{(\bfD_r^{(-\calN_\delta(t))})^{1/2}\bfB^{(-\calN_\delta(t))} - \bfB^{(-\calN_\delta(t))}\bfD_r^{1/2}}_F + \norm{\bfW_r^{(-\calN_\delta(t))}\bfB^{(-\calN_\delta(t))} - \bfW_r}_F \norm{\bfD_r^{1/2}}.
\end{align*}
By Lemma B.3 of \cite{chen2020nonconvex}, we have
$$
\norm{(\bfD_r^{(-\calN_\delta(t))})^{1/2}  \bfB^{(-\calN_\delta(t))} -   \bfB^{(-\calN_\delta(t))} \bfD_r^{1/2}} = O_p \left( \frac{\kappa}{\sqrt{\psi_{\min}}} \right) \norm{(\bar{\bX}^{(-\calN_\delta(t))} - \bar{\bX})\bfW_r^{(-\calN_\delta(t))}}_F .
$$ 
In addition, by Davis-Kahan theorem, we have 
$$
\norm{\bfW_r^{(-\calN_\delta(t))}\bfB^{(-\calN_\delta(t))} - \bfW_r}_F = O_p\left( \frac{\kappa}{\psi_{\min}} \right)\norm{(\bar{\bX}^{(-\calN_\delta(t))} - \bar{\bX})\bfW_r^{(-\calN_\delta(t))}}_F.
$$
So, we have
$$
\norm{\boldsymbol{\calF}_r^{(-\calN_\delta(t))}\bfB^{(-\calN_\delta(t))}-\boldsymbol{\calF}_r }_F = O_p \left( \frac{\kappa}{\sqrt{\psi_{\min}}} \right) \norm{(\bar{\bX}^{(-\calN_\delta(t))} - \bar{\bX})\bfW_r^{(-\calN_\delta(t))}}_F.
$$
Note that
\begin{align*}
&(\bar{\bX}^{(-\calN_\delta(t))} - \bar{\bX})\bfW_r^{(-\calN_\delta(t))} = 
\begin{bmatrix}
0 & \bE^{(\calN_\delta(t))} \\
\bE^{(\calN_\delta(t))\top} & 0
\end{bmatrix}
\frac{1}{\sqrt{2}} \begin{bmatrix}
\bfU^{(-\calN_\delta(t))}_r \\
\bfV^{(-\calN_\delta(t))}_r
\end{bmatrix},\\
&\text{  where  }\ \ \bE^{(\calN_\delta(t))} = \begin{bmatrix}
0 & \cdots & \epsilon_{1,t-\delta} & \cdots & \epsilon_{1,t+\delta} & \cdots & 0 \\
\vdots & \vdots & \vdots  & \vdots & \vdots & \vdots & \vdots \\
 0 & \cdots & \epsilon_{N,t-\delta} & \cdots & \epsilon_{N,t+\delta} & \cdots & 0 \\
\end{bmatrix},\\
&(\bar{\bX}^{(-\calN_\delta(t))} - \bar{\bX})\bfW_r^{(-\calN_\delta(t))} = \frac{1}{\sqrt{2}}
 \begin{bmatrix}
\sum_{s \in \calN_\delta(t)}\epsilon_{1s}  v_{s}^{(-\calN_\delta(t))} \\ \vdots \\ \sum_{s \in \calN_\delta(t)}\epsilon_{Ns}  v_{s}^{(-\calN_\delta(t))} \\ 0 \\ \vdots \\ 0 \\ \sum_{i=1}^N \epsilon_{i,t-\delta} u_{i}^{(-\calN_\delta(t))} \\ \vdots \\ \sum_{i=1}^N \epsilon_{i,t+\delta} u_{i}^{(-\calN_\delta(t))} \\ 0 
\end{bmatrix} .
\end{align*}
Hence, we have
\begin{align*}
 &\norm{(\bar{\bX}^{(-\calN_\delta(t))} - \bar{\bX})\bfW_r^{(-\calN_\delta(t))}}_F \\
 &\lesssim \left\|\underbrace{\left[\sum_{i=1}^N \epsilon_{i,t-\delta} u_{i}^{(-\calN_\delta(t))}, \cdots , \sum_{i=1}^N \epsilon_{i,t+\delta} u_{i}^{(-\calN_\delta(t))}\right]}_{\coloneqq a_1}\right\|_F
  + \left\|\underbrace{\left[\sum_{s \in \calN_\delta(t)}\epsilon_{1s}  v_{s}^{(-\calN_\delta(t))}, \cdots , \sum_{s \in \calN_\delta(t)}\epsilon_{Ns}  v_{s}^{(-\calN_\delta(t))}\right]}_{\coloneqq a_2}\right\|_F.
\end{align*}
We bound $||a_1||_F$ first. Using the decomposition 
\eqref{eq:decomposition_bai} in Lemma \ref{lem:decomposition_bai}, we have
\begin{align*}
& a_1  =(\bfD_r^{(-\calN_\delta(t))})^{-1}\left(\bfH_{\rm BN,1}^{(-\calN_\delta(t))}\right)^{-1} (\bfH^0)^{-1} (\bfD_r^0)^{1/2} \left[\sum_{i=1}^N \epsilon_{i,t-\delta} u_i^0 , \cdots , \sum_{i=1}^N \epsilon_{i,t+\delta} u_i^0 \right]  \ \ \cdots\cdots (a_{1-1})  \\
& +  (\bfD_r^{(-\calN_\delta(t))})^{-1} \left( \bfH_{\rm BN,0}^{(-\calN_\delta(t))} \right)^{\top} (\bfH^0)^{\top} (\bfD_r^0)^{1/2} \left[ \sum_{i=1}^N \sum_{k \in \calN_\delta(t)^c} \epsilon_{i,t-\delta}\epsilon_{ik} v_k^0 , \cdots , \sum_{i=1}^N \sum_{k \in \calN_\delta(t)^c} \epsilon_{i,t+\delta}\epsilon_{ik} v_k^0 \right]  \ \ \cdots (a_{1 - 2}) \\
& +   T^{1/2} (\bfD_r^{(-\calN_\delta(t))})^{-1} \left[ \sum_{i=1}^N \epsilon_{i,t-\delta} R_{\lambda,i}^{(-\calN_\delta(t))} , \dots, \sum_{i=1}^N \epsilon_{i,t+\delta} R_{\lambda,i}^{(-\calN_\delta(t))} \right]  \ \ \cdots\cdots (a_{1 - 3}) 
\end{align*}
Note that
$$
\norm{a_{1-1}}_F^2 \leq \left\|(\bfD_r^{(-\calN_\delta(t))})^{-1}\left(\bfH_{\rm BN,1}^{(-\calN_\delta(t))}\right)^{-1} (\bfH^0)^{-1} (\bfD_r^0)^{1/2} \right\|^2 \sum_{s \in \calN_\delta(t)} \left\| \sum_{i=1}^N \epsilon_{is} u_i^0 \right\|^2.
$$
Because $\left(\bfH_{\rm BN,1}^{(-\calN_\delta(t))}\right)^{-1} = O_p(1)$ and $(\bfH^0)^{-1} = O_p(\psi_{\min}^{1/2})$ by Lemma \ref{lem:limits}, we have 
$$
\left\|(\bfD_r^{(-\calN_\delta(t))})^{-1}\left(\bfH_{\rm BN,1}^{(-\calN_\delta(t))}\right)^{-1} (\bfH^0)^{-1} (\bfD_r^0)^{1/2} \right\| = O_p(\kappa^{1/2}).
$$
Moreover, since $\bbE\left[\left.\left\| \sum_{i=1}^N \epsilon_{is} u_i^0 \right\|^2 \right| \bM^0\right] 
 \leq ||\Cov(\bfe_s)|| ||\bfU_r^0||_F^2 
\lesssim  r$ where $|\calN_\delta(t)| = 2\delta + 1 \asymp (\log N)^{\nu}$, we have $||a_{1-1}||_F^2 = O_p\left( r \kappa (\log N)^{\nu} \right)$. In addition, note that
$$
\norm{a_{1-2}} \leq \left\|(\bfD_r^{(-\calN_\delta(t))})^{-1} \left( \bfH_{\rm BN,0}^{(-\calN_\delta(t))} \right)^{\top} (\bfH^0)^{\top} (\bfD_r^0)^{1/2} \right\|  \sum_{s \in \calN_\delta(t)} \left\| \sum_{i=1}^N \sum_{k \in \calN_\delta(t)^c} \epsilon_{is}\epsilon_{ik} v_k^0 \right\|.
$$
Because $\bfH_{\rm BN,0}^{(-\calN_\delta(t))} = O_p(1)$ and $\bfH^0 = O_p(\psi_{\min}^{-1/2})$ by Lemma \ref{lem:limits}, we have 
$$
\left\|(\bfD_r^{(-\calN_\delta(t))})^{-1} \left( \bfH_{\rm BN,0}^{(-\calN_\delta(t))} \right)^{\top} (\bfH^0)^{\top} (\bfD_r^0)^{1/2} \right\| = O_p(\psi_{\min}^{-1}).
$$
Hence, by claim \ref{clm:doubletimedependence}, we have $||a_{1-2}|| = O_p\left( \frac{ r^{1/2}\sqrt{N}(\log N)^{\nu}}{\psi_{\min}} + \frac{N q_{3,t} (\log N)^{\nu}}{\psi_{\min}} \right)$. 

\begin{claim}\label{clm:doubletimedependence}
We have
$\sum_{s \in \calN_\delta(t)} \left\| \sum_{i=1}^N \sum_{k \in \calN_\delta(t)^c} \epsilon_{is}\epsilon_{ik} v_k^0 \right\| =O_p\left(  r^{1/2}\sqrt{N}(\log N)^{\nu} + N q_{3,t} (\log N)^{\nu} \right).$
\end{claim}

Lastly, we have by Lemma \ref{lem:decomposition_bai} that
\begin{align*}
 \norm{a_{1-3}}_F^2 &\leq T \left\|((\bfD_r^{(-\calN_\delta(t))})^{-1} \right\|^2 \sum_{s \in \calN_\delta(t)} \left\| \sum_{i=1}^N \epsilon_{is} R_{\lambda,i}^{(-\calN_\delta(t))} \right\|^2\\
& \leq T \left\|((\bfD_r^{(-\calN_\delta(t))})^{-1} \right\|^2 \left( \sum_{s \in \calN_\delta(t)}\sum_{i=1}^N \epsilon_{is}^2 \right) \sum_{i=1}^N \left\|R_{\lambda,i}^{(-\calN_\delta(t))} \right\|^2\\
& = O_p\left(\frac{  N\max\{N^2,T^2 \} (\log N)^{\nu}}{\psi_{\min}^4}  \right).
\end{align*}
Therefore, we have
$$
\norm{a_1}_F = O_p\left( r^{1/2} \kappa^{1/2} (\log N)^{\nu/2}
+ \frac{ N q_{3,t} (\log N)^{\nu}}{\psi_{\min}}
+ \frac{ \sqrt{N} \max\{N,T \} (\log N)^{\nu/2}}{\psi_{\min}^2} \right).
$$
In addition, we have by Lemma \ref{lem:loovector_ext} that w.h.p.,
\begin{align*}
\norm{a_2}_F^2 &= \sum_{i=1}^N \left( \sum_{s \in \calN_\delta(t)}\epsilon_{is}  v_{s}^{(-\calN_\delta(t))} \right)^2 \leq \left(\sum_{i=1}^N  \sum_{s \in \calN_\delta(t)}\epsilon_{is}^2 \right) \left(\sum_{s \in \calN_\delta(t)}\norm{v_{s}^{(-\calN_\delta(t))}}^2 \right) \\&\lesssim \kappa^2 \left(\sum_{i=1}^N  \sum_{s \in \calN_\delta(t)}\epsilon_{is}^2 \right)\left(\sum_{s \in \calN_\delta(t)}\norm{v_{s}^{0}}^2 \right).
\end{align*}
Hence, we have $\norm{a_2}_F^2 = O_p\left(  \kappa^2 N (\log N)^{2 \nu} q^2_{2,t}
\right)$. To sum up, we have
\begin{align*}
 \norm{(\bar{\bX}^{(-\calN_\delta(t))} - \bar{\bX})\bfW_r^{(-\calN_\delta(t))}}_F
&= O_p\left(  \kappa \sqrt{N} (\log N)^{\nu} q_{2,t} 
+  r^{1/2} k^{1/2} (\log N)^{\nu/2} \right. \\
&\qquad \quad  \left. + \frac{ N q_{3,t} (\log N)^{\nu}}{\psi_{\min}}
+ \frac{  \sqrt{N} \max\{N,T \} (\log N)^{\nu/2}}{\psi_{\min}^2}
 \right), \\
\norm{
\boldsymbol{\calF}_r
\bfO - 
\boldsymbol{\calF}_r^{(-\calN_\delta(t))}
\bfR^{(-\calN_\delta(t))}
}_F 
&= O_p\left( \frac{ \kappa^2 \sqrt{N} (\log N)^{\nu} q_{2,t} }{\psi^{1/2}_{\min}}
+ \frac{ r^{1/2} \kappa^{3/2} (\log N)^{\nu/2}}{\psi^{1/2}_{\min}} \right. \\
&\qquad \quad  \left. + \frac{ \kappa N q_{3,t} (\log N)^{\nu}}{\psi^{3/2}_{\min}}
+ \frac{  \kappa \sqrt{N} \max\{N,T \} (\log N)^{\nu/2}}{\psi_{\min}^{5/2}}
 \right).
\end{align*}
(iv) This proof is the same as that of Lemma \ref{lem:maintechlem} (iv). Hence, we omit it.\ \ $\square$
\bigskip

\noindent\textbf{Proof of Claim \ref{clm:doubletimedependence}.}
Note that
$$\sum_{s \in \calN_\delta(t)} \left\| \sum_{i=1}^N \sum_{k \in \calN_\delta(t)^c} \epsilon_{is}\epsilon_{ik} v_k^0 \right\|
\lesssim \sum_{s \in \calN_\delta(t)} \left\| \sum_{i=1}^N \sum_{k \in \calN_\delta(t)^c} (\epsilon_{is}\epsilon_{ik} - \bbE[\epsilon_{is}\epsilon_{ik} ] ) v_k^0 \right\| + \sum_{s \in \calN_\delta(t)} \left\| \sum_{i=1}^N \sum_{k \in \calN_\delta(t)^c} \bbE[\epsilon_{is}\epsilon_{ik}] v_k^0 \right\|.
$$
We bound the first term. By the independence of error terms across $i$, we have
\begin{align*}
\bbE\left[ \left. \left\| \sum_{i=1}^N \sum_{k \in \calN_\delta(t)^c} (\epsilon_{is}\epsilon_{ik} - \bbE[\epsilon_{is}\epsilon_{ik} ] ) v_k^0 \right\|^2 
 \right| \bM^0\right] = \sum_{i=1}^N  \bbE\left[ \left. \left\| \sum_{k \in \calN_\delta(t)^c} (\epsilon_{is}\epsilon_{ik} - \bbE[\epsilon_{is}\epsilon_{ik} ] ) v_k^0 \right\|^2 
 \right| \bM^0\right].
\end{align*}
In addition, we have
\begin{align*}
\bbE\left[ \left. \left\| \sum_{k \in \calN_\delta(t)^c} (\epsilon_{is}\epsilon_{ik} - \bbE[\epsilon_{is}\epsilon_{ik} ] ) v_k^0 \right\|^2 
 \right| \bM^0\right] &= \sum_{l=1}^r \left|\bfV^{0(-\calN_\delta(t))\top}_{r,l}\Cov\left(\epsilon_{is} \bfe_i^{(-\calN_\delta(t))}\right) \bfV^{0(-\calN_\delta(t))}_{r,l} \right| \\
 & \leq \sum_{l=1}^r \norm{\bfV^{0(-\calN_\delta(t))}_{r,l} }^2 \left\|\Cov\left(\epsilon_{is} \bfe_i^{(-\calN_\delta(t))}\right)\right\|\\
 & \leq \left\|\Cov\left(\epsilon_{is} \bfe_i^{(-\calN_\delta(t))}\right)\right\| \norm{\bfV_r^0}_F^2 \\
& \lesssim r,
\end{align*}
where $\bfV_r^{0(-\calN_\delta(t))}$ is the matrix derived by zeroing out the columns in $\calN_\delta(t)$ from $\bfV_r^0$ and $\epsilon_{is}\bfe_i^{(-\calN_\delta(t))}$ is the matrix derived by zeroing out the columns in $\calN_\delta(t)$ from $\epsilon_{is}\bfe_i = [\epsilon_{is}\epsilon_{i1}, \cdots, \epsilon_{is}\epsilon_{iT}]^\top$. Here, we use Assumption B'(iii) to have
$$
\left\|\Cov\left(\epsilon_{is} \bfe_i^{(-\calN_\delta(t))}\right)\right\|_1 = \max_{k \in \calN_\delta(t)^c} \sum_{l \in \calN_\delta(t)^c }|\Cov(\epsilon_{is}\epsilon_{ik},\epsilon_{is}\epsilon_{il})| \leq C
$$
for some constant $C>0$ where $s \in \calN_\delta(t)$. Hence, we have 
$$
\sum_{s \in \calN_\delta(t)} \left\| \sum_{i=1}^N \sum_{k \in \calN_\delta(t)^c} (\epsilon_{is}\epsilon_{ik} - \bbE[\epsilon_{is}\epsilon_{ik} ] ) v_k^0 \right\| = O_p(r^{1/2}\sqrt{N}(\log N)^{\nu}) .
$$ 
In addition, by Assumption E'(ii), we have
$$
\sum_{s \in \calN_\delta(t)} \left\| \sum_{i=1}^N \sum_{k \in \calN_\delta(t)^c} \bbE[\epsilon_{is}\epsilon_{ik}] v_k^0 \right\| = O_p\left( N q_{3,t} (\log N)^{\nu} \right). \ \ \square
$$
\bigskip

\begin{lemma}\label{lem:decomposition_bai}
Let $(\widehat{\bLambda}^{(-\calN_\delta(t))},\widehat{\bF}^{(-\calN_\delta(t))})$ be the PC estimators derived from $\bX^{(-\calN_\delta(t))} = \bM^0 + \bE^{(-\calN_\delta(t))}$ where $\bE^{(-\calN_\delta(t))}$ is the matrix which replaces the columns in $\calN_\delta(t)$ of $\bE$ with zeros. Then, we have the following decomposition:
\begin{align}\label{eq:decomposition_bai}
\nonumber u_{i}^{(-\calN_\delta(t))} 
& = (\bfD_r^{(-\calN_\delta(t))})^{-1}\left(\bfH_{\rm BN,1}^{(-\calN_\delta(t))}\right)^{-1} (\bfH^0)^{-1} (\bfD_r^0)^{1/2} u_i^0\\
& + (\bfD_r^{(-\calN_\delta(t))})^{-1} \left( \bfH_{\rm BN,0}^{(-\calN_\delta(t))} \right)^{\top} (\bfH^0)^{\top} (\bfD_r^0)^{1/2}  \bfV_r^{0\top} \bfe^{(-\calN_\delta(t))}_i  + T^{1/2} (\bfD_r^{(-\calN_\delta(t))})^{-1} \calR_{\lambda,i}^{(-\calN_\delta(t))},
\end{align}
where $\bfH_{\rm BN,0}=\bLambda^{0\top}\bLambda^0\bF^{0\top}\widehat{\bF}^{(-\calN_\delta(t))}(\bfD_r^{(-\calN_\delta(t))})^{-2}$ and $\bfH_{\rm BN,1}^{(-\calN_\delta(t))} = \left( \widehat{\bF}^{(-\calN_\delta(t)) \top} \bF^0 / T \right)^{-1}$. In addition,
$$
\frac{1}{N} \sum_{i=1}^N \norm{\calR_{\lambda,i}^{(-\calN_\delta(t))}}^2 = O_p \left( \frac{\max\{N, T\}}{\min\{N, T\}\psi_{\min}^2}  \right). 
$$

\end{lemma}
\bigskip

\noindent\textbf{Proof of Lemma \ref{lem:decomposition_bai}.}
By the equation (15) in \cite{bai2023approximate}, we have the following expansion:
\begin{align*}
&\widehat{\lambda}_i^{(-\calN_\delta(t))} = \left(\bfH_{\rm BN,1}^{(-\calN_\delta(t))}\right)^{-1} \lambda_i^0 + \left( \bfH_{\rm BN,0}^{(-\calN_\delta(t))} \right)^{\top} \bF^{0 \top} \bfe^{(-\calN_\delta(t))}_i / T + \calR_{\lambda,i}^{(-\calN_\delta(t))},
\end{align*}
where
$$
\calR_{\lambda,i}^{(-\calN_\delta(t))} = \left( \widehat{\bF}^{(-\calN_\delta(t))} - \bF^0 \bfH_{\rm BN,0}^{(-\calN_\delta(t))} \right)^{\top} \bfe^{(-\calN_\delta(t))}_i / T .
$$
Then, using the relations 
$$
\widehat{\lambda}_i^{(-\calN_\delta(t))} = T^{-1/2} \bfD_r^{(-\calN_\delta(t))}  u_{i}^{(-\calN_\delta(t))},  \ \ \lambda_i^0 = T^{-1/2} (\bfH^0)^{-1} (\bfD_r^0)^{1/2} u_i^0, \ \ \bF^0 = T^{1/2} \bfV_r^0 (\bfD_r^0)^{1/2} \bfH^0,
$$
we have
\begin{align*}
 T^{-1/2} \bfD_r^{(-\calN_\delta(t))}  u_{i}^{(-\calN_\delta(t))} 
 &= T^{-1/2} \left(\bfH_{\rm BN,1}^{(-\calN_\delta(t))}\right)^{-1} (\bfH^0)^{-1} (\bfD_r^0)^{1/2} u_i^0 \\
  & \ \ + T^{-1/2} \left( \bfH_{\rm BN,0}^{(-\calN_\delta(t))} \right)^{\top} (\bfH^0)^{\top} (\bfD_r^0)^{1/2}  \bfV_r^{0\top} \bfe^{(-\calN_\delta(t))}_i
   + \calR_{\lambda,i}^{(-\calN_\delta(t))},
 \end{align*}
 and
\begin{align*} 
u_{i}^{(-\calN_\delta(t))} 
& = (\bfD_r^{(-\calN_\delta(t))})^{-1}\left(\bfH_{\rm BN,1}^{(-\calN_\delta(t))}\right)^{-1} (\bfH^0)^{-1} (\bfD_r^0)^{1/2} u_i^0\\
&+(\bfD_r^{(-\calN_\delta(t))})^{-1} \left( \bfH_{\rm BN,0}^{(-\calN_\delta(t))} \right)^{\top} (\bfH^0)^{\top} (\bfD_r^0)^{1/2}  \bfV_r^{0\top} \bfe^{(-\calN_\delta(t))}_i  + T^{1/2} (\bfD_r^{(-\calN_\delta(t))})^{-1} \calR_{\lambda,i}^{(-\calN_\delta(t))}.
\end{align*}
We can easily check that $\norm{\left(\bfH_{\rm BN,1}^{(-\calN_\delta(t))}\right)^{-1}},  \norm{\bfH_{\rm BN,0}^{(-\calN_\delta(t))}} = O_p(1)$. In addition, by Proposition 2 of \cite{bai2023approximate}, we have
$$
\frac{1}{N} \sum_{i=1}^N \norm{\calR_{\lambda,i}^{(-\calN_\delta(t))}}^2 = O_p \left( \frac{\max\{N, T\}}{\min\{N, T\}N^\alpha T}  \right).
$$
Moreover, as noted in Section \ref{sec:proof_normality_PC_indp}, $N^\alpha T \asymp \psi_{\min}^2$ with probability converging to 1. Hence, we can say
$$
\frac{1}{N} \sum_{i=1}^N \norm{\calR_{\lambda,i}^{(-\calN_\delta(t))}}^2 = O_p \left( \frac{\max\{N, T\}}{\min\{N, T\}\psi_{\min}^2}  \right). \ \ \square
$$
\bigskip

\begin{lemma}\label{lem:loovector_ext}
When $ \sum_{s \in \calN_\delta(t)} \norm{v_s^0}^2 \ll 1/\kappa^2$ and $\norm{\bE} \ll \psi_{\min} $, we have $||v_{s}^{(-\calN_\delta(t))}|| \lesssim \kappa ||v_s^0||$ for all $s \in \calN_\delta(t)$.
\end{lemma}

\noindent\textbf{Proof of Lemma \ref{lem:loovector_ext}.} 
Fix $s \in \calN_\delta(t)$. Note that $\bX^{(-\calN_\delta(t))} - \bM^0 = \bE^{(-\calN_\delta(t))}$ where $\bE^{(-\calN_\delta(t))}$ replaces the columns in $\calN_\delta(t)$ of $\bE$ with zeros. Denote the matrix derived by zeroing out the $s$-th column of $\bX^{(-\calN_\delta(t))}$ by $\bX^{(-\calN_\delta(t)),zero}$ and 
$$
\bar{\bX}^{(-\calN_\delta(t)),zero} = \begin{bmatrix}
0 & \bX^{(-\calN_\delta(t)),zero} \\
\bX^{(-\calN_\delta(t)),zero\top} & 0
\end{bmatrix}.
$$
In addition, denote the corresponding leading $r$ eigenvectors of $\bar{\bX}^{(-\calN_\delta(t)),zero}$ by $\bfW_r^{(-\calN_\delta(t)),zero} = \frac{1}{\sqrt{2}} \begin{bmatrix}
\bfU^{(-\calN_\delta(t)),zero}_r \\
\bfV^{(-\calN_\delta(t)),zero}_r
\end{bmatrix}$. Here, $\bfU^{(-\calN_\delta(t)),zero}_r \bfD^{(-\calN_\delta(t)),zero}_r \bfV^{(-\calN_\delta(t)),zero\top}_r$ is the top-r singular value decomposition of $\bX^{(-\calN_\delta(t)),zero}$. 

First, we confirm that $||v_{s}^{(-\calN_\delta(t)),zero}|| = 0$. Note that entries on $(N+s)$-th row of $\bar{\bX}^{(-\calN_\delta(t)),zero}$ is zeros. So, if there is an eigenvector whose entries on $(N+s)$-th row is not zero, the corresponding eigenvalue must be zero. However, the top-$r$ eigenvalues of $\bar{\bX}^{(-\calN_\delta(t)),zero}$ are bigger than $\frac{1}{2}\psi_{\min}$ by Weyl's theorem because
$$ 
 \norm{\bar{\bX}^{(-\calN_\delta(t)),zero} - \bar{\bM}^0} = \norm{\bX^{(-\calN_\delta(t)),zero} - \bM^0} \leq \norm{\bM^0_{\cdot,s}} + \norm{\bE^{(-\calN_\delta(t))}} < \frac{1}{4} \psi_{\min}.
$$
It follows from $\norm{\bM^0_{\cdot,s}} = \left(\sum_{i=1}^N (m_{is}^{0})^2\right)^{1/2} \leq \kappa \psi_{\min} \norm{v_s^0} < \frac{1}{8} \psi_{\min}$ by the assumption $\sum_{s \in \calN_\delta(t)} \norm{v_s^0}^2 \ll 1/\kappa^2$,
and $\norm{\bE^{(-\calN_\delta(t))}} \leq  \norm{\bE} < \frac{1}{8} \psi_{\min} $. Then, since $\bfW_r^{(-\calN_\delta(t)),zero}$ is the collection of top-$r$ eigenvectors, we have $W^{(-\calN_\delta(t)),zero}_{N+s} = \frac{1}{\sqrt{2}}v^{(-\calN_\delta(t)),zero}_{s} = 0$.

Next, we bound $|| \bfW_r^{(-\calN_\delta(t))} sgn(\bfW_r^{(-\calN_\delta(t))\top} \bfW_r^{(-\calN_\delta(t)),zero}) - \bfW_r^{(-\calN_\delta(t)),zero}||_F$. By Davis-Kahan theorem, we have
\begin{align*}
 &\norm{ \bfW_r^{(-\calN_\delta(t))} sgn(\bfW_r^{(-\calN_\delta(t))\top} \bfW_r^{(-\calN_\delta(t)),zero}) - \bfW_r^{(-\calN_\delta(t)),zero} }_F\\
& \ \ \leq \frac{2\sqrt{2}}{\psi_{\min}} \norm{ (\bar{\bX}^{(-\calN_\delta(t)),zero} - \bar{\bX}^{(-\calN_\delta(t))}) \bfW_r^{(-\calN_\delta(t)),zero}}_F.   
\end{align*}
For $l \neq N+s$, we have
$$
(\bar{\bX}^{(-\calN_\delta(t)),zero} - \bar{\bX}^{(-\calN_\delta(t))})_{l,\cdot} \bfW_r^{(-\calN_\delta(t)),zero} = (\bar{\bX}^{(-\calN_\delta(t)),zero} - \bar{\bX}^{(-\calN_\delta(t))})_{l,N+s} W^{(-\calN_\delta(t)),zero}_{N+s} = 0
$$
because $W^{(-\calN_\delta(t)),zero}_{N+s} = 0$. So, we have
\begin{align*}
&\norm{(\bar{\bX}^{(-\calN_\delta(t)),zero} - \bar{\bX}^{(-\calN_\delta(t))})_{N+s,\cdot} \bfW_r^{(-\calN_\delta(t)),zero}}_2 \\
&= \norm{\bar{\bM}^{0}_{N+s, \cdot}\bfW_r^{(-\calN_\delta(t)),zero}}_2\leq \norm{ \bar{\bM}^{0}_{N+s, \cdot}}_2 = \left(\sum_{i=1}^N (m_{is}^{0})^2 \right)^{1/2} \leq  \kappa \psi_{\min} \norm{v_s^0}
\end{align*}
and $|| \bfW_r^{(-\calN_\delta(t))} sgn(\bfW_r^{(-\calN_\delta(t))\top} \bfW_r^{(-\calN_\delta(t)),zero}) - \bfW_r^{(-\calN_\delta(t)),zero}||_F \leq 2\sqrt{2}  \kappa ||v_s^0||$. Therefore,
\begin{align*}
\norm{v_{s}^{(-\calN_\delta(t))}} &= \sqrt{2} \norm{e_{N+s}^\top \bfW_r^{(-\calN_\delta(t))}} \\
&= \sqrt{2} \norm{ e_{N+s}^\top \bfW_r^{(-\calN_\delta(t))}sgn(\bfW_r^{(-\calN_\delta(t))\top} \bfW_r^{(-\calN_\delta(t)),zero})}\\
&  \leq \sqrt{2} \norm{e_{N+s}^\top \bfW_r^{(-\calN_\delta(t)),zero}}\\
&\quad + \sqrt{2} \norm{e_{N+s}^\top (\bfW_r^{(-\calN_\delta(t))} sgn(\bfW_r^{(-\calN_\delta(t))\top} \bfW_r^{(-\calN_\delta(t)),zero}) - \bfW_r^{(-\calN_\delta(t)),zero})}\\
&  \leq  \sqrt{2} \norm{ \bfW_r^{(-\calN_\delta(t))} sgn(\bfW_r^{(-\calN_\delta(t))\top} \bfW_r^{(-\calN_\delta(t)),zero}) - \bfW_r^{(-\calN_\delta(t)),zero}}_F \\
&\lesssim  \kappa \norm{v_{s}^0}
\end{align*}
where $e_k$ is the $k$-th column of the $(N + T) \times (N + T)$ identity matrix $I_{(N+T)}$. $\square$

\begin{lemma}\label{lem:conditionalvariance}
$\norm{\Cov(\bfe_t |  (\bfe_s)_{s \in (\calN_\delta(t))^c})} = O_p\left( N^{1/3} \right)$.
\end{lemma}

\noindent\textbf{Proof of Lemma \ref{lem:conditionalvariance}.} 
Let $i \neq j$. Then, by the cross-sectional independence, we have
$$
\Cov(\epsilon_{it},\epsilon_{jt} | (\bfe_s)_{s \in (\calN_\delta(t))^c})  = \Cov(\epsilon_{it},\epsilon_{jt} | \{(\epsilon_{is})_{s \in (\calN_\delta(t))^c}, (\epsilon_{js})_{s \in (\calN_\delta(t))^c}\}).
$$
Note that
\begin{align*}
&\bbE\left[ \epsilon_{it} \epsilon_{jt} | \{(\epsilon_{is})_{s \in (\calN_\delta(t))^c}, (\epsilon_{js})_{s \in (\calN_\delta(t))^c}\} \right] \\
&= \bbE \left[ \left. \bbE\left[ \epsilon_{it} \epsilon_{jt} | \{(\epsilon_{is})_{s \in (\calN_\delta(t))^c}, \bfe_j \} \right]  \right| \{(\epsilon_{is})_{s \in (\calN_\delta(t))^c}, (\epsilon_{js})_{s \in (\calN_\delta(t))^c}\} \right]\\
& = \bbE \left[ \left. \epsilon_{jt} \bbE\left[ \epsilon_{it} | (\epsilon_{is})_{s \in (\calN_\delta(t))^c} \right]  \right| \{(\epsilon_{is})_{s \in (\calN_\delta(t))^c}, (\epsilon_{js})_{s \in (\calN_\delta(t))^c}\} \right]\\
& = \bbE\left[ \epsilon_{it} | (\epsilon_{is})_{s \in (\calN_\delta(t))^c} \right] \bbE\left[ \epsilon_{jt} | (\epsilon_{js})_{s \in (\calN_\delta(t))^c} \right] .
\end{align*}
Then, we have
\begin{align*}
&\Cov(\epsilon_{it},\epsilon_{jt} | \{(\epsilon_{is})_{s \in (\calN_\delta(t))^c}, (\epsilon_{js})_{s \in (\calN_\delta(t))^c}\}) \\
& = \bbE\left[ \left( \epsilon_{it} - \bbE[\epsilon_{it}| \{(\epsilon_{is})_{s \in (\calN_\delta(t))^c}] \right)\left( \epsilon_{jt} - \bbE[\epsilon_{jt}| \{(\epsilon_{js})_{s \in (\calN_\delta(t))^c}] \right) | \{(\epsilon_{is})_{s \in (\calN_\delta(t))^c}, (\epsilon_{js})_{s \in (\calN_\delta(t))^c}\} \right] \\
&= \bbE\left[ \epsilon_{it} \epsilon_{jt} | \{(\epsilon_{is})_{s \in (\calN_\delta(t))^c}, (\epsilon_{js})_{s \in (\calN_\delta(t))^c}\} \right]
-  \bbE\left[ \epsilon_{it} | (\epsilon_{is})_{s \in (\calN_\delta(t))^c} \right] \bbE\left[ \epsilon_{jt} | (\epsilon_{js})_{s \in (\calN_\delta(t))^c} \right]\\
&=0.
\end{align*}
Hence, we have
$$
\norm{\Cov(\bfe_t |  (\bfe_s)_{s \in (\calN_\delta(t))^c})} = \max_{1\leq i \leq N} \Var(\epsilon_{it} | (\epsilon_{is})_{s \in (\calN_\delta(t))^c}) \leq \max_{1\leq i \leq N} \bbE[\epsilon_{it}^2 | (\epsilon_{is})_{s \in (\calN_\delta(t))^c}].
$$
Let $w_{it} = \bbE[\epsilon_{it}^2 | (\epsilon_{is})_{s \in (\calN_\delta(t))^c}]$. By Jensen's inequality, 
\begin{align*}
\bbE\left[ w_{it}^3 \right] = \bbE\left[ \left( \bbE \left[ \epsilon_{it}^2 | (\epsilon_{is})_{s \in (\calN_\delta(t))^c}\right] \right)^3 \right] \leq \bbE\left[ \bbE\left[ \epsilon_{it}^6 | (\epsilon_{is})_{s \in (\calN_\delta(t))^c}\right] \right] = \bbE \left[ \epsilon_{it}^6 \right] .
\end{align*}
Then, because for $a>0$,
$$
P(w_{it} \geq a ) \leq \frac{\bbE \left[ w_{it}^3 \right]}{a^3} \leq \frac{\bbE \left[ \epsilon_{it}^6 \right]}{a^3},
$$
we have for any $l_{NT} \rightarrow \infty$,
$$
P\left(\max_{1\leq i \leq N}w_{it} \geq N^{1/3}l_{NT} \right) \leq  \sum_{ 1 \leq i \leq N} P(w_{it} \geq N^{1/3}l_{NT} ) \leq \frac{ \max_{1\leq i \leq N} \bbE \left[ \epsilon_{it}^6 \right]}{ l_{NT}^3} \rightarrow 0 .
$$
Therefore, we have $\max_{1\leq i \leq N}w_{it} = O_p(N^{1/3})$ and so, $\norm{\Cov(\bfe_t |  (\bfe_s)_{s \in (\calN_\delta(t))^c})} = O_p(N^{1/3})$. $\square$

\subsection{Proof of Proposition \ref{pro:pre_cross}}

\paragraph{Proof of Proposition \ref{pro:pre_cross} (i).}

Here, $R_{1,i}$ and $R_{2,i}$ are defined in \eqref{eq:Ydecomposition}.

\begin{lemma}\label{lem:bound_r1_cross}
Under the assumption for Proposition \ref{pro:pre_cross} (i),
\begin{align*}
\norm{R_{1,i}} & = O_p\left( 
 \frac{ \kappa^{5/2} \sqrt{T} \max\{\sqrt{N},\sqrt{T}\} (\log N)^{\omega} \rho_{2,i}}{\psi_{\min}^{3/2}} + \frac{r^{1/2}\kappa^{5/2} \max\{\sqrt{N}, \sqrt{T} \} T^{1/6} }{\psi_{\min}^{3/2}} \right. \\
& \qquad \qquad  \left.  + \frac{ \kappa^{2} T \max\{\sqrt{N},\sqrt{T} \} (\log N)^{\omega} \rho_{3,i} }{\psi_{\min}^{5/2}} + \frac{ \kappa^{2} \sqrt{T} \max\{N^{3/2},T^{3/2} \} (\log N)^{\omega/2}}{\psi_{\min}^{7/2}}
 \right).
\end{align*}

\end{lemma}

\begin{lemma}\label{lem:bound_r2_cross}
Under the assumption for Proposition \ref{pro:pre_cross} (i),
\begin{align*}
&\norm{R_{2,i}} =O_p\left( \frac{ \kappa^{11/2} \rho_{1,i} \max\{N,T\}}{\psi_{\min}^{3/2}} + \frac{ r \kappa^{3/2} \rho_{1,i} }{\sqrt{\psi_{\min}}} \right).
\end{align*}
\end{lemma}

\noindent \textbf{Proof of Lemmas \ref{lem:bound_r1_cross} and \ref{lem:bound_r2_cross}.} 
Basically, the proof is symmetric to those of Lemmas \ref{lem:bound_r1_dp} and \ref{lem:bound_r2_dp}. Hence, we omit it.

\paragraph{Proof of Proposition \ref{pro:pre_cross} (ii).}

The proof is symmetric to that of Proposition \ref{pro:pre_dp} (i). Hence,  it simply follows from Lemmas \ref{lem:bound_r1} and \ref{lem:bound_r2}. $\square$

\subsection{Proof of Proposition \ref{pro:pre_general}}

We show Proposition \ref{pro:pre_general} (ii) first.

\paragraph{Proof of Proposition \ref{pro:pre_general} (ii).}

Note that we have
\begin{align*}
\bfO^\top Z_{t} - Z^0_{t} = (\bfY_r^{0\top}\bfY_r^0)^{-1}\bfY_r^{0\top} \bfe_t + \underbrace{\left((\widetilde{\bfY}_r^\top\widetilde{\bfY}_r)^{-1}\widetilde{\bfY}_r - (\bfY_r^{0\top} \bfY_r^0)^{-1} \bfY_r^0 \right)\bfe_t}_{\coloneqq R_{1,t}} + \underbrace{\left( (\widetilde{\bfY}_r^\top\widetilde{\bfY}_r)^{-1}\widetilde{\bfY}^{\top}_r  \bfY^{0}_r - I_r \right)Z_t^0 }_{\coloneqq R_{2,t}},
\end{align*}
where $\bfe_t = [\epsilon_{1t}, \dots, \epsilon_{Nt}]$.

\begin{lemma}\label{lem:bound_r1_general}
Under the assumption for Proposition \ref{pro:pre_general} (ii),
\begin{align*}
\norm{R_{1,t}} & = O_p\left( 
 \frac{ \kappa^{5/2} \sqrt{N} \max\{\sqrt{N},\sqrt{T}\} (\log N)^{\nu}q_{2,t}}{\psi_{\min}^{3/2}} + \frac{r^{1/2}\kappa^{5/2} \max\{\sqrt{N},\sqrt{T}\} N^{1/6}}{\psi_{\min}^{3/2}} \right. \\
& \qquad \qquad  \left.  + \frac{ \kappa^{2} N \max\{\sqrt{N},\sqrt{T} \} (\log N)^{\nu} q_{3,t} }{\psi_{\min}^{5/2}} + \frac{ \kappa^{2} \sqrt{N} \max\{N^{3/2},T^{3/2} \} (\log N)^{\nu/2}}{\psi_{\min}^{7/2}}
 \right).
\end{align*}

\end{lemma}

\begin{lemma}\label{lem:bound_r2_general}
Under the assumption for Proposition \ref{pro:pre_general} (ii),
\begin{align*}
&\norm{R_{2,t}} =O_p\left( \frac{ \kappa^{11/2} q_{1,t} \max\{N,T\}}{\psi_{\min}^{3/2}} + \frac{ r \kappa^{3/2} q_{1,t} }{\sqrt{\psi_{\min}}} \right).
\end{align*}
\end{lemma}

\paragraph{Proof of Lemma \ref{lem:bound_r1_general}.}

Basically, the proof is similar to that of Lemma \ref{lem:bound_r1_dp}. Here, we only prove the parts where the proofs are different from those of Lemma \ref{lem:bound_r1_dp}. First, we check the part deriving the bound of $\norm{\Cov(\bfe_t | (\bfe_s)_{s\in (\calN_{\delta_2}(t))^c})}$. In the general dependence case, we have by the conditional weak dependence assumption that
\begin{align*}
\norm{\Cov(\bfe_t | (\bfe_s)_{s\in (\calN_{\delta_2}(t))^c})} &\leq \max_{1\leq l \leq N} \sum_{j=1}^N  \abs{\Cov (\epsilon_{lt}, \epsilon_{jt} | (\bfe_s)_{s\in (\calN_{\delta_2}(t))^c})} \\
&= O_p \left( \max_{1 \leq i \leq N} \Var (\epsilon_{it}| (\bfe_s)_{s\in (\calN_{\delta_2}(t))^c}) \right) \\
& = O_p (N^{1/3}).
\end{align*}
The last equation comes from the proof of Lemma \ref{lem:conditionalvariance}. Next, we show that
\begin{align*}
\bbE\left[ \left. \left\| \sum_{i=1}^N \sum_{k \in \calN_{\delta_2}(t)^c} (\epsilon_{is}\epsilon_{ik} - \bbE[\epsilon_{is}\epsilon_{ik} ] ) v_k^0 \right\|^2 
 \right| \bM^0\right] \lesssim rN .
\end{align*}
Define $\gamma_{ik} = \epsilon_{is}\epsilon_{ik} - \bbE[\epsilon_{is}\epsilon_{ik} ] $. In addition, let $ \bGamma =(\gamma_{it})_{i \leq N, t \leq T}$ and $\bGamma^{(-\calN_{\delta_2}(t))} $ be the matrix which replaces the columns in $\calN_{\delta_2}(t)$ of $\bGamma$ with zeros. Then, we have
\begin{align*}
\norm{ \sum_{i=1}^N \sum_{k \in \calN_{\delta_2}(t)^c} \gamma_{ik}  v_k^0 }^2 
&= \sum_{l=1}^r \left( \sum_{i=1}^N \sum_{k \in \calN_{\delta_2}(t)^c} \gamma_{ik} v_{kl}^0  \right)^2 \\
&= \sum_{l=1}^r \bfP_l^\top vec\left( \left( \bGamma^{(-\calN_{\delta_2}(t))}\right)^{\top} \right) vec\left( \left( \bGamma^{(-\calN_{\delta_2}(t))}\right)^{\top} \right)^{\top} \bfP_l,
\end{align*}
where $\bfP_l = \left[(\bfV_{r,l}^{0})^{\top}, \cdots  , (\bfV_{r,l}^{0})^{\top} \right]^{\top}$ is the $NT \times 1$ vector which piles up the $N$ number of $\bfV_{r,l}^{0}$s and $\bfV_{r,l}^{0}$ is the $l$-th column of $\bfV_{r}^{0}$. Then, we have
\begin{align*}
\bbE \left[ \left. \norm{ \sum_{i=1}^N \sum_{k \in \calN_{\delta_2}(t)^c} \gamma_{ik}  v_k^0 }^2 \right\| \bM^0 \right] 
&= \sum_{l=1}^r \bfP_l^\top \Cov \left( vec\left( \left( \bGamma^{(-\calN_{\delta_2}(t))}\right)^{\top} \right)\right) \bfP_l \\
& \leq \sum_{l=1}^r \norm{\bfP_l}^2 \norm{Cov \left( vec\left( \left( \bGamma^{(-\calN_{\delta_2}(t))}\right)^{\top} \right)\right)}.
\end{align*}
Note that
$$
\sum_{l=1}^r \norm{\bfP_l}^2 = \sum_{l=1}^r N \sum_{s=1}^T (v_{ls}^0)^2 = N \norm{\bfV_r^0}_F^2 \leq rN,
$$
and
$$
\norm{Cov \left( vec\left( \left( \bGamma^{(-\calN_{\delta_2}(t))}\right)^{\top} \right)\right)} \leq \max_{1\leq i\leq N} \max_{k \in \calN_{\delta_2}(t)^c} \sum_{1\leq j \leq N} \sum_{q \in \calN_{\delta_2}(t)^c} \abs{\Cov(\epsilon_{is}\epsilon_{ik},\epsilon_{js}\epsilon_{jq} )} \leq C
$$
for some constant $C>0$ by Assumption B'''. Hence, we have the desired result. Except for the above parts, the proof is same as that of Lemma \ref{lem:bound_r1_dp}.

\paragraph{Proof of Lemma \ref{lem:bound_r2_general}.}
This proof is symmetric to the proof of Lemma \ref{lem:bound_r2} except for the part where we show $\norm{\bfY_r^{0\top} \bE \bfZ_r^{0\top}}_F = O_p\left( r \kappa \psi_{\min} \right)$. Note that
\begin{align*}
\bbE\left[ \left. \norm{\bfY_r^{0\top} \bE \bfZ_r^{0\top}}_F^2 \right| \bM^0 \right] &= \sum_{k=1}^r \sum_{l=1}^r \sum_{i=1}^N \sum_{j=1}^N \sum_{t=1}^T \sum_{s=1}^T \bbE \left[ \epsilon_{it} \epsilon_{js} \right] Y_{i,k}^0 Z_{t,l}^0 Y_{j,k}^0 Z_{s,l}^0 \\
&= \sum_{k=1}^r \sum_{l=1}^r \bfA_{k,l}^\top \Cov\left( vec(\bE ) \right) \bfA_{k,l}\\
&\leq \sum_{k=1}^r \sum_{l=1}^r \norm{\bfA_{k,l}}^2 \norm{\Cov\left( vec(\bE ) \right)} \\
& = \norm{\bfY_{r}^0}_F^2 \norm{\bfZ_{r}^0}_F^2 \norm{\Cov\left( vec(\bE ) \right)}\\
& \lesssim  r^2 \kappa^2 \psi_{\min}^2 ,
\end{align*}
where 
$\bfA_{k,l} = \left[Y_{1,k}^0 Z_{1,l}^0,Y_{2,k}^0 Z_{1,l}^0,\cdots,Y_{N,k}^0 Z_{1,l}^0 , Y_{1,k}^0 Z_{2,l}^0 , \cdots,  ,Y_{N,k}^0 Z_{2,l}^0 , Y_{1,k}^0 Z_{3,l}^0 , \cdots, Y_{N,k}^0 Z_{T,l}^0\right]$ is the $NT \times 1 $ vector. Here, we use the fact that
$$
\sum_{k=1}^r \sum_{l=1}^r \norm{\bfA_{k,l}}^2 = \sum_{k=1}^r \sum_{l=1}^r \sum_{i=1}^N \sum_{t=1}^T  (Y_{i,k}^0)^2 (Z_{t,l}^0)^2 = \norm{\bfY_{r}^0}_F^2 \norm{\bfZ_{r}^0}_F^2.
$$

\paragraph{Proof of Proposition \ref{pro:pre_general} (i).}

The proof is symmetric to that of Proposition \ref{pro:pre_general} (ii). Hence, we omit it. $\square$

\section{Proofs of Theorems and Lemma in the Main Text}

\subsection{Proofs of Theorems \ref{thm:convergence_indp} and \ref{thm:clt_indp}}\label{sec:proof_normality_PC_indp}

\paragraph{Proof of Theorem \ref{thm:clt_indp}.}

First, we prove Theorem \ref{thm:clt_indp}. Note that by Assumptions A(i) - (ii), $||T^{-1/2} \bF^0||$, $\psi_{r}(T^{-1/2}\bF^0)$, $||N^{-\alpha/2}\bLambda^0||$, and $\psi_{r}(N^{-\alpha/2}\bLambda^0)$ converge (in probability) to certain deterministic positive constants, respectively. Hence, w.h.p., there are constants $c,C > 0$ such that
\begin{align*}
 c &< \psi_{r}(N^{-\alpha/2}\bLambda^0)\psi_{r}(T^{-1/2}\bF^0) \leq N^{-\alpha/2} T^{-1/2} \psi_{r}(\bLambda^0 \bF^{0\top}) = N^{-\alpha/2} T^{-1/2}  \psi_{\min} \\
 &\leq N^{-\alpha/2} T^{-1/2} \psi_{\max} = N^{-\alpha/2} T^{-1/2}\norm{\bLambda^0 \bF^{0\top}} \leq \norm{N^{-\alpha/2}\bLambda^0} \norm{T^{-1/2}\bF^0} < C .  
\end{align*}
Then, because we have $c N^{\alpha/2} T^{1/2} \leq \psi_{\min} \leq \psi_{\max} \leq C N^{\alpha/2} T^{1/2}$ w.h.p., we can replace $\psi_{\min}$ in Assumption F with $N^{\alpha/2} T^{1/2}$, and $\kappa$ is bounded. In addition, since $u_i^0 = \bfG^\top (\bLambda^{0\top}\bLambda^0)^{-1/2} \lambda^0_i$ where $\bfG$ is an eigenvectors matrix of $\left( \frac{\bLambda^{0\top} \bLambda^0 }{N^\alpha} \right)^{1/2} \left( \frac{\bF^{0 \top} \bF^0 }{T} \right) \left( \frac{\bLambda^{0\top} \bLambda^0 }{N^\alpha} \right)^{1/2}$ as noted in the proof of Lemma \ref{lem:rotation}, we have
$$
\norm{u_i^0} \leq \norm{(\bLambda^{0\top}\bLambda^0)^{-1/2}} \norm{\lambda^0_i} = O_p\left( N^{(\alpha_i - \alpha  -1 )/2} \right) 
$$
by the assumption that $\norm{\lambda^0_i} \leq C N^{\frac{\alpha_i -1}{2}}$. So, we can set $\rho_i$ in Assumptions E(i) and F(i) to $N^{(\alpha_i - \alpha  -1 )/2}$. Then, Assumptions E(i) and F(i) are satisfied under Assumptions A and D(i). Similarly, because
$$
\norm{v_t^0} \leq \norm{(\bF^{0\top}\bF^0)^{-1/2}} \norm{f^0_t} = O_p\left( T^{  -1 /2} \right),
$$
we can set $q_t$ in Assumptions E(ii) and F(ii) to $T^{-1/2}$. Then, Assumptions E(ii) and F(ii) are satisfied under Assumptions A and D(ii).

\paragraph{Proof of Theorem \ref{thm:clt_indp} (i) and (ii).}

Using Lemma \ref{lem:rotation}, we can get the following decomposition from Proposition \ref{pro:pre_indp} (i):
\begin{gather}\label{eq:lambda_decomp}
 \sqrt{T}(\widehat{\lambda}_i - \bfH^{-1} \lambda_i^0 ) = \sqrt{T} \bfH^{-1} (\bF^{0\top} \bF^{0})^{-1} \bF^{0\top} \bfe_i + \bfD_r^{1/2} \bfO \calR_{y,i}    
\end{gather}
where $\bfH = (\bfD_r^{1/2} \bfO \bfH^0)^{-1}$. Then, by Assumptions A, C, Lemmas \ref{lem:limits}, and \ref{lem:Olimits}, we have
\begin{align*}
\sqrt{T} \bfH^{-1} (\bF^{0\top} \bF^{0})^{-1} \bF^{0\top} \bfe_i
 &= \left( \frac{\bfD_r^{1/2}}{N^{\alpha/4}T^{1/4}} \right)
 \bfO \left( N^{\alpha/4}T^{1/4} \bfH^0 \right) \left(\frac{\bF^{0\top}\bF^0}{T}\right)^{-1} \frac{1}{\sqrt{T}} \sum_{t=1}^T \epsilon_{it}f_t^0\\ 
&\conD \calN \left( 0,  \calD^{1/2} \calI_{sgn} \calD^{1/2}\calG^{*\top }\bSigma_{\bLambda}^{-1/2} \bSigma_{\bF}^{-1}\bPhi_{\bF,i} \bSigma_{\bF}^{-1} \bSigma_{\bLambda}^{-1/2}\calG^{*} \calD^{1/2}\calI_{sgn}\calD^{1/2}\right).
\end{align*}
Since $\calI_{sgn}$ and $\calD^{1/2}$ are diagonal matrices, we know $\calI_{sgn}\calD^{1/2} =\calD^{1/2}\calI_{sgn}$. In addition, $\calG^* \calI_{sgn}$ is also an eigenvector of $\bSigma_{\bLambda}^{1/2} \bSigma_{\bF} \bSigma_{\bLambda}^{1/2}$ which has a different column sign from $\calG^*$ because the diagonal elements of $\calI_{sgn}$ consists of $\pm 1$. In fact, as noted in the proofs of Lemmas \ref{lem:limits} and \ref{lem:Olimits}, the column sign of $\calG^*$ is determined by the sign alignment between $\bLambda^0$ and $\bfU_r^0$, and the sign of $\calI_{sgn}$ is determined by the sign alignment between $\bfU_r^0$ and $\bfU_r$. Hence, $\calG \coloneqq\calG^* \calI_{sgn}$ is the eigenvector of $\bSigma_{\bLambda}^{1/2} \bSigma_{\bF} \bSigma_{\bLambda}^{1/2}$ whose column sign is determined by the sign alignment between $\bLambda^0$ and $\bfU_r$. Hence, given $\bLambda^0$ and $\bfU_r$, it is determined. Then, we have
\begin{align*}
&\calN \left( 0,  \calD^{1/2} \calI_{sgn} \calD^{1/2}\calG^{*\top }\bSigma_{\bLambda}^{-1/2} \bSigma_{\bF}^{-1}\bPhi_{\bF,i} \bSigma_{\bF}^{-1} \bSigma_{\bLambda}^{-1/2}\calG^{*} \calD^{1/2}\calI_{sgn}\calD^{1/2}\right)\\
&\overset{d}{=}\calN \left( 0, \calD \calG^{\top }\bSigma_{\bLambda}^{-1/2} \bSigma_{\bF}^{-1}\bPhi_{\bF,i} \bSigma_{\bF}^{-1} \bSigma_{\bLambda}^{-1/2}\calG \calD \right).
\end{align*}
In addition, from the relation $\calD^{2} = \calG^\top \bSigma_{\bLambda}^{1/2} \bSigma_{\bF} \bSigma_{\bLambda}^{1/2}  \calG$, we have $\calG^\top\bSigma_{\bLambda}^{-1/2} \bSigma_{\bF}^{-1} = \calD^{-2} \calG^\top \bSigma_{\bLambda}^{1/2} $. Hence,
\begin{align*}
\calN \left( 0, \calD \calG^{\top }\bSigma_{\bLambda}^{-1/2} \bSigma_{\bF}^{-1}\bPhi_{\bF,i} \bSigma_{\bF}^{-1} \bSigma_{\bLambda}^{-1/2}\calG \calD \right) 
\overset{d}{=} \calN \left( 0, \calD^{-1} \calG^\top\bSigma_{\bLambda}^{1/2} \bPhi_{\bF,i} \bSigma_{\bLambda}^{1/2}  \calG \calD^{-1} \right)
\overset{d}{=} \calN \left( 0, \calQ^{- \top} \bPhi_{\bF,i} \calQ^{-1} \right),
\end{align*}
where $\calQ^{- \top} = \calD^{-1} \calG^\top \bSigma_{\bLambda}^{1/2}$. In addition, we have $||\bfD_r^{1/2}||||\bfO||||\calR_{y,i}|| = o_p(1)$ from Proposition \ref{pro:pre_indp} (i) since $\psi_{\max} \asymp N^{\alpha/2}T^{1/2}$ and $||\bfD_r^{1/2}||  \asymp N^{\alpha/4}T^{1/4}$.

Next, we show the asymptotic normality of $\widehat{f}_t$. Using Lemma \ref{lem:rotation}, we can get the following decomposition from Proposition \ref{pro:pre_indp} (ii):
\begin{gather}\label{eq:F_decomp}
 \sqrt{N^\alpha}\left( \widehat{f}_t - \bfH^{\top} f_t^0 \right) = \sqrt{N^\alpha} \bfH^{\top} (\bLambda^{0\top}\bLambda^0)^{-1} \bLambda^{0\top} \bfe_t + \sqrt{N^{\alpha}T} \bfD_r^{-1/2} \bfO \calR_{z,t}   
\end{gather}
where $\bfH = (\bfD_r^{1/2} \bfO \bfH^0)^{-1}$. Then, by Assumptions A, C, Lemmas \ref{lem:limits}, and \ref{lem:Olimits}, we have
\begin{align*}
 \sqrt{N^\alpha} \bfH^{\top} (\bLambda^{0\top}\bLambda^0)^{-1} \bLambda^{0\top} \bfe_t 
 &= \left(N^{\alpha/4}T^{1/4} \bfD_r^{-1/2} \right)  \bfO \left( \frac{1}{N^{\alpha/4}T^{1/4}} (\bfH^{0})^{- \top} \right)\left(\frac{\bLambda^{0\top}\bLambda^0}{N^\alpha}\right)^{-1} \frac{1}{\sqrt{N^{\alpha}}} \sum_{i=1}^N \epsilon_{is}\lambda^0_i\\ 
&\conD \calN \left( 0, \calD^{-1/2} \calI_{sgn}\calD^{-1/2} \calG^{*\top }\bSigma_{\bLambda}^{-1/2} \bPhi_{\bLambda,t} \bSigma_{\bLambda}^{-1/2}\calG^{*} \calD^{-1/2}\calI_{sgn} \calD^{-1/2} \right)\\
&\overset{d}{=} \calN \left( 0,  \calD^{-1}\calG^{\top }\bSigma_{\bLambda}^{-1/2} \bPhi_{\bLambda,t} \bSigma_{\bLambda}^{-1/2}\calG \calD^{-1} \right)\\
&\overset{d}{=} \calN \left( 0,  \calD^{-2} \calQ \bPhi_{\bLambda,t}  \calQ^\top \calD^{-2} \right).
\end{align*}
In addition, we have $\sqrt{N^{\alpha}T} ||\bfD_r^{-1/2}|| ||\bfO|| ||\calR_{z,t}||  = o_p(1)$ from Proposition \ref{pro:pre_indp} (ii) since $\psi_{\max} \asymp \sqrt{N^{\alpha}T}$ and $||\bfD_r^{1/2}||  \asymp N^{\alpha/4}T^{1/4}$. \ \ $\square$

\paragraph{Proof of Theorem \ref{thm:clt_indp} (iii).}

We have the following decomposition:
\begin{align*}
\widehat{m}_{it} - m^0_{it} &= \widehat{f}_{t}^{\top} \widehat{\lambda}_{i} - m^0_{it} \\
& = f_t^{0 \top} \left( \bfH \widehat{\lambda}_{i} - \lambda_i^0 \right) + \left( \bfH^{-\top}\widehat{f}_{t}  - f_t^0 \right)^{\top} \lambda_i^0 +  \left( \bfH^{-\top}\widehat{f}_{t}  - f_t^0 \right)^{\top} \left( \bfH \widehat{\lambda}_{i} - \lambda_i^0 \right) \\
&= f_t^{0 \top} \left(\bF^{0\top} \bF^0 \right)^{-1} \sum_{s=1}^{T} f_s^0 \epsilon_{is}  + \lambda_i^{0\top} \left(\bLambda^{0\top} \bLambda^0 \right)^{-1} \sum_{j=1}^{N} \lambda^0_j \epsilon_{jt} + \sum_{k=1}^{3} \Delta_{k,it},
\end{align*}
where the residual terms are
\begin{align*}
&\Delta_{1,it} = \frac{1}{\sqrt{T}} f_t^{0 \top} (\bfH^{0})^{-1} \calR_{y,i}, \ \
\Delta_{2,it} = \sqrt{T} \lambda_i^{0\top} (\bfH^{0})^{\top} \calR_{z,t},\ \
 \Delta_{3,it} = \left( \bfH^{-\top}\widehat{f}_{t}  - f_t^0 \right)^{\top} \left( \bfH \widehat{\lambda}_{i} - \lambda_i^0 \right).
\end{align*}
Furthermore, we have $\norm{\calR_{y,i}} = O_p\left(\frac{N^{(\alpha_i - \alpha - 1 )/2}\max\{N,T\}}{(N^\alpha T)^{3/4}} + \frac{\max\{\sqrt{N},\sqrt{T}\}}{(N^\alpha T)^{3/4}} \right) $ by Proposition \ref{pro:pre_indp} (i) since w.h.p., $\psi_{\min} \asymp N^{\alpha/2} T^{1/2}$ and $\kappa = O_p(1)$. So, we have by Lemma \ref{lem:limits} (iii) that
$$
\norm{\Delta_{1,it}} \leq \frac{1}{\sqrt{T}} \norm{f_t^0} \norm{(\bfH^{0})^{-1}} \norm{\calR_{y,i}} = O_p\left( \frac{ N^{\frac{\alpha_i -1}{2}}  \max\{N, T \}}{N^\alpha T } + \frac{ \max\{\sqrt{N}, \sqrt{T} \}}{N^{\alpha/2} T } \right) .
$$
Similarly, we have $\norm{\calR_{z,t}} =  O_p\left(\frac{\max\{N,T\}}{(N^\alpha T)^{3/4}\sqrt{T}} + \frac{\max\{\sqrt{N},\sqrt{T}\}}{(N^\alpha T)^{3/4}} \right) $ from Proposition \ref{pro:pre_indp} (ii). Hence, we have by Lemma \ref{lem:limits} (iii) that
$$
\norm{\Delta_{2,it}} \leq \sqrt{T} \norm{\lambda_i^0} \norm{\bfH^{0}} \norm{\calR_{z,t}} = O_p\left( \frac{ N^{\frac{\alpha_i - 1}{2}}  \max\{N, T \}}{N^\alpha T } + \frac{ N^{\frac{\alpha_i -1}{2}} \max\{\sqrt{N}, \sqrt{T} \}}{N^\alpha T^{1/2}  } \right) .
$$
In addition, the terms in $\Delta_{3,it}$ can be bounded like
\begin{align*}
&\norm{\left((\bLambda^{0\top}\bLambda^0)^{-1} \sum_{i=1}^{N} \lambda_i^0 \epsilon_{it}\right)^{\top}(\bF^{0\top} \bF^0)^{-1} \sum_{t=1}^{T} f_t^0 \epsilon_{it}} \leq \norm{(\bLambda^{0\top}\bLambda^0)^{-1}} \norm{(\bF^{0\top} \bF^0)^{-1}} \norm{ \sum_{i=1}^{N} \lambda_i^0 \epsilon_{it}} \norm{\sum_{t=1}^{T} f_t^0 \epsilon_{it}} \\
&  \quad \qquad\qquad\qquad\qquad\qquad\qquad\qquad\qquad\qquad\qquad = O_p\left( \frac{1}{N^{\alpha/2} T^{1/2}} \right) ,\\
& \norm{(\bLambda^{0\top}\bLambda^0)^{-1} \sum_{j=1}^{N} \lambda^0_j \epsilon_{jt}} \frac{1}{\sqrt{T}} \norm{(\bfH^{0})^{-1} \calR_{y,i}} 
 = O_p\left( \frac{ N^{\frac{\alpha_i -1}{2}} \max\{N, T \}}{N^{3\alpha/2} T } + \frac{ \max\{\sqrt{N}, \sqrt{T} \}}{N^{\alpha} T } \right),\\
& \norm{(\bF^{0\top} \bF^0)^{-1} \sum_{s=1}^{T} f_s^0 \epsilon_{is}} \sqrt{T} \norm{(\bfH^{0})^{\top}  \calR_{z,t}} 
 = O_p\left( \frac{ \max\{N, T \}}{N^\alpha T^{3/2} } + \frac{ \max\{\sqrt{N}, \sqrt{T} \}}{N^{\alpha} T } \right).
\end{align*}
Next, we derive the bound of $\calV_{it}^{-1}$. Note that w.h.p.,
$$
\norm{ N^{-\alpha}  \lambda_i^{0\top} \bSigma_{\bLambda}^{-1} \bPhi_{\bLambda,t} \bSigma_{\bLambda}^{-1} \lambda_i^0}  
\geq N^{-\alpha} \norm{ \lambda_{i}^0}^2 \lambda_{\min} \left(  \bSigma_{\bLambda}^{-1} \right)^2 \lambda_{\min}(\bPhi_{\bLambda,t}) \geq   \frac{c}{N^{(1 + \alpha - \alpha_i)} },
$$
for some constant $c > 0$, because $\norm{\lambda_i^0} \geq c N^{(\alpha_i - 1)/2}$. Similarly, we have $\norm{ T^{-1} f_t^{0 \top} \bSigma_{\bF}^{-1} \bPhi_{\bF,i} \bSigma_{\bF}^{-1} f_t^0}
\geq \frac{c}{T}$ for some constant $c>0$ w.h.p. Therefore, we have
$$
\calV_{it}^{-1} = \frac{1}{N^{-\alpha}  \lambda_i^{0\top} \bSigma_{\bLambda}^{-1} \bPhi_{\bLambda,t} \bSigma_{\bLambda}^{-1} \lambda_i^0 + T^{-1} f_t^{0 \top} \bSigma_{\bF}^{-1} \bPhi_{\bF,i} \bSigma_{\bF}^{-1} f_t^0} = O_p\left( \min\{ N^{(1 + \alpha - \alpha_i)} ,T \} \right),
$$
and we can check that $\calV_{it}^{-1/2}\norm{\sum_{k=1}^3 \Delta_{k,it}} = o_p(1)$ under our assumptions. Lastly, we can show that $\calV_{it}^{-1/2}\left(f_t^{0 \top} \left(\bF^{0\top} \bF^0 \right)^{-1} \sum_{s=1}^{T} f_s^0 \epsilon_{is}  + \lambda_i^{0\top} \left(\bLambda^{0\top} \bLambda^0 \right)^{-1} \sum_{j=1}^{N} \lambda_j^0 \epsilon_{jt}\right) \conD \calN(0,1)$ by the same assertion as in the proof of Theorem 3 in \cite{bai2003inferential}. It completes the proof. $\square$
\bigskip

\paragraph{Proof of Theorem \ref{thm:convergence_indp}.}

First, we check the following conditions of Proposition \ref{pro:pre_indp}:
$$
\kappa \rho_i \conP 0, \qquad \kappa q_t \conP 0, \qquad \text{and} \qquad  \frac{ \max\{\sqrt{N},\sqrt{T}\}}{ \psi_{\min}} \conP 0.
$$
As noted above, we can replace $\psi_{\min}$, $\rho_i$, $q_{t}$ with $N^{\alpha/2}T^{1/2}$, $N^{(\alpha_i - \alpha - 1)/2}$, $T^{-1/2}$, respectively. In addition, w.h.p., $\kappa$ is bounded. Hence, we have w.h.p.,
\begin{gather*}
\kappa \rho_i \lesssim N^{(\alpha_i - \alpha - 1)/2} \leq N^{- \alpha/2} \rightarrow 0, \quad \kappa q_t \lesssim T^{- 1/2} \rightarrow 0, \quad 
\frac{ \max\{\sqrt{N},\sqrt{T}\}}{ \psi_{\min}} \asymp \frac{ \max\{\sqrt{N},\sqrt{T}\}}{ N^{\alpha/2}T^{1/2}} \rightarrow  0.  
\end{gather*}
Hence, the above conditions of Proposition \ref{pro:pre_indp} are satisfied, and we can use the bounds of $\norm{\calR_{y,i}}$ and $\norm{\calR_{z,t}}$ in  Proposition \ref{pro:pre_indp} by replacing $\psi_{\min}$, $\rho_i$, $q_{t}$ with $N^{\alpha/2}T^{1/2}$, $N^{(\alpha_i - \alpha - 1)/2}$, $T^{-1/2}$. Note that
$$
\widehat{\lambda}_i - \bfH^{-1} \lambda_i^0  =  \bfH^{-1} (\bF^{0\top} \bF^{0})^{-1} \bF^{0\top} \bfe_i + \frac{1}{\sqrt{T}} \bfD_r^{1/2} \bfO \calR_{y,i} .
$$
By Lemma \ref{lem:limits}, $\bfH^{-1} = \bfD_r^{1/2} \bfO \bfH^0 = O_p(1)$, $\bfD_r^{1/2} = O_p(N^{\alpha/4}T^{1/4})$, and $(\bF^{0\top} \bF^{0})^{-1} \bF^{0\top} \bfe_i = O_p \left( \frac{1}{\sqrt{T}} \right) $. Hence, we have from Proposition \ref{pro:pre_indp} (i) that
\begin{gather*}
\norm{\widehat{\lambda}_i - \bfH^{-1} \lambda_i^0 } = O_p \left( \frac{1}{\sqrt{T}} + \sqrt{\frac{N^{\alpha_i}}{N}} \frac{\max\{N, T \}}{N^\alpha T} \right).
\end{gather*}
Note that
\begin{gather*}
\widehat{f}_t - \bfH^{\top} f_t^0  =  \bfH^{\top} (\bLambda^{0\top}\bLambda^0)^{-1} \bLambda^{0\top} \bfe_t  + \sqrt{T} \bfD_r^{-1/2} \bfO \calR_{z,t} .
\end{gather*}
By Lemma \ref{lem:limits}, $\bfH^{\top} = \bfD_r^{-1/2} \bfO (\bfH^0)^{-\top} = O_p(1)$, $\bfD_r^{-1/2} = O_p(1 / N^{\alpha/4}T^{1/4})$, and $(\bLambda^{0\top}\bLambda^0)^{-1} \bLambda^{0\top} \bfe_t = O_p \left( \frac{1}{\sqrt{T}} \right) $. Similarly, we can have from Proposition \ref{pro:pre_indp} (ii) that
\begin{gather*}
\norm{ \widehat{f}_t - \bfH^{\top} f_t^0 } = O_p \left( \frac{1}{\sqrt{N^\alpha}} + \frac{\max\{N, T \}}{N^\alpha T} \right).
\end{gather*}
In addition, the bound of $||\widehat{m}_{it} - m^0_{it}||$ can be easily derived from the relation
$$
\norm{\widehat{m}_{it} - m^0_{it}} \leq \norm{\widehat{\lambda}_i - \bfH^{-1} \lambda_i^0} \norm{\bfH^{\top} f_t^0} + \norm{\widehat{f}_t - \bfH^{\top} f_t^0} \norm{\bfH^{-1} \lambda_i^0} + \norm{\widehat{\lambda}_i - \bfH^{-1} \lambda_i^0} \norm{\widehat{f}_t - \bfH^{\top} f_t^0}.
$$
Hence, we omit it here. $\square$

\subsection{Proof of Theorems \ref{thm:convergence_dp} and \ref{thm:clt_dp}}\label{sec:proof_normality_PC_dp}

\paragraph{Proof of Theorem \ref{thm:clt_dp} (i) and (ii).}

First, we prove Theorem \ref{thm:clt_dp}. As noted in Section \ref{sec:proof_normality_PC_indp}, we can replace $\psi_{\min}$, $\rho_i$, $q_{1,t}$ in Assumptions E' and F' with $N^{\alpha/2}T^{1/2}$, $N^{(\alpha_i - \alpha - 1)/2}$, $T^{- 1/2}$, respectively. In addition, because $\norm{v_s^0} \leq \norm{(\bF^{0 \top} \bF^0)^{-1/2}}\norm{f_s^0}$, we have
\begin{align*}
\frac{1}{|\calN_\delta(t)|} \sum_{s \in \calN_\delta(t)} \norm{v_s^0}^2 \leq \norm{(\bF^{0 \top} \bF^0)^{-1/2}}^2 \frac{1}{|\calN_\delta(t)|} \sum_{s \in \calN_\delta(t)} \norm{f_s^0}^2 = O_p\left( T^{- 1} \right)
\end{align*}
by Assumption A(iv). So, we can set $q_{2,t}$ in Assumptions E'(ii) and F'(ii) to $T^{- 1/2}$. Moreover, we have
\begin{align*}
&\frac{1}{|\calN_\delta(t)| N} \sum_{s \in \calN_\delta(t)} \sum_{i=1}^N \sum_{k \in \calN_\delta (t)^c}|\Cov(\epsilon_{is},\epsilon_{ik})| \norm{v_k^0} \\ &\leq 
\norm{(\bF^{0 \top} \bF^0)^{-1/2}} \frac{1}{|\calN_\delta(t)| N} \sum_{s \in \calN_\delta(t)} \sum_{i=1}^N \sum_{k \in \calN_\delta (t)^c}|\Cov(\epsilon_{is},\epsilon_{ik})| \norm{f_k^0} = O_p\left( T^{-1/2} \right),
\end{align*}
because by Assumption B',
\begin{align*}
&\frac{1}{|\calN_\delta(t)| N} \sum_{s \in \calN_\delta(t)} \sum_{i=1}^N \sum_{k \in \calN_\delta (t)^c}|\Cov(\epsilon_{is},\epsilon_{ik})| \bbE[\norm{f_k^0}] \\
&\ \ \leq \frac{C_1}{|\calN_\delta(t)| N} \sum_{s \in \calN_\delta(t)} \sum_{i=1}^N \sum_{k \in \calN_\delta (t)^c}|\Cov(\epsilon_{is},\epsilon_{ik})|
\leq C_1 \max_{s \in \calN_\delta(t)} \max_{i \leq N} \sum_{k \in \calN_\delta (t)^c}|\Cov(\epsilon_{is},\epsilon_{ik})| \leq C_2 ,
\end{align*}
for some constants $C_1,C_2 > 0$. Hence, we can set $q_{3,t}$ in Assumptions E'(ii) and F'(ii) to $T^{ - 1/2}$. Then, the assumptions of Proposition \ref{pro:pre_dp} are satisfied under the assumptions of Theorem \ref{thm:clt_dp}, once we replace $\psi_{\min}$, $\rho_i$, $q_{1,t}$, $q_{2,t}$, $q_{3,t}$ with $N^{\alpha/2}T^{1/2}$, $N^{(\alpha_i - \alpha - 1)/2}$, $T^{- 1/2}$, $T^{- 1/2}$, $T^{ - 1/2}$, respectively. Therefore, we can use the result of Proposition \ref{pro:pre_dp}. Except for this discussion, the proofs of Theorem \ref{thm:clt_dp} (i) and (ii) are the same as those of Theorem \ref{thm:clt_indp} (i) and (ii) in Section \ref{sec:proof_normality_PC_indp}. Hence, we omit it. $\square$

\paragraph{Proof of Theorem \ref{thm:clt_dp} (iii).}

The way of proof is basically the same as that of Theorem \ref{thm:clt_indp} (iii). By the same token as in the proof of Theorem \ref{thm:clt_indp} (iii) with the aid of Proposition \ref{pro:pre_dp}, we can derive
\begin{align*}
& \norm{\Delta_{1,it}}=  O_p\left( \frac{ N^{\frac{\alpha_i -1}{2}} \max\{N, T \}}{N^\alpha T } + \frac{ \max\{\sqrt{N}, \sqrt{T} \}}{N^{\alpha/2} T } \right) , \\  
&\norm{\Delta_{2,it}} = O_p\left( 
 \frac{N^{\frac{\alpha_i }{2}}  \max\{\sqrt{N},\sqrt{T}\} (\log N)^{\nu} }{N^\alpha T } 
 +  \frac{N^{\frac{\alpha_i -1}{2}}  \max\{N,T\}  }{N^\alpha T } \right. \\
& \qquad \qquad \qquad  \left. + \frac{N^{\frac{\alpha_i -1}{2}} \max\{\sqrt{N},\sqrt{T}\} N^{1/6}}{N^\alpha T^{1/2}  } 
  + \frac{N^{\frac{\alpha_i }{2}} \max\{N^{3/2},T^{3/2} \} (\log N)^{\nu/2}}{N^{2 \alpha} T^{3/2} }
 \right), \\ 
&\norm{\Delta_{3,it}} = O_p \left( \frac{1}{N^{\alpha/2} T^{1/2} }
+ \frac{ N^{\frac{\alpha_i -1}{2}} \max\{N, T \}}{N^{3\alpha/2} T  } + \frac{ \max\{N, T \} N^{1/6}}{N^\alpha T^{3/2}  }
\right. \\
&  \qquad \qquad \qquad \left.   + \frac{ \sqrt{N} \max\{N^{3/2},T^{3/2} \} (\log N)^{\nu/2}}{N^{2\alpha} T^{2} }  + \frac{ N^{\frac{\alpha_i -1}{2}} \max\{N^2, T^2 \}N^{1/6}}{N^{2\alpha} T^2  } \right. \\
&  \qquad \qquad \qquad \left.
+ \frac{ N^{\frac{\alpha_i }{2}} \max\{N^{5/2}, T^{5/2} \}(\log N)^{\nu/2}}{N^{3\alpha} T^{5/2}  }
\right) .
\end{align*}
In addition, we have from the proof of Theorem \ref{thm:clt_indp} (iii) that
$$
\calV_{it}^{-1} = \frac{1}{N^{-\alpha}  \lambda_i^{0\top} \bSigma_{\bLambda}^{-1} \bPhi_{\bLambda,t} \bSigma_{\bLambda}^{-1} \lambda_i^0 + T^{-1} f_t^{0 \top} \bSigma_{\bF}^{-1} \bPhi_{\bF,i} \bSigma_{\bF}^{-1} f_t^0} = O_p\left( \min\{ N^{(1 + \alpha - \alpha_i)} ,T \} \right).
$$
So, $\calV_{it}^{-1/2}\sum_{k=1}^3 \Delta_{k,it} = o_p(1)$ under our assumptions. Lastly, we can show that 
$$
\calV_{it}^{-1/2}\left(f_t^{0\top} \left(\bF^{0\top} \bF^{0} \right)^{-1} \sum_{s=1}^{T} f_s^0 \epsilon_{is}  + \lambda_i^{0\top} \left(\bLambda^{0\top} \bLambda^0 \right)^{-1} \sum_{j=1}^{N} \lambda_j^0 \epsilon_{jt}\right) \conD \calN(0,1)
$$
by the same assertion as in the proof of Theorem 3 in \cite{bai2003inferential}. It completes the proof. $\square$
\bigskip

\paragraph{Proof of Theorem \ref{thm:convergence_dp}.}

First, we check the following conditions of Proposition \ref{pro:pre_dp}:
$$
\kappa \rho_i \conP 0, \qquad (\log N)^{\nu/2} \kappa q_{2,t} \conP 0, \qquad \frac{ \max\{\sqrt{N},\sqrt{T}\}}{ \psi_{\min}} \conP 0.
$$
As noted above, we can replace $\psi_{\min}$, $\rho_i$, $q_{2,t}$ with $N^{\alpha/2}T^{1/2}$, $N^{(\alpha_i - \alpha - 1)/2}$, $T^{ - 1/2}$, respectively. In addition, w.h.p., $\kappa$ is bounded. Hence, we have w.h.p.,
\begin{gather*}
\kappa \rho_i \lesssim N^{(\alpha_i - \alpha - 1)/2} \leq N^{( - \alpha/2)} \rightarrow 0, \qquad \frac{ \max\{\sqrt{N},\sqrt{T}\}}{ \psi_{\min}} \asymp \frac{ \max\{\sqrt{N},\sqrt{T}\}}{ N^{\alpha/2}T^{1/2}} \rightarrow 0,\\ 
(\log N)^{\nu/2} \kappa q_{2,t} \lesssim (\log N)^{\nu/2} 
 T^{- 1 /2} \rightarrow 0 .  
\end{gather*}
Then, the above conditions of Proposition \ref{pro:pre_dp} are satisfied. In addition, as noted above, we can set $\rho_i$, $q_{1,t}$, $q_{2,t}$, $q_{3,t}$ in Assumption E' to $N^{\alpha/2}T^{1/2}$, $N^{(\alpha_i - \alpha - 1)/2}$, $T^{ - 1/2}$, $T^{ - 1/2}$, $T^{ - 1/2}$, respectively. Hence, we can use the bounds of $\norm{\calR_{y,i}}$ and $\norm{\calR_{z,t}}$ in Proposition \ref{pro:pre_dp} after replacing $\psi_{\min}$, $\rho_i$, $q_{1,t}$, $q_{2,t}$, $q_{3,t}$ with $N^{\alpha/2}T^{1/2}$, $N^{(\alpha_i - \alpha - 1)/2}$, $T^{- 1/2}$, $T^{ - 1/2}$, $T^{ - 1/2}$, respectively. Note that
$$
\widehat{\lambda}_i - \bfH^{-1} \lambda_i^0  =  \bfH^{-1} (\bF^{0\top} \bF^{0})^{-1} \bF^{0\top} \bfe_i + \frac{1}{\sqrt{T}} \bfD_r^{1/2} \bfO \calR_{y,i} .
$$
By Lemma \ref{lem:limits}, $\bfH^{-1} = \bfD_r^{1/2} \bfO \bfH^0 = O_p(1)$, $\bfD_r^{1/2} = O_p(N^{\alpha/4}T^{1/4})$, and $(\bF^{0\top} \bF^{0})^{-1} \bF^{0\top} \bfe_i = O_p \left( \frac{1}{\sqrt{T}} \right) $. Hence, we have from Proposition \ref{pro:pre_dp} (i) that
\begin{gather*}
\norm{\widehat{\lambda}_i - \bfH^{-1} \lambda_i^0 } = O_p \left( \frac{1}{\sqrt{T}} + \sqrt{\frac{N^{\alpha_i}}{N}} \frac{\max\{N, T \}}{N^\alpha T} \right).
\end{gather*}
In addition, because $\widehat{f}_t - \bfH^{\top} f_t^0  =  \bfH^{\top} (\bLambda^{0\top}\bLambda^0)^{-1} \bLambda^{0\top} \bfe_t  + \sqrt{T} \bfD_r^{-1/2} \bfO \calR_{z,t}$, we can similarly have from Proposition \ref{pro:pre_dp} (ii) that
\begin{gather*}
\norm{ \widehat{f}_t - \bfH^{\top} f_t^0 } = O_p \left( \frac{1}{\sqrt{N^\alpha}} + \frac{\max\{N, T \}N^{1/6}}{N^\alpha T}   + \frac{\sqrt{N} \max\{N^{3/2}, T^{3/2} \}(\log N)^{\nu}}{N^{2\alpha} T^{3/2}}\right).
\end{gather*}
In addition, the bound of $||\widehat{m}_{it} - m^0_{it}||$ can easily be derived from the bounds of $||\widehat{\lambda}_i - \bfH^{-1} \lambda_i^0 ||$ and $|| \widehat{f}_t - \bfH^{\top} f_t^0||$. Hence, we omit it. $\square$

\subsection{Proof of Theorems \ref{thm:convergence_cross} and \ref{thm:clt_cross}}\label{sec:proof_normality_PC_cross}

\paragraph{Proof of Theorem \ref{thm:clt_cross} (i) and (ii).}

As noted in Section \ref{sec:proof_normality_PC_indp}, we can replace $\psi_{\min}$, $\rho_{1,i}$, $q_{t}$ in Assumptions E'' and F'' with $N^{\alpha/2}T^{1/2}$, $N^{(\alpha_{1,i} - \alpha - 1)/2}$, $T^{ - 1/2}$, respectively. In addition, we have
\begin{align*}
\frac{1}{|\calN_\delta(i)|} \sum_{j \in \calN_\delta(i)} \norm{u_j^0}^2 \leq \norm{(\bLambda^{0 \top} \bLambda^0)^{-1/2}}^2 \frac{1}{|\calN_\delta(i)|} \sum_{j \in \calN_\delta(i)} \norm{\lambda_j^0}^2 = O_p\left( N^{\alpha_{2,i} - \alpha  - 1} \right)
\end{align*}
by Assumption A'. So, we can set $\rho_{2,i}$ in Assumptions E''(i) and F''(i) to $N^{(\alpha_{2,i} - \alpha  - 1)/2}$. Moreover, we have
\begin{align*}
&\frac{1}{|\calN_\delta(i)| T} \sum_{j \in \calN_\delta(i)} \sum_{t=1}^T \sum_{k \in \calN_\delta (i)^c}|\Cov(\epsilon_{jt},\epsilon_{kt})| \norm{u_k^0} \\ &\leq 
\norm{(\bLambda^{0 \top} \bLambda^0)^{-1/2}} \frac{1}{|\calN_\delta(i)| T} \sum_{j \in \calN_\delta(i)} \sum_{t=1}^T \sum_{k \in \calN_\delta (i)^c}|\Cov(\epsilon_{jt},\epsilon_{kt})| \norm{\lambda_k^0} = O_p\left( N^{-\alpha/2} \right),
\end{align*}
because by Assumption B'',
\begin{align*}
&\frac{1}{|\calN_\delta(i)| T} \sum_{j \in \calN_\delta(i)} \sum_{t=1}^T \sum_{k \in \calN_\delta (i)^c}|\Cov(\epsilon_{jt},\epsilon_{kt})| \norm{\lambda_k^0}  \\
&\ \ \leq \frac{C_1}{|\calN_\delta(i)| T} \sum_{j \in \calN_\delta(i)} \sum_{t=1}^T \sum_{k \in \calN_\delta (i)^c}|\Cov(\epsilon_{jt},\epsilon_{kt})| 
\leq C_1 \max_{j \in \calN_\delta(i)} \max_{t \leq T} \sum_{k \in \calN_\delta (i)^c}|\Cov(\epsilon_{jt},\epsilon_{kt})| \leq C_2 ,
\end{align*}
for some constants $C_1,C_2 > 0$. Hence, we can set $\rho_{3,i}$ in Assumptions E''(i) and F''(i) to $N^{ - \alpha/2}$. Then, the assumptions of Proposition \ref{pro:pre_cross} are satisfied under the assumptions of Theorem \ref{thm:clt_cross}, once we replace $\psi_{\min}$, $q_t$, $\rho_{1,i}$, $\rho_{2,i}$, $\rho_{3,i}$ with $N^{\alpha/2}T^{1/2}$, $T^{- 1/2}$, $N^{(\alpha_{1,i} - \alpha - 1)/2}$, $N^{(\alpha_{2,i} - \alpha - 1)/2}$, $N^{ - \alpha/2}$, respectively. Therefore, we can use the result of Proposition \ref{pro:pre_cross}. Except for this discussion, the proofs of Theorem \ref{thm:clt_cross} (i) and (ii) are the same as those of Theorem \ref{thm:clt_indp} (i) and (ii) in Section \ref{sec:proof_normality_PC_indp}. Hence, we omit it. $\square$

\paragraph{Proof of Theorem \ref{thm:clt_cross} (iii).}

The way of proof is basically the same as that of Theorem \ref{thm:clt_indp} (iii). By the same token as in the proof of Theorem \ref{thm:clt_indp} (iii) with the aid of Proposition \ref{pro:pre_cross}, we can derive
\begin{align*}
&\norm{\Delta_{1,it}} = O_p\left( 
 \frac{ N^{\frac{\alpha_{2,i} -1}{2}} \max\{\sqrt{N},\sqrt{T}\} (\log N)^{\omega} }{N^\alpha T^{1/2} } 
 +  \frac{N^{\frac{\alpha_{1,i} -1}{2}} \max\{N,T\}  }{N^\alpha T } \right. \\
& \qquad \qquad \qquad  \left. + \frac{ \max\{\sqrt{N},\sqrt{T} \} T^{1/6}}{N^{\alpha/2} T  } 
  + \frac{ \max\{N^{3/2},T^{3/2} \} (\log N)^{\omega/2}}{ N^{ 3\alpha/2} T^{3/2} }
 \right) \coloneqq O_p(A), \\ 
& \norm{\Delta_{2,it}}=  O_p\left( \frac{ N^{\frac{\alpha_{1,i} -1}{2}} \max\{N, T \}}{N^\alpha T } + \frac{ N^{\frac{\alpha_{1,i} -1}{2}} \max\{\sqrt{N}, \sqrt{T} \}}{N^{\alpha} T^{1/2} } \right) , \\  
&\norm{\Delta_{3,it}} = O_p \left( \frac{1}{N^{\alpha/2} T^{1/2} }
+ \frac{ \max\{N, T \}}{N^\alpha T^{3/2}  } + A \right) .
\end{align*}
In addition, we have from the proof of Theorem \ref{thm:clt_indp} (iii) that
$$
\calV_{it}^{-1} = \frac{1}{N^{-\alpha}  \lambda_i^{0\top} \bSigma_{\bLambda}^{-1} \bPhi_{\bLambda,t} \bSigma_{\bLambda}^{-1} \lambda_i^0 + T^{-1} f_t^{0 \top} \bSigma_{\bF}^{-1} \bPhi_{\bF,i} \bSigma_{\bF}^{-1} f_t^0} = O_p\left( \min\{ N^{(1 + \alpha - \alpha_{1,i})} ,T  \} \right).
$$
So, $\calV_{it}^{-1/2}\sum_{k=1}^3 \Delta_{k,it} = o_p(1)$ under our assumptions. Lastly, we can show that 
$$
\calV_{it}^{-1/2}\left(f_t^{0\top} \left(\bF^{0\top} \bF^{0} \right)^{-1} \sum_{s=1}^{T} f_s^0 \epsilon_{is}  + \lambda_i^{0\top} \left(\bLambda^{0\top} \bLambda^0 \right)^{-1} \sum_{j=1}^{N} \lambda_j^0 \epsilon_{jt}\right) \conD \calN(0,1)
$$
by the same assertion as in the proof of Theorem 3 in \cite{bai2003inferential}. It completes the proof. $\square$
\bigskip

\paragraph{Proof of Theorems \ref{thm:convergence_cross}.}

First, we check the following conditions of Proposition \ref{pro:pre_cross}:
$$
\kappa q_t \conP 0, \qquad (\log N)^{\nu/2} \kappa \rho_{2,i} \conP 0, \qquad \frac{ \max\{\sqrt{N},\sqrt{T}\}}{ \psi_{\min}} \conP 0 .
$$
As noted above, we can replace $\psi_{\min}$, $q_t$, $\rho_{2,i}$ with $N^{\alpha/2}T^{1/2}$, $T^{ - 1/2}$, $N^{(\alpha_{2,i} - \alpha - 1)/2}$, respectively. In addition, w.h.p., $\kappa$ is bounded. Hence, we have w.h.p.,
\begin{gather*}
\kappa q_t \lesssim  T^{ - 1/2} \rightarrow 0, \qquad \frac{ \max\{\sqrt{N},\sqrt{T}\}}{ \psi_{\min}} \asymp \frac{ \max\{\sqrt{N},\sqrt{T}\}}{ N^{\alpha/2}T^{1/2}} \rightarrow 0,\\ 
(\log N)^{\nu/2} \kappa \rho_{2,i} \lesssim (\log N)^{\nu/2} 
 N^{(\alpha_{2,i} - \alpha - 1)/2} \leq (\log N)^{\nu/2} N^{- \alpha/2} \rightarrow 0 .  
\end{gather*}
Then, the above conditions of Proposition \ref{pro:pre_cross} are satisfied. In addition, as noted above, we can set $q_t$, $\rho_{1,i}$, $\rho_{2,i}$, $\rho_{3,i}$ in Assumption E'' to $N^{\alpha/2}T^{1/2}$, $T^{- 1/2}$, $N^{(\alpha_{1,i} - \alpha - 1)/2}$, $N^{(\alpha_{2,i} - \alpha - 1)/2}$, $N^{ - \alpha/2}$, respectively. Hence, we can use the bounds of $\norm{\calR_{y,i}}$ and $\norm{\calR_{z,t}}$ in Proposition \ref{pro:pre_cross} by replacing $\psi_{\min}$, $q_t$, $\rho_{1,i}$, $\rho_{2,i}$, $\rho_{3,i}$ with $N^{\alpha/2}T^{1/2}$, $T^{ - 1 / 2}$, $N^{(\alpha_{1,i} - \alpha - 1)/2}$, $N^{(\alpha_{2,i} - \alpha - 1)/2}$, $N^{ - \alpha/2}$, respectively. Then, by using the similar steps in the Proof of Theorem \ref{thm:convergence_indp}, we can derive the desired results from Proposition \ref{pro:pre_cross}. $\square$

\subsection{Proof of Theorems \ref{thm:convergence_general} and \ref{thm:clt_general}}\label{sec:proof_normality_PC_general}

The convergence rate and the asymptotic normality of $\widehat{\lambda}_i$ can be derived by the same token as in the cross-sectional dependence case in Section \ref{sec:proof_normality_PC_cross} from Proposition \ref{pro:pre_general} (i). Similarly, the convergence rate and the asymptotic normality of $\widehat{f}_t$ can be derived by the same way as in the temporal dependence case in Section \ref{sec:proof_normality_PC_dp} from Proposition \ref{pro:pre_general} (ii). In addition, the bound of $||\widehat{m}_{it} - m^0_{it}||$ can easily be derived from the bounds of $||\widehat{\lambda}_i - \bfH^{-1} \lambda_i^0 ||$ and $|| \widehat{f}_t - \bfH^{\top} f_t^0||$. Lastly, for the asymptotic normality of $\widehat{m}_{it}$, the way of proof is basically the same as that of Theorem \ref{thm:clt_indp} (iii). By the same token as in the proof of Theorem \ref{thm:clt_indp} (iii) with the aid of Proposition \ref{pro:pre_general}, we can derive
\begin{align*}
&\norm{\Delta_{1,it}} = O_p\left( 
 \frac{ N^{\frac{\alpha_{2,i} -1}{2}} \max\{\sqrt{N},\sqrt{T}\} (\log N)^{\omega} }{N^\alpha T^{1/2} } 
 +  \frac{N^{\frac{\alpha_{1,i} -1}{2}} \max\{N,T\}  }{N^\alpha T } \right. \\
& \qquad \qquad \qquad  \left. + \frac{ \max\{\sqrt{N},\sqrt{T} \} T^{1/6}}{N^{\alpha/2} T  } 
  + \frac{ \max\{N^{3/2},T^{3/2} \} (\log N)^{\omega/2}}{ N^{ 3\alpha/2} T^{3/2} }
 \right) \coloneqq O_p(A), \\ 
&\norm{\Delta_{2,it}} = O_p\left( 
 \frac{N^{\frac{\alpha_{1,i}}{2}}  \max\{\sqrt{N},\sqrt{T}\} (\log N)^{\nu} }{N^\alpha T } 
 +  \frac{N^{\frac{\alpha_{1,i} -1}{2}}  \max\{N,T\}  }{N^\alpha T } \right. \\
& \qquad \qquad \qquad  \left. + \frac{N^{\frac{\alpha_{1,i} -1}{2}} \max\{\sqrt{N},\sqrt{T}\} N^{1/6}}{N^\alpha T^{1/2}  } 
  + \frac{N^{\frac{\alpha_{1,i}}{2}} \max\{N^{3/2},T^{3/2} \} (\log N)^{\nu/2}}{N^{2 \alpha} T^{3/2} }
 \right)\\
& \norm{\Delta_{3,it}} = O_p \left( A +  \frac{1}{N^{\alpha/2} T^{1/2} }
   +  \frac{\max\{N, T \}N^{1/6}}{N^\alpha T^{3/2}} + \left[ \sqrt{\frac{N^{\alpha_{1,i}}}{N}} + \sqrt{\frac{N^{\alpha_{2,i}}}{N}} \right] \frac{\max\{N^2, T^2 \}N^{1/6}(\log N)^{\omega}}{N^{2\alpha} T^2 } \right.
 \\
  &  \qquad\quad +  \frac{\max\{N^{3/2}, T^{3/2} \}N^{1/6}}{N^{3\alpha/2} T^{11/6}} +  \frac{\max\{N^{5/2}, T^{5/2} \}N^{1/6}(\log N)^{\omega/2}}{N^{5\alpha/2} T^{5/2}}   \\
   &   \qquad\quad  
   +  \frac{\sqrt{N}\max\{N^{3/2}, T^{3/2} \}(\log N)^{\nu/2}}{N^{2\alpha} T^{2}} +  \left( \sqrt{N^{\alpha_{1,i}}} + \sqrt{N^{\alpha_{2,i}}} \right)  \frac{\max\{N^{5/2}, T^{5/2} \}(\log N)^{\nu/2+\omega}}{N^{3\alpha} T^{5/2}} \\
   &  \left. \qquad\quad  
   +  \frac{\sqrt{N}\max\{N^{2}, T^{2} \}(\log N)^{\nu/2}}{N^{5\alpha/2} T^{7/3}} +  \frac{\sqrt{N}\max\{N^{3}, T^{3} \}(\log N)^{(\omega+\nu)/2}}{N^{7\alpha/2} T^{3}}
   \right) .
 \end{align*}
Then, $\calV_{it}^{-1/2}\sum_{k=1}^3 \Delta_{k,it} = o_p(1)$ where $\calV_{it}^{-1} = O_p\left( \min\{ N^{(1 + \alpha - \alpha_{1,i})} ,T  \} \right)$ under our assumptions. Lastly, we can show that 
$$
\calV_{it}^{-1/2}\left(f_t^{0\top} \left(\bF^{0\top} \bF^{0} \right)^{-1} \sum_{s=1}^{T} f_s^0 \epsilon_{is}  + \lambda_i^{0\top} \left(\bLambda^{0\top} \bLambda^0 \right)^{-1} \sum_{j=1}^{N} \lambda_j^0 \epsilon_{jt}\right) \conD \calN(0,1)
$$
by the same assertion as in the proof of Theorem 3 in \cite{bai2003inferential}. It completes the proof. $\square$

\subsection{Technical Lemmas}

\begin{lemma}\label{lem:rotation}
Let $\bfH^0 = \frac{1}{\sqrt{T}}((\bLambda^{0\top}\bLambda^0)^{1/2}\bfG (\bfD_r^{0})^{-1/2})^{-1}$ where $\bfG$ is an eigenvectors matrix of $\left( \frac{\bLambda^{0\top} \bLambda^0 }{N^\alpha} \right)^{1/2} \left( \frac{\bF^{0 \top} \bF^0 }{T} \right) \left( \frac{\bLambda^{0\top} \bLambda^0 }{N^\alpha} \right)^{1/2}$. Then, $\bLambda^0 = T^{-1/2}\bfY_r^0 (\bfH^0)^{-\top}$ and $\bF^0 = T^{1/2} \bfZ_r^0 \bfH^0$.
\end{lemma}

\noindent\textbf{Proof of Lemma \ref{lem:rotation}.} Let $\boldsymbol{\Theta} = \left(\frac{\bLambda^{0\top}\bLambda^0}{N^\alpha}\right)^{1/2}\left(\frac{\bF^{0\top} \bF^0}{T}\right)\left(\frac{\bLambda^{0\top}\bLambda^0}{N^\alpha}\right)^{1/2}$ and $\bfG$ be a $r \times r$ matrix whose columns are the eigenvectors of $\boldsymbol{\Theta}$ such that $\Delta_\Theta = \bfG^{\top} \boldsymbol{\Theta} \bfG$ is a descending order diagonal matrix of the eigenvalues of $\boldsymbol{\Theta}$. Define $\bfH_{\bLambda} = (\bLambda^{0\top} \bLambda^0)^{-1/2}\bfG$. Then, we have
\begin{align*}
(\bLambda^0 \bF^{0\top} \bF^0 \bLambda^{0\top}) \bLambda^0 \bfH_{\bLambda} &= \bLambda^0 (\bLambda^{0\top} \bLambda^0)^{-1/2} (\bLambda^{0\top} \bLambda^0)^{1/2} \bF^{0\top} \bF^0 (\bLambda^{0\top} \bLambda^0)^{1/2} (\bLambda^{0\top} \bLambda^0)^{1/2} \bfH_{\bLambda}\\
&= \bLambda^0 (\bLambda^{0\top} \bLambda^0)^{-1/2} \left[(\bLambda^{0\top} \bLambda^0)^{1/2} \bF^{0\top} \bF^0 (\bLambda^{0\top} \bLambda^0)^{1/2} \bfG \right]\\
&=\bLambda^0  (\bLambda^{0\top} \bLambda^0)^{-1/2} (N^\alpha T) \boldsymbol{\Theta} \bfG \\
&=  \bLambda^0  (\bLambda^{0\top} \bLambda^0)^{-1/2} \bfG (N^\alpha T) \Delta_\Theta \\
&= \bLambda^0  \bfH_{\bLambda}  (N^\alpha T) \Delta_\Theta.
\end{align*}
In addition, note that $(\bLambda^0  \bfH_{\bLambda})^\top \bLambda^0  \bfH_{\bLambda} = \bfH_{\bLambda}^\top \bLambda^{0\top} \bLambda^0  \bfH_{\bLambda} = \bfG^\top \bfG = I_r$. Therefore, the column of $\bLambda^0  \bfH_{\bLambda}$ are the eigenvectors of $\bLambda^0 \bF^{0\top} \bF^0 \bLambda^{0\top}$ and the left singular vectors of $\bM^0 = \bLambda^0 \bF^{0\top}$. So, $\bfU_r^0 = \bLambda^0  \bfH_{\bLambda}$ and $\bfY_r^0 = \bLambda^0  \bfH_{\bLambda}(\bfD_r^{0})^{1/2} = \sqrt{T} \bLambda^0  (\bfH^0)^{\top}$. Then, $\bfZ_r^0 = T^{-1/2} \bF^0 (\bfH^0)^{-1}$ follows from $\bfY_r^0 = \sqrt{T} \bLambda^0  (\bfH^0)^{\top}$. $\square$
\bigskip

\begin{lemma}\label{lem:limits}
(i) $\bfG \conP \calG^*$ where $\calG^*$ is the eigenvector of $\bSigma_{\bLambda}^{1/2} \bSigma_{\bF} \bSigma_{\bLambda}^{1/2}$ corresponding to the sign of $\bfG$.\\
(ii) $\frac{1}{N^{\alpha/2} T^{1/2}} \bfD_r^0 \conP \calD$ and $\frac{1}{N^{\alpha/2} T^{1/2}} \bfD_r \conP \calD$.\\
(iii) $ (N^{\alpha/4} T^{1/4}) \bfH^0 \conP \calD^{1/2} \calG^{*\top} \bSigma_{\bLambda}^{-1/2}   $.
\end{lemma}

\noindent\textbf{Proof of Lemma \ref{lem:limits}.} 
(i) Since $\boldsymbol{\Theta}$ converges to $\bSigma_{\bLambda}^{1/2} \bSigma_{\bF} \bSigma_{\bLambda}^{1/2}$ by Assumption A, the eigenvalues of $\boldsymbol{\Theta}$ will be distinct for large $N$ and $T$. So, the eigenvector matrix of $\boldsymbol{\Theta}$ is unique except that each column can be replaced by the negative of itself. As we can know in the proof of Lemma \ref{lem:rotation}, the sign of $\bfG$ is determined by the sign alignment between $\bfU_r^0$ and $\bLambda^0$. Then, by the eigenvector perturbation theory, there is a unique eigenvector of $\bSigma_{\bLambda}^{1/2} \bSigma_{\bF} \bSigma_{\bLambda}^{1/2}$, says, $\calG^*$, such that $||\bfG - \calG^*||= o_p(1)$ and the sign of $\calG^*$ corresponds to that of $\bfG$.\\
(ii) Note that the square roots of eigenvalues of $\boldsymbol{\Theta}$ are the same as the singular values of $\frac{1}{N^{\alpha/2} T^{1/2}} \bLambda^0 \bF^{0\top}$, that is, $\frac{1}{N^{\alpha/2} T^{1/2}} \bfD_r^0$. Then, by the continuity of eigenvalues, we have $\frac{1}{N^{\alpha/2} T^{1/2}} \bfD_r^0 \conP \calD$. In addition, note that
$$
\frac{1}{N^\alpha T} \bX \bX^\top = \frac{\bLambda^0 (\bF^{0\top} \bF^0)\bLambda^{0\top}}{N^\alpha T} + \frac{\bLambda^0\bF^{0\top}\bE^\top}{N^\alpha T} + \frac{\bE \bF^0 \bLambda^{0\top}}{N^\alpha T} + \frac{\bE \bE^\top}{N^\alpha T} =\frac{\bLambda^0 (\bF^{0\top} \bF^0)\bLambda^{0\top}}{N^\alpha T} + o_p(1)
$$
since $||\bE|| = O_p(\max\{\sqrt{N},\sqrt{T} \})$, $||\bLambda^0|| = O_p(N^{\alpha/2})$, and $||\bF^0|| = O_p(T^{1/2})$. By the matrix perturbation theorem, the $r$ largest eigenvalues of $\frac{1}{N^\alpha T} \bX \bX^\top$, that is, $\frac{1}{N^\alpha T} \bfD_r^2$, are determined by $\frac{\bLambda^0 (\bF^{0\top} \bF^0)\bLambda^{0\top}}{N^\alpha T}$ whose eigenvalues are $\frac{1}{N^\alpha T}(\bfD_r^{0})^2$. Hence, $\frac{1}{N^{\alpha/2} T^{1/2}} \bfD_r \conP \calD$.\\
(iii) Note that 
$$
(N^{\alpha/4} T^{1/4}) \bfH^0 = \left(  \frac{\bfD_r^0}{N^{\alpha/2} T^{1/2} }\right)^{1/2}\bfG^\top \left(\frac{\bLambda^{0\top}\bLambda^0}{N^\alpha} \right)^{-1/2} .
$$
Hence, the result follows from Assumption A, Lemmas \ref{lem:limits} (i), and (ii). $\square$
\bigskip

\begin{lemma}\label{lem:Olimits}
$\bfO \conP \calI_{sgn}$ where $\calI_{sgn}$ is a $r \times r$ diagonal matrix consisting of the diagonal elements of $\pm 1$ and the signs are determined by the sign alignment between $\bfU_r$ and $\bfU_r^0$.
\end{lemma}

\noindent\textbf{Proof of Lemma \ref{lem:Olimits}.} 
First, we derive the limit of $\bfU_r^{0\top} \bfU_r$. For simplicity, let $K_{NT} = N^{\alpha/2}T^{1/2}$. From the relations $\bX \bX^\top \bfU_r = \bfU_r \bfD_r^2$ and $\bX = \bfY_r^{0} \bfZ_r^{0\top} + \bE$, we have
\begin{align*}
&\left(\frac{\bfZ_r^{0\top}\bfZ_r^0}{K_{NT}}\right)^{1/2} \frac{\bfY_r^{0\top}}{\sqrt{K_{NT}}}\frac{\bX \bX^\top }{K_{NT}^2} \bfU_r = \left(\frac{\bfZ_r^{0\top}\bfZ_r^0}{K_{NT}}\right)^{1/2} \frac{\bfY_r^{0\top} \bfU_r}{\sqrt{K_{NT}}} \left(\frac{\bfD_r^2}{K_{NT}^2} \right),\\
&\left(\frac{\bfZ_r^{0\top}\bfZ_r^0}{K_{NT}}\right)^{1/2} \left(\frac{\bfY_r^{0\top}\bfY_r^0}{K_{NT}}\right) \left(\frac{\bfZ_r^{0\top}\bfZ_r^0}{K_{NT}}\right) \left(\frac{\bfY_r^{0\top} \bfU_r}{\sqrt{K_{NT}}}\right) 
+ d_{NT} 
= \left(\frac{\bfZ_r^{0\top}\bfZ_r^0}{K_{NT}}\right)^{1/2} \frac{\bfY_r^{0\top} \bfU_r}{\sqrt{K_{NT}}} \left(\frac{\bfD_r^2}{K_{NT}^2} \right)
\end{align*}
where
$$
d_{NT} = \left(\frac{\bfZ_r^{0\top}\bfZ_r^0}{K_{NT}}\right)^{1/2} 
\left[ \frac{\bfY_r^{0\top}\bfY_r^0}{K_{NT}} \frac{\bfZ_r^{0\top}}{\sqrt{K_{NT}}} \frac{\bE^{\top}}{K_{NT}} + \frac{\bfY_r^{0\top}}{\sqrt{K_{NT}}} \frac{\bE}{K_{NT}} \frac{\bfZ_r^{0}\bfY_r^{0\top}}{K_{NT}} + \frac{\bfY_r^{0\top}}{\sqrt{K_{NT}}} \frac{\bE}{K_{NT}} \frac{\bE^\top}{K_{NT}} 
\right] \bfU_r.
$$
Because $\norm{\frac{\bfY_r^{0\top}\bfY_r^0}{\psi_{\min}}},\norm{\frac{\bfZ_r^{0\top}\bfZ_r^0}{\psi_{\min}}} = O_p (\kappa)$, $\norm{\frac{\bfY_r^0}{\sqrt{\psi_{\min}}}},\norm{\frac{\bfZ_r^0}{\sqrt{\psi_{\min}}}} = O_p (\kappa^{1/2})$, $\norm{\frac{\bE}{\psi_{\min}}} = O_p \left( \frac{\max\{\sqrt{N}, \sqrt{T} \}}{\psi_{\min}}\right)$, $\psi_{\min} \asymp K_{NT}$, and $\kappa$ is bounded under our assumptions, we have $d_{NT} = o_p(1)$. Additionally, define
\begin{align*}
&B_{NT} = \left(\frac{\bfZ_r^{0\top}\bfZ_r^0}{ K_{NT}}\right)^{1/2} \left(\frac{\bfY_r^{0\top}\bfY_r^0}{ K_{NT}}\right) \left(\frac{\bfZ_r^{0\top}\bfZ_r^0}{ K_{NT}}\right)^{1/2} = \left( \frac{\bfD_r^0}{ K_{NT}} \right)^2,\\
&R_{NT} = \left(\frac{\bfZ_r^{0\top}\bfZ_r^0}{K_{NT}}\right)^{1/2} \frac{\bfY_r^{0\top}\bfU_r}{\sqrt{K_{NT}}} = \frac{\bfD_r^0}{K_{NT}} \bfU_r^{0\top} \bfU_r,\\
&V_{NT} = \frac{\bfD_r}{K_{NT}}.
\end{align*}
Then, since $R_{NT}$ is invertible by Claim \ref{clm:invertible}, we have
$$
[B_{NT} + d_{NT}R_{NT}^{-1}]R_{NT} = R_{NT} V_{NT}^2.
$$

\begin{claim}\label{clm:invertible}
$\bfU_r^{0\top} \bfU_r$ is invertible and $||(\bfU_r^{0\top} \bfU_r)^{-1}||$ is bounded with probability converging to 1.
\end{claim}

In addition, to normalize $R_{NT}$, define $\bar{R}_{NT} = R_{NT} V_{NT}^{*-1}$ where $V_{NT}^{*2}$ is a diagonal matrix consisting of the diagonal elements of $R_{NT}^\top R_{NT} = (\bfU_r^\top  \bfU_r^0)\left( \frac{\bfD_r^0}{K_{NT}}\right)^2 (\bfU_r^{0\top} \bfU_r)$. Then, we have
$$
[B_{NT} + d_{NT}R_{NT}^{-1}]\bar{R}_{NT} = \bar{R}_{NT} V_{NT}^2
$$
and $\bar{R}_{NT}$ is the eigenvector of $B_{NT} + d_{NT}R_{NT}^{-1}$. Because $B_{NT} \conP \calD^2$ by Lemma \ref{lem:limits} (ii) and $d_{NT}R_{NT}^{-1} = o_p(1)$, $B_{NT} + d_{NT}R_{NT}^{-1} \conP \calD^2$. Since the eigenvalues of $\calD^2$ are distinct, those of $B_{NT} + d_{NT}R_{NT}^{-1}$ will be distinct for large $N$ and $T$. So, the eigenvector of $B_{NT} + d_{NT}R_{NT}^{-1}$ is unique up to the column sign, and the sign of $\bar{R}_{NT}$ is determined by the sign alignment between $\bfU_r^0$ and $\bfU_r$. By the eigenvector perturbation theory, there is a unique eigenvector of $\calD^2$, says, $\calI_{sgn}$, such that $||\bar{R}_{NT} - \calI_{sgn}|| =o_p(1)$ and the sign of $\calI_{sgn}$ corresponds to that of $\bar{R}_{NT}$. In addition, since $\calI_{sgn}$ is the eigenvector of the diagonal matrix $\calD^2$, it is a $r \times r$ diagonal matrix consisting of the diagonal elements of $\pm 1$.

Note that $\bfU_r^{0\top} \bfU_r = \left( \frac{\bfD_r^0}{K_{NT}} \right)^{-1} \bar{R}_{NT} V^{*}_{NT}$. Then, By Lemma \ref{lem:limits} (ii) and Claim \ref{clm:Vstar} with the above result, we have
$$
\bfU_r^{0\top} \bfU_r \conP \calD^{-1} \calI_{sgn} \calD = \calI_{sgn}.
$$
Here, $\calI_{sgn}$ is the $r \times r$ diagonal matrix consisting of the diagonal elements of $\pm 1$ and the sign of these are determined by the sign alignment between $\bfU_r^{0}$ and $\bfU_r$.

\begin{claim}\label{clm:Vstar}
$R_{NT}^\top R_{NT} = (\bfU_r^\top \bfU_r^0)\left( \frac{\bfD_r^0}{K_{NT}}\right)^2 (\bfU_r^{0\top} \bfU_r) \conP \calD^2$.
\end{claim}

Lastly, we show that $\bfO \conP \calI_{sgn}$. Let $\bfQ = \argmin_{\bfR \in \calO^{r\times r}} \left\|
\begin{bmatrix}
\bfU_r \\
\bfV_r
\end{bmatrix}
\bfR - 
\begin{bmatrix}
\bfU_r^0 \\
\bfV_r^0
\end{bmatrix}
\right\|_F$. Then, as noted in the proof of Lemma \ref{lem:maintechlem}, by Lemma B.4 of \cite{chen2020nonconvex}, $||\bfO - \bfQ || = o_p(1)$. In addition, we have
$$
\bfU_r^\top \bfU_r^0 = \bfU_r^\top \bfU_r \bfQ + \bfU_r^\top (\bfU_r^0 - \bfU_r\bfQ) = \bfQ + \bfU_r^\top (\bfU_r^0 - \bfU_r\bfQ) .
$$
As noted in the proof of Lemma \ref{lem:maintechlem}, by Lemma B.2 of \cite{chen2020nonconvex}, we have $||\bfU_r^0 - \bfU_r\bfQ|| = O_p \left( \frac{ \max\{\sqrt{N},\sqrt{T}\}} {K_{NT}} \right) = o_p(1)$. Hence, we have
$$
\norm{\bfU_r^\top \bfU_r^0 - \bfQ} \leq \norm{\bfU_r} \norm{\bfU_r^0 - \bfU_r\bfQ} = o_p(1).
$$
Then, because $\norm{\bfO - \bfU_r^\top \bfU_r^0} = o_p(1)$, we have $\bfO \conP \calI_{sgn}$. \ \ $\square$
\bigskip

\begin{comment}
Define $A = \begin{bmatrix}
U_r^\top  &
V_r^\top 
\end{bmatrix} \begin{bmatrix}
\bfU_r^0 \\
\bfV_r^0
\end{bmatrix} = U_r^\top \bfU_r^0 + V_r^\top \bfV_r^0$. We know that $U_r^\top \bfU_r^0 \conP \calI_{sgn}$. In addition, by using the same method, we can derive $V_r^\top \bfV_r^0 \conP \calI_{sgn}$. Here, we use the fact that the sign alignment between $\bfU_r^0$ and $U_r$ is the same as that between $\bfV_r^0$ and $V_r$. Hence, $A \conP 2\calI_{sgn}$. In addition, by using the similar method to Claim \ref{clm:invertible}, we can show that $A$ is invertible. So, $\bfQ = A(A^\top A)^{-1/2}$ (e.g., Lemma 35 of \cite{ma2020implicit}). 
\end{comment}

\noindent\textbf{Proof of Claim \ref{clm:invertible}.} 
As noted above, by Lemma B.2 of \cite{chen2020nonconvex}, we have $||\bfU_r^0 - \bfU_r\bfQ|| = O_p \left( \frac{ \max\{\sqrt{N},\sqrt{T}\}} {K_{NT}} \right) = o_p(1)$. Hence, by Weyl's inequality, we have w.h.p., that
$$
\psi_{r}(\bfU_r^\top \bfU_r^0) \geq \psi_{r}(\bfQ) - \norm{\bfU_r^\top (\bfU_r^0 - \bfU_r\bfQ)} \geq \frac{1}{2}. \ \ \square
$$

\noindent\textbf{Proof of Claim \ref{clm:Vstar}.} 
We have by Lemma \ref{lem:limits} (ii) that
$$
\frac{\bfU_r^\top \bX \bX^\top \bfU_r}{K_{NT}^2} = \frac{\bfD_r^2}{K_{NT}^2} \conP \calD^2 .
$$
In addition, by the same token as in the proof of Lemma \ref{lem:limits} (ii), we can show that 
$$
R_{NT}^\top R_{NT} = \frac{\bfU_r^\top \bfY_r^0\bfZ_r^{0\top}\bfZ_r^0\bfY_r^{0\top}\bfU_r}{K_{NT}^2} = \frac{\bfU_r^\top \bX\bX^\top \bfU_r}{K_{NT}^2} + o_p(1).
$$
Hence, $R_{NT}^\top R_{NT} \conP \calD^2$. $\square$

\subsection{Proof of Lemma \ref{lem:example}}\label{sec:proof_example}

\paragraph{Proof of Lemma \ref{lem:example} (i).} Let $\epsilon_{it}$ be a MA(q) process with $q \leq C_1 (\log N)^\nu$ for some constant $C_1 > 0$. Then, we set $\delta = C_1 \lceil (\ln N)^\nu\rceil$. Let $a \geq 1$ be a natural number. Note that $\epsilon_{it}$ is a function of $(u_{i,t-q}, \dots, u_{it})$ and $\epsilon_{i,t + \delta + a}$ is a function of $(u_{i,t + \delta + a - q}, \dots, u_{i, t + \delta + a})$. Because $\delta - q \geq 0$, there is no intersection between $(u_{i,t-q}, \dots, u_{it})$ and $(u_{i,t + \delta + a - q}, \dots, u_{i, t + \delta + a})$. Hence, $\epsilon_{i,t + \delta + a}$ is independent of $\epsilon_{it}$. Similarly, $\epsilon_{i,t - \delta - a}$ is independent of $\epsilon_{it}$. Hence, $(\epsilon_{is})_{s \in \calN_\delta(t)^c}$ is independent of $\epsilon_{it}$ and $\bbE[\epsilon_{it}| (\epsilon_{is})_{s \in \calN_\delta(t)^c}] = \bbE[\epsilon_{it}]$. $\square$

\paragraph{Proof of Lemma \ref{lem:example} (ii).} It is instructive to start with the case of AR(1) process.

\begin{lemma}\label{lem:ar1}
For each $i \in [N]$, let $\epsilon_{it}$ be a stationary AR(1) process such that
$$
\epsilon_{it} = \phi_{(i)} \epsilon_{i,t-1} + u_{it}, \ \ \text{ where } \max_i \abs{\phi_{(i)}} < \vartheta < 1, \ \  u_{it} \sim i.i.d. \ \  \calN (0,\sigma_{u,i}^2).
$$
Define $\epsilon_{i,pre} = ( \epsilon_{i1}, \dots, \epsilon_{i,t-\delta-1} )$ and $\epsilon_{i,post} = (\epsilon_{i,t+\delta+1}, \dots , \epsilon_{iT}  )$. If $\delta = C \lceil \ln N \rceil$ for some constant $C > 0$, we have
$$
\max_i \bbE\left[ \bbE\left[\epsilon_{it} |\epsilon_{i,pre}, \epsilon_{i,post} \right]^2  \right] \lesssim N^{-1} .
$$
\end{lemma}

\noindent\textbf{Proof of Lemma \ref{lem:ar1}.} 

\noindent\textbf{Step 1.} First, we show that for each $i$,
$$
\bbE\left[\epsilon_{it} |\epsilon_{i,pre}, \epsilon_{i,post} \right] = \bbE\left[\epsilon_{it} |\epsilon_{i,t-\delta-1},\epsilon_{i,t+\delta+1} \right] .
$$
Fix $i = i_o$. For notational simplicity, let $\epsilon_{i_o,t} = \epsilon_{o,t}$, $\epsilon_{i_o,pre} = \epsilon_{o,pre}$, $\epsilon_{i_o,post} = \epsilon_{o,post}$, $u_{i_o,t} = u_{o,t}$, and $\phi_{(i_o)} = \phi_o$. Because, for all $s \geq 1$,
$$
\epsilon_{o,t+\delta+1+s} = \phi_{o}^s \epsilon_{o,t+\delta+1} + \sum_{k=0}^{s-1} \phi_{o}^k u_{o,t+\delta+1+s-k},
$$
conditioning on $\{\epsilon_{o,t+\delta +1}, \epsilon_{o,pre} \}$, $(\epsilon_{o,t+\delta+1+s})_{s \geq 1}$ is independent of $\epsilon_{o,t}$. Hence, we can say
\begin{gather}\label{eq:ar1_step1}
\bbE\left[\epsilon_{o,t} |\epsilon_{o,pre}, \epsilon_{o,post} \right]
= \bbE\left[\epsilon_{o,t} |\epsilon_{o,pre}, \epsilon_{o,t+\delta+1}, \epsilon_{o,t+\delta+2}, \dots , \epsilon_{o,T} \right]
= \bbE\left[\epsilon_{o,t} |\epsilon_{o,pre}, \epsilon_{o,t+\delta+1} \right] .
\end{gather}
In addition, because $\epsilon_{o,t} = \phi_{o}^{\delta+1} \epsilon_{o,t-\delta-1}  + \sum_{k=0}^{\delta} \phi_{o}^{k} u_{o,t-k}$, we have by Claim \ref{clm:ar1} that
\begin{align*}
\bbE\left[\epsilon_{o,t} |\epsilon_{o,pre}, \epsilon_{o,t+\delta+1} \right] 
&= \phi_{o}^{\delta+1} \bbE\left[\epsilon_{o,t-\delta-1} |\epsilon_{o,pre}, \epsilon_{o,t+\delta+1} \right] + \sum_{k=0}^{\delta} \phi_{o}^{k} \bbE\left[u_{o,t-k}| \epsilon_{o,pre}, \epsilon_{o,t+\delta+1} \right] \\
&= \phi_{o}^{\delta+1} \epsilon_{o,t-\delta-1} + \sum_{k=0}^{\delta} \phi_{o}^{k} \bbE\left[u_{o,t-k}| \epsilon_{o,t-\delta-1}, \epsilon_{o,t+\delta+1} \right].
\end{align*}

\begin{claim}\label{clm:ar1}
For $t - \delta \leq s \leq t$, we have
$\bbE\left[u_{o,s}| \epsilon_{o,pre}, \epsilon_{o,t+\delta+1} \right]  = \bbE\left[u_{o,s}| \epsilon_{o,t-\delta-1}, \epsilon_{o,t+\delta+1} \right]$.
\end{claim}

Hence, we know
\begin{align*}
 \bbE\left[\epsilon_{o,t} | \epsilon_{o,t-\delta-1}, \epsilon_{o,t+\delta+1} \right] &= \bbE\left[ \bbE\left[\epsilon_{o,t} |\epsilon_{o,pre}, \epsilon_{o,t+\delta+1} \right] | \epsilon_{o,t-\delta-1}, \epsilon_{o,t+\delta+1} \right]\\
&= \phi_{o}^{\delta+1} \epsilon_{o,t-\delta-1} + \sum_{k=0}^{\delta} \phi_{o}^{k} \bbE\left[u_{o,t-k}| \epsilon_{o,t-\delta-1}, \epsilon_{o,t+\delta+1} \right],   
\end{align*}
and so, $\bbE\left[\epsilon_{o,t} |\epsilon_{o,pre}, \epsilon_{o,t+\delta+1} \right] = \bbE\left[\epsilon_{o,t} | \epsilon_{o,t-\delta-1}, \epsilon_{o,t+\delta+1} \right]$. Then, with \eqref{eq:ar1_step1}, it shows that
$$
\bbE\left[\epsilon_{o,t} |\epsilon_{o,pre}, \epsilon_{o,post} \right] = \bbE\left[\epsilon_{o,t} |\epsilon_{o,t-\delta-1},\epsilon_{o,t+\delta+1} \right] .
$$
Therefore, for any $i$,
$$
\bbE\left[\epsilon_{it} |\epsilon_{i,pre}, \epsilon_{i,post} \right] = \bbE\left[\epsilon_{it} |\epsilon_{i,t-\delta-1},\epsilon_{i,t+\delta+1} \right] .
$$

\noindent\textbf{Proof of Claim \ref{clm:ar1}.} Note that
$$
\epsilon_{o,t+\delta+1} = \phi_o^{(2\delta+2)} \epsilon_{o,t-\delta-1} + Q, \ \ \text{  where  } Q = \sum_{k=0}^{2\delta+1} \phi_o^{k} u_{o,t+\delta+1-k}.
$$
Because $Q = \epsilon_{o,t+\delta+1} - \phi_o^{(2\delta+2)} \epsilon_{o,t-\delta-1} $ and $\epsilon_{o,t+\delta+1} = Q + \phi_o^{(2\delta+2)} \epsilon_{o,t-\delta-1}$, there are continuous functions $h_1$ and $h_2$ such that $h_1(\epsilon_{o,t+\delta+1},\epsilon_{o,pre}) = (Q,\epsilon_{o,pre})$ and $h_2(Q,\epsilon_{o,pre}) = (\epsilon_{o,t+\delta+1},\epsilon_{o,pre})$. Hence, the sigma-algebra generated by $(Q,\epsilon_{o,pre})$ is the same as that by $(\epsilon_{o,t+\delta+1},\epsilon_{o,pre})$. So, we have
$$
\bbE\left[u_{o,s}| \epsilon_{o,t+\delta+1},\epsilon_{o,pre}  \right] = 
\bbE\left[u_{o,s}| Q, \epsilon_{o,pre} \right].
$$
In addition, because $\epsilon_{o,pre}$ is a function of $\{u_{o,k} : k \leq t-\delta-1\}$ while $Q$ is a function of $\{u_{o,k} : t-\delta \leq k \leq t+\delta+1\}$, $\epsilon_{o,pre}$ is independent of $Q$ and $u_{o,s}$ where $t - \delta \leq s \leq t$. So, we have
$$
\bbE\left[u_{o,s}| Q, \epsilon_{o,pre} \right] = \bbE\left[u_{o,s}| Q \right] = \bbE\left[u_{o,s}| \epsilon_{o,t+\delta+1} - \phi_o^{(2\delta+2)} \epsilon_{o,t-\delta-1} \right] = f(\epsilon_{o,t-\delta-1},\epsilon_{o,t+\delta+1})
$$
for some function $f$. Then, since
$$
\bbE\left[ u_{o,s} | \epsilon_{o,t-\delta-1}, \epsilon_{o,t+\delta+1} \right] = \bbE\left[\bbE\left[u_{o,s}| \epsilon_{o,t+\delta+1},\epsilon_{o,pre} \right] |\epsilon_{o,t-\delta-1},\epsilon_{o,t+\delta+1} \right] = f(\epsilon_{o,t-\delta-1},\epsilon_{o,t+\delta+1}),
$$
we have $\bbE\left[u_{o,s}| \epsilon_{o,pre}, \epsilon_{o,t+\delta+1} \right]  = \bbE\left[u_{o,s}| \epsilon_{o,t-\delta-1}, \epsilon_{o,t+\delta+1} \right]$. $\square$

\bigskip

\noindent\textbf{Step 2.} Because $(u_{is})_{ s \leq T}$ is a multivariate normal, $(\epsilon_{i,t-\delta-1},\epsilon_{i,t+\delta+1},\epsilon_{it})$ has the following multivariate normal distribution:
\begin{align*}
\begin{pmatrix}
\epsilon_{i,t-\delta-1} \\
\epsilon_{i,t+\delta+1} \\
\epsilon_{it}
\end{pmatrix}
\sim \calN \left( \begin{pmatrix}
0 \\
0 \\
0
\end{pmatrix},
\frac{\sigma_{u,i}^2}{1 - \phi_{(i)}^2}
\begin{pmatrix}
1 & \phi_{(i)}^{2\delta + 2} & \phi_{(i)}^{\delta + 1} \\
\phi_{(i)}^{2\delta + 2} & 1 & \phi_{(i)}^{\delta + 1} \\
\phi_{(i)}^{\delta + 1} & \phi_{(i)}^{\delta + 1} & 1
\end{pmatrix}\right) .
\end{align*}
Here, the variance form comes from the auto-covariance of the stationary AR(1) process. Then, since the conditional expectation of the multivariate normal distribution has a linear form, we have
\begin{align*}
\bbE\left[\epsilon_{it} |\epsilon_{i,t-\delta-1},\epsilon_{i,t+\delta+1} \right] &= 
\begin{pmatrix}
\phi_{(i)}^{\delta + 1} & \phi_{(i)}^{\delta + 1} 
\end{pmatrix}
\begin{pmatrix}
1 & \phi_{(i)}^{2\delta + 2} \\
\phi_{(i)}^{2\delta + 2} & 1 
\end{pmatrix}^{-1}
\begin{pmatrix}
\epsilon_{t-\delta-1} \\
\epsilon_{t+\delta+1} 
\end{pmatrix} \\
& = \frac{1}{1 - \phi_{(i)}^{4\delta + 4}} \left( \phi_{(i)}^{\delta + 1} - \phi_{(i)}^{3\delta + 3}  \right) \left( \epsilon_{t-\delta-1} + \epsilon_{t+\delta+1}   \right)   \\
& =  \frac{\phi_{(i)}^{\delta + 1}}{1 + \phi_{(i)}^{2\delta + 2}} \left( \epsilon_{t-\delta-1} + \epsilon_{t+\delta+1}   \right) .
\end{align*}
Hence, we have for all $i$,
\begin{align*}
\bbE\left[ \bbE\left[\epsilon_{it} |\epsilon_{i,pre}, \epsilon_{i,post} \right]^2  \right] & =  \bbE\left[ \bbE\left[\epsilon_{it} |\epsilon_{i,t-\delta-1},\epsilon_{i,t+\delta+1} \right]^2 \right]\\
& = 2\sigma_{u,i}^2 \left( \frac{\phi_{(i)}^{\delta + 1}}{1 + \phi_{(i)}^{2\delta + 2}} \right)^2 \left( \frac{1}{1-\phi_{(i)}^2} + \frac{\phi_{(i)}^{\delta+1}}{1-\phi_{(i)}^2} \right)\\
& \lesssim \left( \frac{\phi_{(i)}^{\delta + 1}}{1 + \phi_{(i)}^{2\delta + 2}} \right)^2 \lesssim \phi_{(i)}^{2\delta} \lesssim \vartheta^{2\delta} .
\end{align*}
Note that $\delta = C \lceil \ln N \rceil \geq C \ln N$. Then, we have
\begin{align*}
\vartheta^{2\delta} = e^{2\delta \ln \vartheta} \leq e^{2 C \ln N \ln \vartheta} = \exp \left(\ln N^{(2C \ln \vartheta) } \right) = N^{(2C \ln \vartheta) } .
\end{align*}
In addition, if $C \geq \frac{1}{2\ln (\vartheta^{-1})}$, we have $2C \ln \vartheta \leq -1 $. Hence, when $C \geq \frac{1}{2\ln (\vartheta^{-1})}$, we have
$$
\max_i \bbE\left[ \bbE\left[\epsilon_{it} |\epsilon_{i,pre}, \epsilon_{i,post} \right]^2  \right] \lesssim N^{-1}. \ \ \square
$$
\bigskip

Then, we can generalize the above result to the case of AR(p) process.

\begin{lemma}\label{lem:arp}
For each $i \in [N]$, $\epsilon_{it}$ is a stationary AR(p) process such that
$$
\epsilon_{it} = \phi^{(i)}_1 \epsilon_{i,t-1} + \cdots + \phi^{(i)}_p \epsilon_{i,t-p} + u_{it}, \ \ \text{ where } u_{it} \sim i.i.d. \ \  \calN (0,\sigma_{u,i}^2),
$$
and there is a constant $0< \vartheta <1$ such that $\max_{1\leq i \leq N,1\leq k\leq p}\abs{\psi^{(i)}_k}< \vartheta $, where $( \psi^{(i)}_1,\dots,\psi^{(i)}_p )$ are the roots of the characteristic polynomial
$$
\psi^{p} - \phi_1^{(i)} \psi^{p-1} - \cdots - \phi^{(i)}_{p-1} \psi -  \phi^{(i)}_{p} = 0.
$$
Then, if $\delta = C \lceil \ln N \rceil$ for some large constant $C>0$, we have for sufficiently large $\delta$,
$$
\max_i \bbE\left[ \bbE\left[\epsilon_{it} |\epsilon_{i,pre}, \epsilon_{i,post} \right]^2  \right] \leq C N^{-1}
$$
for some constant $C>0$.
\end{lemma}

\noindent\textbf{Proof of Lemma \ref{lem:arp}.} 

\noindent\textbf{Step 1.} First, we show that for each $i$,
$$
\bbE\left[\epsilon_{it} |\epsilon_{i,pre}, \epsilon_{i,post} \right] = \bbE\left[\epsilon_{it} |\epsilon_{i,t-\delta-p},\dots,\epsilon_{i,t-\delta-1},\epsilon_{i,t+\delta+1},\dots,\epsilon_{i,t+\delta+p} \right] .
$$
Fix $i = i_o$. For notational simplicity, let $\epsilon_{i_o,t} = \epsilon_{o,t}$, $\epsilon_{i_o,pre} = \epsilon_{o,pre}$, $\epsilon_{i_o,post} = \epsilon_{o,post}$, $u_{i_o,t} = u_{o,t}$, and $\phi^{(i_o)}_{k} = \phi_{o,k}$ where $1 \leq k \leq p$. Note that, for all $s \geq 1$, there are constants $a_{1,s} , \cdots, a_{p,s}, b_0, \dots, b_{s-1}$ depending on $\{\phi_{o,1},\dots,\phi_{o,p}\}$ such that
$$
\epsilon_{o,t} = a_{1,s} \epsilon_{o,t-s} + \cdots + a_{p,s} \epsilon_{o,t-s-p+1} +  \sum_{k=0}^{s-1} b_k u_{o,t-k}.
$$
Then, because we have for all $s \geq 1$,
$$
\epsilon_{o,t+\delta+p+s} = a_{1,s} \epsilon_{o,t+\delta+p} + \cdots + a_{p,s} \epsilon_{o,t+\delta+1} +  \sum_{k=0}^{s-1} b_k u_{o,t+\delta+p+s-k},
$$
conditioning on $\{\epsilon_{o,t+\delta +1}, \dots, \epsilon_{o,t+\delta +p}, \epsilon_{o,pre} \}$, $(\epsilon_{o,t+\delta+p+s})_{s \geq 1}$ is independent of $\epsilon_{o,t}$. Hence, we can say
\begin{gather}\label{eq:arp_step1}
\bbE\left[\epsilon_{o,t} |\epsilon_{o,pre}, \epsilon_{o,post} \right]
= \bbE\left[\epsilon_{o,t} |\epsilon_{o,pre}, \epsilon_{o,t+\delta+1}, \epsilon_{o,t+\delta+2}, \dots , \epsilon_{o,T} \right]
= \bbE\left[\epsilon_{o,t} |\epsilon_{o,pre}, \epsilon_{o,t+\delta +1}, \dots, \epsilon_{o,t+\delta +p} \right] .
\end{gather}
In addition, because 
$$
\epsilon_{o,t} = a_{1,(\delta+1)} \epsilon_{o,t-\delta-1} + \cdots + a_{p,(\delta+1)} \epsilon_{o,t-\delta-p} +  \sum_{k=0}^{\delta } b_k u_{o,t-k},
$$
we have by Claim \ref{clm:arp} that
\begin{align*}
\bbE\left[\epsilon_{o,t} |\epsilon_{o,pre}, \epsilon_{o,t+\delta+1}, \dots, \epsilon_{o,t+\delta+p} \right] 
&=  \bbE\left[ a_{1,(\delta+1)} \epsilon_{o,t-\delta-1} + \cdots + a_{p,(\delta+1)} \epsilon_{o,t-\delta-p} |\epsilon_{o,pre}, \epsilon_{o,t+\delta+1}, \dots, \epsilon_{o,t+\delta+p} \right]\\
&\ \ + \sum_{k=0}^{\delta} b_k \bbE\left[u_{o,t-k}| \epsilon_{o,pre}, \epsilon_{o,t+\delta+1}, \dots, \epsilon_{o,t+\delta+p}  \right] \\
&=   a_{1,(\delta+1)} \epsilon_{o,t-\delta-1} + \cdots + a_{p,(\delta+1)} \epsilon_{o,t-\delta-p}\\
&\ \ + \sum_{k=0}^{\delta} b_k \bbE\left[u_{o,t-k}| \epsilon_{o,t-\delta-p},\dots,\epsilon_{o,t-\delta-1}, \epsilon_{o,t+\delta+1}, \dots, \epsilon_{o,t+\delta+p}  \right].
\end{align*}

\begin{claim}\label{clm:arp}
For $t - \delta \leq s \leq t$, we have
$$
\bbE\left[u_{o,s}| \epsilon_{o,pre}, \epsilon_{o,t+\delta+1}, \dots, \epsilon_{o,t+\delta+p} \right]  = \bbE\left[u_{o,s}| \epsilon_{o,t-\delta-p},\dots,\epsilon_{o,t-\delta-1}, \epsilon_{o,t+\delta+1}, \dots, \epsilon_{o,t+\delta+p}  \right].
$$
\end{claim}

Hence, we know
\begin{align*}
&\bbE\left[\epsilon_{o,t} | \epsilon_{o,t-\delta-p},\dots,\epsilon_{o,t-\delta-1}, \epsilon_{o,t+\delta+1}, \dots, \epsilon_{o,t+\delta+p} \right] \\
&= \bbE\left[ \bbE\left[\epsilon_{o,t} |\epsilon_{o,pre}, \epsilon_{o,t+\delta+1}, \dots, \epsilon_{o,t+\delta+p} \right] | \epsilon_{o,t-\delta-p},\dots,\epsilon_{o,t-\delta-1}, \epsilon_{o,t+\delta+1}, \dots, \epsilon_{o,t+\delta+p} \right] \\
& = a_{1,(\delta+1)} \epsilon_{o,t-\delta-1} + \cdots + a_{p,(\delta+1)} \epsilon_{o,t-\delta-p} + \sum_{k=0}^{\delta} b_k \bbE\left[u_{o,t-k}| \epsilon_{o,t-\delta-p},\dots,\epsilon_{o,t-\delta-1}, \epsilon_{o,t+\delta+1}, \dots, \epsilon_{o,t+\delta+p}  \right],    
\end{align*}
and so, 
$$
\bbE\left[\epsilon_{o,t} |\epsilon_{o,pre}, \epsilon_{o,t+\delta+1}, \dots, \epsilon_{i,t+\delta+p} \right]  = \bbE\left[\epsilon_{o,t} | \epsilon_{o,t-\delta-p},\dots,\epsilon_{o,t-\delta-1}, \epsilon_{o,t+\delta+1}, \dots, \epsilon_{o,t+\delta+p} \right].
$$
Then, with \eqref{eq:arp_step1}, it shows that
$$
\bbE\left[\epsilon_{o,t} |\epsilon_{o,pre}, \epsilon_{o,post} \right] = \bbE\left[\epsilon_{o,t} | \epsilon_{o,t-\delta-p},\dots,\epsilon_{o,t-\delta-1}, \epsilon_{o,t+\delta+1}, \dots, \epsilon_{o,t+\delta+p} \right].
$$
Therefore, for any $i$,
$$
\bbE\left[\epsilon_{it} |\epsilon_{i,pre}, \epsilon_{i,post} \right] = \bbE\left[\epsilon_{it} | \epsilon_{i,t-\delta-p},\dots,\epsilon_{i,t-\delta-1}, \epsilon_{i,t+\delta+1}, \dots, \epsilon_{i,t+\delta+p} \right] .
$$

\noindent\textbf{Proof of Claim \ref{clm:arp}.} Note that
\begin{align*}
&\epsilon_{o,t+\delta+1} =  a_{1,(2\delta+2)} \epsilon_{o,t-\delta-1} + \cdots + a_{p,(2\delta+2)} \epsilon_{o,t-\delta-p} + Q_1, \ \ \text{  where  } Q_1 = \sum_{k=0}^{2\delta + 1 } b_k u_{o,t+\delta+1-k},\\   
&\epsilon_{o,t+\delta+2} =  a_{1,(2\delta+3)} \epsilon_{o,t-\delta-1} + \cdots + a_{p,(2\delta+3)} \epsilon_{o,t-\delta-p} + Q_2, \ \ \text{  where  } Q_2 = \sum_{k=0}^{2\delta + 2 } b_k u_{o,t+\delta+2-k},\\  
&\qquad \qquad \qquad  \qquad \qquad \qquad \qquad \qquad \qquad \vdots \\
&\epsilon_{o,t+\delta+p} =  a_{1,(2\delta+p+1)} \epsilon_{o,t-\delta-1} + \cdots + a_{p,(2\delta+p+1)} \epsilon_{o,t-\delta-p} + Q_p, \ \ \text{  where  } Q_p = \sum_{k=0}^{2\delta + p } b_k u_{o,t+\delta+p-k}.
\end{align*}
Because for all $1\leq l \leq p$,
\begin{align*}
&Q_l = \epsilon_{o,t+\delta+l} - (a_{1,(2\delta+l+1)} \epsilon_{o,t-\delta-1} + \cdots + a_{l,(2\delta+l+1)} \epsilon_{o,t-\delta-p})  ,\\
&\epsilon_{o,t+\delta+l} = Q_l +  a_{1,(2\delta+l+1)} \epsilon_{o,t-\delta-1} + \cdots + a_{l,(2\delta+l+1)} \epsilon_{o,t-\delta-p}  ,
\end{align*}
there are continuous functions $h_{1}$ and $h_{2}$ such that $h_{1}(\epsilon_{o,t+\delta+1},\dots,\epsilon_{o,t+\delta+p},\epsilon_{o,pre}) = (Q_1,\dots, Q_p,\epsilon_{o,pre})$ and $h_{2}(Q_1,\dots, Q_p,\epsilon_{o,pre}) = (\epsilon_{o,t+\delta+1},\dots,\epsilon_{o,t+\delta+p},\epsilon_{o,pre})$. Hence, the sigma-algebra generated by $(Q_1,\dots, Q_p,\epsilon_{o,pre}) $ is the same as that by $(\epsilon_{o,t+\delta+1},\dots,\epsilon_{o,t+\delta+p},\epsilon_{o,pre})$. So, we have
$$
\bbE\left[u_{o,s}| \epsilon_{o,t+\delta+1},\dots,\epsilon_{o,t+\delta+p},\epsilon_{o,pre} \right] = 
\bbE\left[u_{o,s}| Q_1,\dots, Q_p,\epsilon_{o,pre} \right].
$$
In addition, because $\epsilon_{o,pre}$ is a function of $\{u_{o,k} : k \leq t-\delta-1\}$ while $Q_l$ is a function of $\{u_{o,k} : t-\delta \leq k \leq t+\delta+l\}$, $\epsilon_{o,pre}$ is independent of $(Q_1,\dots,Q_p)$ and $u_{o,s}$ where $t - \delta \leq s \leq t$. So, we have
$$
\bbE\left[u_{o,s}| Q_1,\dots, Q_p, \epsilon_{o,pre} \right] = \bbE\left[u_{o,s}| Q_1,\dots, Q_p \right] = f(\epsilon_{o,t-\delta-p},\dots,\epsilon_{o,t-\delta-1}, \epsilon_{o,t+\delta+1}, \dots, \epsilon_{o,t+\delta+p} )
$$
for some function $f$, since $Q_{l}$ is a function of $(\epsilon_{o,t-\delta-p},\dots,\epsilon_{o,t-\delta-1},\epsilon_{o,t+\delta+l})$. Then, because
\begin{align*}
 &\bbE\left[ u_{o,s} | \epsilon_{o,t-\delta-p},\dots,\epsilon_{o,t-\delta-1}, \epsilon_{o,t+\delta+1}, \dots, \epsilon_{o,t+\delta+p} \right] \\
 &= \bbE\left[\bbE\left[u_{o,s}| \epsilon_{o,pre} , \epsilon_{o,t+\delta+1}, \dots, \epsilon_{o,t+\delta+p} \right] |\epsilon_{o,t-\delta-p},\dots,\epsilon_{o,t-\delta-1}, \epsilon_{o,t+\delta+1}, \dots, \epsilon_{o,t+\delta+p} \right] \\
 &= f(\epsilon_{o,t-\delta-p},\dots,\epsilon_{o,t-\delta-1}, \epsilon_{o,t+\delta+1}, \dots, \epsilon_{o,t+\delta+p} ),   
\end{align*}
we have $
\bbE\left[u_{o,s}| \epsilon_{o,pre}, \epsilon_{o,t+\delta+1}, \dots, \epsilon_{o,t+\delta+p} \right]  = \bbE\left[u_{o,s}| \epsilon_{o,t-\delta-p},\dots,\epsilon_{o,t-\delta-1}, \epsilon_{o,t+\delta+1}, \dots, \epsilon_{o,t+\delta+p}  \right]$. $\square$

\bigskip

\noindent\textbf{Step 2.} Let $\calE^{(i)}_{x} = [  \epsilon_{i,t-\delta-1},\dots,\epsilon_{i,t-\delta-p}, \epsilon_{i,t+\delta+1}, \dots, \epsilon_{i,t+\delta+p} ]^\top$ and $\calE^{(i)}_{y} = [\epsilon_{it}]$. Because $( u_{is} )_{ s \leq T}$ is a multivariate normal, $(\calE^{(i)}_{x},\calE^{(i)}_{y})$ has the following multivariate normal distribution:
\begin{align*}
&\begin{pmatrix}
\calE_x^{(i)} \\
\calE_y^{(i)}
\end{pmatrix}
\sim \calN \left( 
\begin{pmatrix}
0 \\
0
\end{pmatrix},
\begin{pmatrix}
\Sigma_{xx}^{(i)} & \Sigma_{xy}^{(i)}  \\
\Sigma_{yx}^{(i)} & \Sigma_{yy}^{(i)}
\end{pmatrix}
\right) , \quad \text{where} \\
&\underbrace{\Sigma_{xx}^{(i)}}_{2p \times 2p} =
\begin{pmatrix}
\Sigma_{loc}^{(i)} & \Sigma_{int}^{(i)}  \\
\Sigma_{int}^{(i)} & \Sigma_{loc}^{(i)}
\end{pmatrix}, \ \
\underbrace{\Sigma_{loc}^{(i)}}_{p \times p} =
\begin{pmatrix}
\gamma^{(i)}_0 & \cdots & \gamma^{(i)}_{p-1}  \\
\vdots & \ddots & \vdots  \\
\gamma^{(i)}_{p-1} & \cdots & \gamma^{(i)}_{0}
\end{pmatrix}, \ \
\underbrace{\Sigma_{int}^{(i)}}_{p \times p} =
\begin{pmatrix}
\gamma^{(i)}_{(2\delta+2)} & \cdots & \gamma^{(i)}_{(2\delta+p+1)}  \\
\vdots & \ddots & \vdots  \\
\gamma^{(i)}_{(2\delta+p+1)} & \cdots & \gamma^{(i)}_{(2\delta+2p)}
\end{pmatrix},\\
&\underbrace{\Sigma_{yx}^{(i)}}_{1 \times 2p} =
\begin{pmatrix}
\gamma^{(i)}_{\delta + 1} & \cdots & \gamma^{(i)}_{\delta + p} &  \gamma^{(i)}_{\delta + 1} & \cdots & \gamma^{(i)}_{\delta + p}
\end{pmatrix},  \ \
\Sigma_{xy}^{(i)} = \Sigma_{yx}^{(i)\top}, \ \
\Sigma_{yy}^{(i)} = \gamma^{(i)}_{0}.
\end{align*}
Here, $\gamma^{(i)}_s$ is the auto-covariance such that $\gamma^{(i)}_s = \bbE[\epsilon_{it} \epsilon_{i,t-s}]$. Then, since the conditional expectation of the multivariate normal distribution has a linear form, we have
$$
\bbE\left[\epsilon_{it} | \epsilon_{i,t-\delta-p},\dots,\epsilon_{i,t-\delta-1}, \epsilon_{i,t+\delta+1}, \dots, \epsilon_{i,t+\delta+p} \right] =\bbE\left[\calE^{(i)}_y | \calE^{(i)}_x \right] = \Sigma^{(i)}_{yx} \Sigma_{xx}^{(i)-1} \calE^{(i)}_x .
$$
To derive the bound of $\norm{\Sigma_{xx}^{(i)-1}}$, we first get the bound of $\norm{\Sigma_{loc}^{(i)-1}}$. Since $\Sigma_{loc}^{(i)}$ is an auto-covariance matrix, by Proposition 1 of \cite{lin2008simplified}, there is a constant $C_{loc}>0$ such that, for all $i$, $\norm{\Sigma_{loc}^{(i)-1}} \leq C_{loc} \frac{1}{\sigma_{u,i}^2}$. In addition, since $p$ is finite, we have 
$\norm{\Sigma_{int}^{(i)}} \lesssim \norm{\Sigma_{int}^{(i)}}_\infty \lesssim \abs{\gamma^{(i)}_{(2\delta+2)}}$. Then, by Claim \ref{clm:autocovariance}, there is a constant $C>0$ such that
$$
\max_i \norm{\Sigma_{int}^{(i)}}^{2} \norm{\Sigma_{loc}^{(i)-1}}^2  \leq  C  \frac{\vartheta^{(4\delta+4)}}{(1 - \vartheta^2)^2}.
$$
Note that, because $\vartheta < 1$, there is $\delta_o>0$ such that for all $\delta > \delta_o$, we have $\max_i \norm{\Sigma_{int}^{(i)}}^{2} \norm{\Sigma_{loc}^{(i)-1}}^2 \leq 1/2$. Then, when $\delta > \delta_o$, we have for all $i$,
$$
\norm{\Sigma_{int}^{(i)} \Sigma_{loc}^{(i)-1}\Sigma_{int}^{(i)} } \leq  \norm{\Sigma_{int}^{(i)}}^{2} \norm{\Sigma_{loc}^{(i)-1}} \leq \frac{1}{2\norm{\Sigma_{loc}^{(i)-1}}} = \frac{1}{2} \psi_{p} \left(\Sigma_{loc}^{(i)} \right),
$$
and so, by Weyl's theorem, when $\delta > \delta_o$, we have for all $i$ that
\begin{align*}
\psi_{p} \left(\Sigma_{loc}^{(i)} - \Sigma_{int}^{(i)} \Sigma_{loc}^{(i)-1}\Sigma_{int}^{(i)}  \right) \geq  \psi_{p} \left(\Sigma_{loc}^{(i)} \right) -  \norm{\Sigma_{int}^{(i)} \Sigma_{loc}^{(i)-1}\Sigma_{int}^{(i)} } \geq \frac{1}{2} \psi_{p} \left(\Sigma_{loc}^{(i)} \right) \geq \frac{1}{2C_{loc}} \sigma_{u,i}^2  .
\end{align*}
Hence, when $\delta > \delta_o$, we have for all $i$, $ \norm{\left(\Sigma^{(i)} _{loc} - \Sigma^{(i)} _{int} \Sigma_{loc}^{(i)-1}\Sigma^{(i)} _{int}  \right)^{-1} } \leq \frac{2C_{loc}}{\sigma_{u,i}^2}$. Moreover, since $\Sigma_{xx}^{(i)-1}$ consists of the block matrices $\left(\Sigma_{loc}^{(i)} - \Sigma_{int}^{(i)} \Sigma_{loc}^{(i)-1}\Sigma_{int}^{(i)}  \right)^{-1}$ and $\left(\Sigma_{loc}^{(i)} - \Sigma_{int}^{(i)} \Sigma_{loc}^{(i)-1}\Sigma_{int}^{(i)}  \right)^{-1} \Sigma_{int}^{(i)} \Sigma_{loc}^{(i)-1}$, when $\delta > \delta_o$, we have for all $i$,
\begin{align*}
\norm{\Sigma_{xx}^{(i)-1}} &\leq 2 \norm{\left(\Sigma_{loc}^{(i)} - \Sigma_{int}^{(i)} \Sigma_{loc}^{(i)-1}\Sigma_{int}^{(i)}  \right)^{-1} } + 2\norm{\left(\Sigma_{loc}^{(i)} - \Sigma_{int}^{(i)} \Sigma_{loc}^{(i)-1}\Sigma_{int}^{(i)}  \right)^{-1} \Sigma_{int}^{(i)} \Sigma_{loc}^{(i)-1}} \\
& \leq 2\norm{\left(\Sigma_{loc}^{(i)} - \Sigma_{int}^{(i)} \Sigma_{loc}^{(i)-1}\Sigma_{int}^{(i)}  \right)^{-1} } \left(1 + \norm{\Sigma_{int}^{(i)}} \norm{\Sigma_{loc}^{(i)-1}} \right) \\
& \leq 2\norm{\left(\Sigma_{loc}^{(i)} - \Sigma_{int}^{(i)} \Sigma_{loc}^{(i)-1}\Sigma_{int}^{(i)}  \right)^{-1} } \left(1 + \frac{1}{\sqrt{2}} \right) \\
&\leq \frac{8C_{loc}}{\sigma_{u,i}^2}.
\end{align*}
In addition, since $p$ is finite, we have
$\norm{\Sigma_{yx}^{(i)}} \lesssim \norm{\Sigma_{yx}^{(i)}}_\infty \lesssim \abs{\gamma^{(i)}_{\delta+1}}$. Therefore, we have by Claim \ref{clm:autocovariance} that, whenever $\delta > \delta_o$,
\begin{align*}
\max_i \bbE[\bbE\left[\epsilon_{it} | \epsilon_{i,t-\delta-p},\dots,\epsilon_{i,t-\delta-1}, \epsilon_{i,t+\delta+1}, \dots, \epsilon_{i,t+\delta+p} \right]^2] &=
\max_i\Sigma_{yx}^{(i)} \Sigma_{xx}^{(i)-1} \bbE[ \calE_x^{(i)} \calE_x^{(i)\top}] \Sigma_{xx}^{(i)-1} \Sigma_{yx}^{(i)\top}  \\
&\leq \max_i \norm{\Sigma_{xx}^{(i)-1}}^2 \norm{\Sigma_{yx}^{(i)}}^2 \norm{\Sigma_{xx}^{(i)}} \\
 & \leq C_1 \max_i \sigma_{u,i}^2 \frac{\vartheta^{2\delta+2}}{(1 - \vartheta^2)^2} \leq C_2  \vartheta^{2\delta},
\end{align*}
for some constants $C_1, C_2 > 0$. Then, because $\delta = C \lceil \ln N \rceil \geq C \ln N $, we have
\begin{align*}
\vartheta^{2\delta} = e^{2\delta \ln \vartheta} \leq e^{2C \ln N \ln \vartheta} = \exp \left(\ln N^{(2C \ln \vartheta) } \right) = N^{(2C \ln \vartheta) } .
\end{align*}
In addition, if $C \geq \frac{1}{2\ln (\vartheta^{-1})}$, we have $2C \ln \vartheta \leq -1 $. Hence, when $C \geq \frac{1}{2\ln (\vartheta^{-1})}$, we have
$$
\max_i \bbE\left[ \bbE\left[\epsilon_{it} |\epsilon_{i,pre}, \epsilon_{i,post} \right]^2  \right] \lesssim N^{-1}. \ \ \square
$$

\begin{claim}\label{clm:autocovariance}
There is a constant $C > 0$ such that for all $i$,
$$
\abs{\gamma_s^{(i)}} \leq C \sigma_{u,i}^2 \frac{\vartheta^s}{1 - \vartheta^2}.
$$
\end{claim}
\noindent\textbf{Proof of Claim \ref{clm:autocovariance}.} Because $ \epsilon_{it} $ is a causal AR process, we have the representation
$\epsilon_{it} = \sum_{k=0}^\infty b^{(i)}_k u_{i,t-k}$ where $\abs{b^{(i)}_k} \leq c \vartheta^k$ for some bounded constant $c>0$. Then, because
$$
\gamma_s^{(i)} = \Cov(\epsilon_{it}, \epsilon_{i,t+s}) = \Cov\left(\sum_{k=0}^\infty b^{(i)}_k u_{i,t-k}, \sum_{k=0}^\infty b^{(i)}_k u_{i,t+s-k} \right) = \sigma_{u,i}^2 \sum_{k=0}^\infty b^{(i)}_k b^{(i)}_{k+s} ,
$$
we have
$$
\abs{\gamma_s^{(i)}} \leq c^2 \sigma_{u,i}^2 \sum_{k=0}^\infty \vartheta^k \vartheta^{k+s} \leq c^2 \sigma_{u,i}^2 \vartheta^{s} \sum_{k=0}^\infty \vartheta^{2k}  \leq C \sigma_{u,i}^2 \frac{\vartheta^s}{1 - \vartheta^2}. \ \ \square
$$

\end{document}

%% file: MyPack.tex
\usepackage{amssymb,amsfonts,mathrsfs,amsmath,amsthm,mathtools,bm,dsfont, thmtools,bbm}\allowdisplaybreaks
\usepackage[dvipsnames]{xcolor}
\usepackage{array, graphicx, graphics,float,multirow,tabularx,tikz,pgfplots, longtable,threeparttable}
\usepackage{setspace}
\usepackage[colorlinks,citecolor=blue,urlcolor=blue, runcolor=blue, filecolor=cyan]{hyperref}
\usepackage{footnotebackref}
\usepackage{bibentry, natbib} % reference
\usepackage{paralist}  % intermize and enumerate as paragraph
\usepackage{appendix} % appendix
\usepackage{mathrsfs}
\usepackage{enumitem}
\usepackage{authblk}
\usepackage{caption}
\usepackage{titlesec}
\usepackage{textcase,relsize}
\usepackage{booktabs}

\declaretheoremstyle[notefont=\bfseries,notebraces={}{},%
    headpunct={},postheadspace=1em]{mystyle}
\declaretheorem[style=mystyle,numbered=no,name=Assumption]{asmp-hand}
\declaretheorem[style=mystyle,numbered=no,name=Condition]{cond-hand}
\declaretheorem[style=mystyle,numbered=no,name=Example]{exmp-hand}

%% file: MyMath.tex
			\def \bfA {\mathbf{A}}
			\def \bfB {\mathbf{B}}
			
	\def \calD {\mathcal{D}}		\def \bfD {\mathbf{D}}
\def \bbE {\mathbb{E}}	\def \calE {\mathcal{E}}		
	\def \calF {\mathcal{F}}		
	\def \calG {\mathcal{G}}		\def \bfG {\mathbf{G}}
			\def \bfH {\mathbf{H}}
	\def \calI {\mathcal{I}}

	\def \calN {\mathcal{N}}		
	\def \calO {\mathcal{O}}		\def \bfO {\mathbf{O}}
			\def \bfP {\mathbf{P}}
	\def \calQ {\mathcal{Q}}		\def \bfQ {\mathbf{Q}}
\def \bbR {\mathbb{R}}	\def \calR {\mathcal{R}}		\def \bfR {\mathbf{R}}

			\def \bfU {\mathbf{U}}
	\def \calV {\mathcal{V}}		\def \bfV {\mathbf{V}}
			\def \bfW {\mathbf{W}}
			
			\def \bfY {\mathbf{Y}}
			\def \bfZ {\mathbf{Z}}

			\def \bfe {\mathbf{e}}

\def \Var {\text{Var}}
\def \Cov {\text{Cov}}

\def \conP {\overset{\bbP}\longrightarrow}
\def \conD {\overset{D}\longrightarrow}

\def \bfY {\bm{Y}}

\newcommand{\norm}[1]{\left\Vert#1\right\Vert}
\newcommand{\abs}[1]{\left\vert#1\right\vert}

\DeclareMathOperator*{\argmin}{arg\,min}